\documentclass[12pt]{article}
\newcommand{\blind}{1}

\usepackage{amsthm,amsmath,amsfonts,amssymb,statmath}
\usepackage[authoryear]{natbib}
\usepackage{graphicx,color}
\usepackage{booktabs,longtable,multirow,makecell}
\usepackage{enumitem}
\usepackage{bm,bbm,bigints}
\usepackage{setspace,etoolbox}
\usepackage{docmute}
\usepackage[colorlinks,citecolor=blue,urlcolor=blue]{hyperref}

\def\spacingset#1{\renewcommand{\baselinestretch}{2}\small\normalsize}
\spacingset{1}

\newcommand{\matrixspacingfix}{\renewcommand{\arraystretch}{0.53}\baselineskip=14.5pt\lineskip=2pt\lineskiplimit=2pt\relax}
\AtBeginEnvironment{bmatrix}{\matrixspacingfix}
\AtBeginEnvironment{pmatrix}{\matrixspacingfix}
\AtBeginEnvironment{vmatrix}{\matrixspacingfix}
\AtBeginEnvironment{Bmatrix}{\matrixspacingfix}
\AtBeginEnvironment{matrix}{\matrixspacingfix}
\AtBeginEnvironment{cases}{\matrixspacingfix}

\DeclareMathOperator{\sign}{sign}
\DeclareMathOperator{\Exp}{Exp}

\newcommand{\Dg}{\text{Diag}}
\newcommand{\Cov}{\mathrm{Cov}}
\newcommand{\Ee}{{\mathbb E}}
\newcommand{\indep}{\perp \!\!\! \perp}
\def\argmin{\mathop{\rm argmin}\limits}
\newcommand{\nbracket}[1]{\left( #1 \right)}
\newcommand{\cbracket}[1]{\left\{ #1 \right\}}
\newcommand{\rbracket}[1]{\left[ #1 \right]}
\newcommand{\norm}[1]{\left\vert \left\vert #1 \right\vert \right\vert}

\newcommand{\lasso}{\widehat{\beta}^{(\lambda)}_{n}}
\newcommand{\lassoE}{\widehat{\beta}^{(\lambda)}_{n,E}}
\newcommand{\lassoj}{\widehat{\beta}^{(\lambda)}_{n,j}}
\newcommand{\lP}{\mathcal{P}^{\lambda}}
\newcommand{\elasso}{\widehat{\beta}^{(\mathcal{P}, \lambda)}_{n}}
\newcommand{\elassoE}{\widehat{\beta}^{(\mathcal{P}, \lambda)}_{n,E}}
\newcommand{\elassoj}{\widehat{\beta}^{(\mathcal{P}, \lambda)}_{n,j}}
\newcommand{\targetE}{\beta^{E}_{n}}
\newcommand{\targetj}{\beta^{E\cdot j}_{n}}
\newcommand{\gammaej}{\gamma^{E\cdot j}_{n}}
\newcommand{\sigmaj}{(\sigma^{E\cdot j})^{2}}
\newcommand{\sigmaE}{\Sigma^{E\cdot j}}

\newcommand{\estE}{\widehat{b}^{E}_{n}}
\newcommand{\estj}{\widehat{b}^{E\cdot j}_{n}}
\newcommand{\gammaj}{\widehat{g}^{E\cdot j}_{n}}
\newcommand{\extracond}{\widehat{V}^{E\cdot j}_{n}}
\newcommand{\unicond}{\widehat{U}^{E\cdot j}_{n}}
\def\Figref#1{Figure~\ref{#1}}
\def\Thmref#1{Theorem~\ref{#1}}

\newtheorem{proposition}{Proposition}[section]
\newtheorem{theorem}{Theorem}[section]
\newtheorem{lemma}[theorem]{Lemma}
\newtheorem{corollary}{Corollary}
\newtheorem{remark}{Remark}
\newtheorem{assumption}{Assumption}

\begin{document}

\let\arxivbibliography\bibliography
\let\arxivbibliographystyle\bibliographystyle
\renewcommand{\bibliography}[1]{}
\renewcommand{\bibliographystyle}[1]{}

\date{}


\if1\blind
{
  \title{\bf Selective Inference for Time-Varying Moderated Effects}
  \author{Soham Bakshi \\
    Department of Statistics,
		University of Michigan,
         MI, USA.\\
          and \\
          Walter Dempsey\\
    Department of Biostatistics,
		University of Michigan,
         MI, USA.\\
    and  \\
    Snigdha Panigrahi \\
    Department of Statistics,
		University of Michigan,
         MI, USA.\\}
  \maketitle
} \fi

\if0\blind
{
  \title{\bf Selective Inference for Time-Varying Moderated Effects}
  \maketitle
} \fi

\begin{abstract}
Causal effect moderation investigates how the effect of interventions (or treatments) on outcome variables changes based on observed characteristics of individuals, known as potential effect moderators. With advances in data collection, datasets containing many observed features as potential moderators have become increasingly common. 
High-dimensional analyses often lack interpretability, with important moderators masked by noise, while low-dimensional, marginal analyses yield many false positives due to strong correlations with true moderators. 
In this paper, we propose a two-step method for selective inference on time-varying causal effect moderation that addresses the limitations of both high-dimensional and marginal analyses.
Our method first selects a relatively smaller, more interpretable model to estimate a linear causal effect moderation using a Gaussian randomization approach. 
We then condition on the selection event to construct a pivot, enabling uniformly asymptotic semi-parametric inference in the selected model. 
Our numerical results show that our method achieves valid coverage rates, even when existing conditional methods and common sample splitting techniques fail. 
Moreover, our method yields shorter, bounded intervals, unlike existing methods that may produce infinitely long intervals.
\end{abstract}

\noindent%

\spacingset{1.9}

\section{Introduction}
\label{sec:intro}

Causal effect moderation investigates how the effect of an intervention on an outcome varies with observed individual characteristics, known as potential effect moderators. Our motivating example is VALENTINE, a mobile-health (mHealth) trial in which individuals in a cardiac rehabilitation program were repeatedly randomized to receive either a digital intervention (e.g., a push notification) promoting physical activity or nothing, with step count in the hour after each treatment occasion as the outcome. The scientific goal is to understand how features such as individual traits, external context, and past treatment responses moderate the time-varying effect of these interventions.
Current methods consider low-dimensional moderation analysis, in which the moderators are fixed beforehand and effects are marginal over observed and unobserved variables outside that set. Advances in data acquisition increasingly yield many candidate moderators — in VALENTINE, individual traits (demographics, enrollment time, activity summaries), external context (time and day of week, weather, location type), and past treatment responses — so researchers must identify the important ones from a large pool, a high-dimensional moderation problem. Penalized regression such as the lasso offers one route to selecting relevant moderators.

While such a data-driven approach yields an interpretable analysis, naive inference on the selected moderators produces overly optimistic p-values and confidence intervals narrower than warranted.
Conditional selective inference provides a promising framework for valid inference after model selection; first introduced for the lasso by \cite{Lee_2016}, termed the polyhedral method, it uses a truncated normal distribution for exact inference.
Suppose the lasso selects a set of variables $\widehat{E}$. 
Our goal is then inference on the post-selection parameters $\left\{\beta^{\widehat{E}\cdot j}: j \in \widehat{E}\right\}$,
which represent the effects of the selected variables.
A selection-adjusted distribution is obtained by conditioning on the selection event $\{\widehat{E}=E\}$, where $E$ is the realized value of the (random) set $\widehat{E}$ on the observed samples. 
This conditional distribution enables the construction of selective confidence intervals $\{\widehat{C}^{\widehat{E}\cdot j} : j \in \widehat{E}\}$, where $\widehat{C}^{\widehat{E}\cdot j}$ is the interval constructed for $\beta^{\widehat{E}\cdot j}$.
This conditioning yields valid inference for each individual selected coefficient:
\begin{equation}
\label{cond:guarantee}
\mathbb{P}_{\mathbb{F}_n}\left[\beta^{\widehat{E}\cdot j} \in \widehat{C}^{\widehat{E}\cdot j} \; \Big\lvert \; \widehat{E}=E \right] \geq 1-\alpha.
\end{equation}

As shown by \cite{Lee_2016} (rederived in Lemma~\ref{lemma:fcr}, Supplement~\ref{app:FCR}), it also controls the false coverage rate (FCR), defined in \cite{benjamini2005false} as
\begin{equation}
\label{eqn:FCR}
\text{FCR}= \mathbb{E}\left[\dfrac{\Big|\left\{j\in \widehat{E}: \beta^{\widehat{E}\cdot j} \notin \widehat{C}^{\widehat{E}\cdot j}\right\}\Big|}{\max(|\widehat{E}|,1)} \right],
\end{equation}
representing the expected proportion of miscoverage for the $|\widehat{E}|$ post-selection parameters.

Building on this polyhedral method, initially proposed for a normal response, \cite{zhao2021selective} developed asymptotic selective inference for effect-modification problems. However, this approach can fail to remain valid for certain parameter values, producing infinitely long intervals~\citep{kivaranovic2019length}. Our method (`SI') instead adds independent Gaussian noise to the penalized estimation objective, as done in the line of work \cite{panigrahi2023approximate, panigrahi2023integrative}. As with data splitting, the added randomization allows the analyst to trade off information used for selection against information retained for inference. Unlike data splitting, however, our method conditions on substantially less information and performs inference using the full data set, yielding narrower intervals.

\begin{figure}[ht]
\centering
\includegraphics[width=\textwidth]{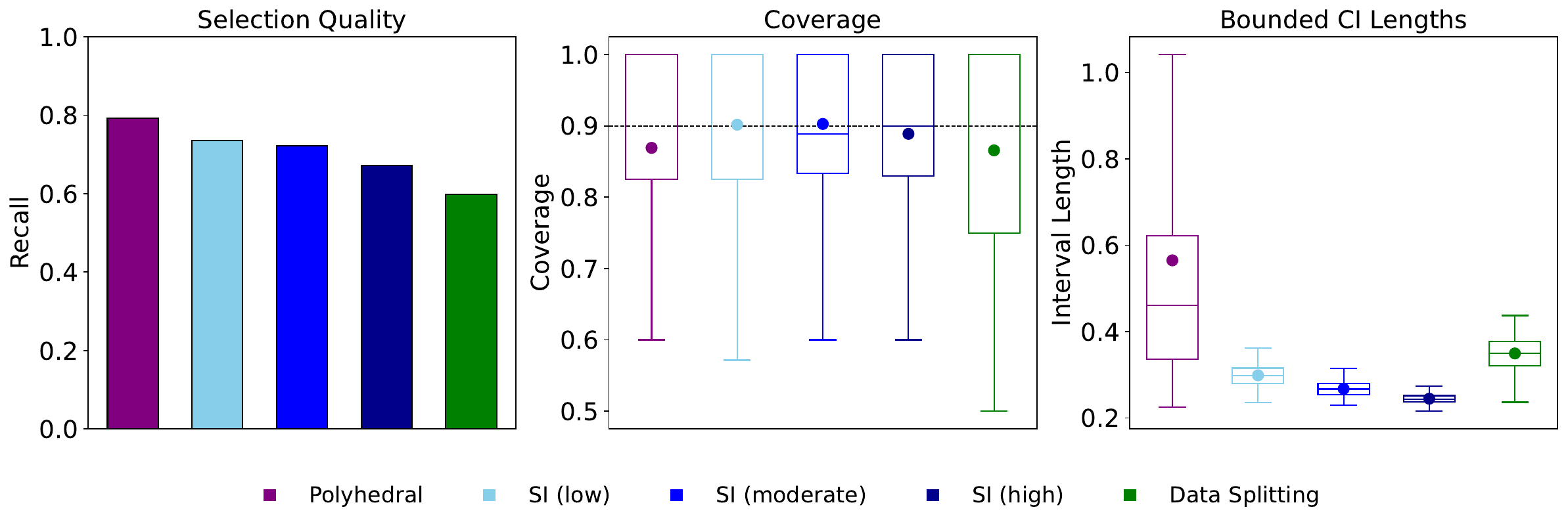}
\caption{Comparison of data splitting, the polyhedral method \citep{zhao2021selective}, and the proposed selective-inference method (`SI') under low, moderate, and high randomization  (see Section~\ref{sec:sim} for details of the full simulation setup). \emph{Left}: recall (proportion of true signals selected). \emph{Center}: empirical coverage rate (1-empirical FCR) (dashed line: nominal level of $0.9$; dots: mean coverage) over $500$ Monte-Carlo replications. \emph{Right}: average lengths of the bounded confidence intervals. `SI' attains nominal coverage at all randomization levels while always yielding bounded intervals that are substantially narrower than both the polyhedral method and data splitting. In contrast, the polyhedral method not only undercovers, but also yields infinitely long intervals in \textbf{\textcolor{red}{12\%}} of the replications (omitted from the length panel). Data splitting yields lower recall than even `SI (high)' and achieves coverage slightly below the target level.}
\label{plot1}
\end{figure}

Figure~\ref{plot1} reports comparison of the proposed selective-inference method (`SI') under low, moderate, and high randomization with data splitting, and the polyhedral method \citep{zhao2021selective}. With a small amount of randomization, selection with randomized lasso closely matches the standard lasso in recall, whereas higher levels of randomization result in lower recall. We use the low level of randomization in all the experiments in this work to ensure that improved inferential performance of our method does not come at the expense of selection quality, particularly because signals missed during selection cannot be recovered through post-selection inference.

Relative to prior work by \cite{panigrahi2023carving} and \cite{panigrahi2024exact},  our approach enables uniformly asymptotic semi-parametric inference for a general class of M-estimators with selected variables, valid under any zero-mean Gaussian randomization scheme with arbitrary covariance. \cite{wang2025asymptotically} develop a selective inference method for quantile regression, a specific instance of M-estimation in which the subgradient of the loss function is bounded. The problems studied in these earlier papers can thus be viewed as special cases of the more general theory developed here; summarized in Table~\ref{tab:comparison}.


\renewcommand{\arraystretch}{1.4}
\begin{table}[h!]
\centering
\begin{tabular}{lccc}
\toprule
 & \makecell{Asymptotic \\ theory} & \makecell{M-estimators\\ post selection} & \makecell{General Gaussian\\ randomization} \\
\midrule
Exact~\citep{panigrahi2024exact}   & $\times$ & $\times$ & $\checkmark$ \\
Carving~\citep{panigrahi2023carving} & $\checkmark$ & $\times$ & $\times$ \\
SI (Proposed Method)     & $\checkmark$ & $\checkmark$ & $\checkmark$ \\
\bottomrule
\end{tabular}
\caption{Comparison with existing works}\label{tab:comparison}
\end{table}

\section{Preliminaries}
\label{sec:preliminaries}


\paragraph{Notations and framework}
\label{sub:notations}

Consider $n$ i.i.d.\ samples, where sample $i \in [n]$ contains longitudinal observations $\{X_{i,t}, A_{i,t}, Y_{i,t}\}_{t\in[T]}$ over decision times $t\in[T]$, where $[m] := \{1,\ldots,m\}$ for any positive integer $m$. Here $Y_{i,t} \in \mathbb{R}$ is the response, observed just after the treatment at time $t$; $A_{i,t} \in \{0,1\}$ is the binary treatment; and $X_{i,t} \in \mathbb{R}^{p}$ collects covariates measured in $(t-1,t]$, prior to treatment assignment.
The overbar denotes a variable's history through time, e.g., $\bar{A}_{i,t}=\{A_{i,1},\ldots, A_{i,t}\}$, so the complete history $H_{i,t} = \{\bar{X}_{i,t},\bar{A}_{i,t-1},\bar{Y}_{i,t-1}\}$ is all information prior to the $t$-th treatment.
The treatment may depend on $H_{i,t}$ through known randomization probabilities $\mathbf{p}_{i} = \left\{p_{i,t}\left(A_{i,t} \mid H_{i,t}\right)\right\}_{t=1}^T$.
This is a typical set-up for examining time-varying causal effect moderation in micro-randomized trials (MRTs), particularly relevant in mHealth applications.

We adopt the standard potential outcomes framework~\citep{potentialoutcome}.
For the $i$-th individual, let $X_{i,t}(\bar{a}_{i,t-1})$ be the potential information observed under the treatment sequence $\bar{a}_{i,t-1} \in \{0,1\}^{\otimes (t-1)}$, and $Y_{i,t}(\bar{a}_{i,t-1}, a_{i,t})$ the potential outcome at time $t$ under $\{\bar{a}_{i, t-1}, a_{i,t}\}$; write $H_{i,t}(\bar{a}_{i,t-1})$ for the corresponding potential history.
Let $M_{i,t}(\bar{a}_{t-1})$ be a vector of deterministic summaries of $H_{i,t}(\bar{a}_{t-1})$, which we consider as the potential time-varying effect moderators. In the rest of the paper, we assume that the potential outcomes are i.i.d.~over individuals, and define the causal moderated effect on $Y_{i,t}$ as:
\begin{equation}
\beta^*(t ; m)=\mathbb{E}_{\mathbf{p}}\left[Y_{i,t}\left(\bar{A}_{i,t-1}, 1\right)-Y_{i,t}\left(\bar{A}_{i,t-1}, 0\right) \mid M_{i,t}\left(\bar{A}_{i,t-1}\right)=m\right].
\label{defn:causal:excursion}
\end{equation}
The expectation in this estimand is taken with respect to the joint distribution of the treatment sequence $\bar{A}_{i,t-1}=\left\{A_{i,1}, A_{i,2}, \ldots, A_{i,t-1}\right\}$.
This is emphasized through the subscript $\mathbf{p} = \left\{p_t\left(A_{i,t} \mid H_{i,t}\right)\right\}_{t=1}^T$, which is the full sequence of randomization probabilities in the MRT.
The causal estimand in \eqref{defn:causal:excursion}, first introduced in \cite{MRT},
has been termed a causal excursion effect in the literature.

Under standard causal inference assumptions of consistency, positivity, and sequential ignorability~\citep{robins97} (formally stated in Supplement~\ref{app:causalassumptions}), the causal excursion effect can be expressed as
\begin{equation}
\label{defn:cee:observed}
\beta^*(t ; m)=\mathbb{E}\left[\mathbb{E}_{\mathbf{p}}\left[Y_{i,t} \mid A_{i,t}=1, H_{i,t}\right]-\mathbb{E}_{\mathbf{p}}\left[Y_{i,t}  \mid A_{i,t}=0, H_{i,t}\right] \mid M_{i,t}=m\right],
\end{equation}
where the inner expectations are over the outcome conditional on the history $H_{i,t}$ and current treatment~$A_{i,t}$, and the outer expectation marginalizes over the history except for those variables included in~$M_{i,t}$.

\subsection{Excursion effect estimation}
We assume a linear working model for the causal excursion effect, i.e.,
$\beta^*(t ; m)= f_t(m)^{\top} \beta$, where $f_t(m) \in \mathbb{R}^p$ is a feature vector constructed from the effect moderators $M_{t} =m$ and time $t$---typically linear and non-linear transformations of each moderator and their interactions, including with time. A consistent estimator of $\beta$ follows from minimizing a weighted, centered least-squares (WCLS) objective~\citep{MRT, walterMRT}.

Following~\cite{shi2023metalearning}, we eliminate the nuisance parameters by
partitioning individuals into two independent parts: the first estimates the
outcome model $\mu_t(H_{i,t},a):=\mathbb{E}[Y_{i,t}\mid H_{i,t},A_{i,t}=a]$, with
estimate $\widehat\mu_t(H_{i,t},a)$. Throughout, $\widetilde p_t(a\mid m)\in(0,1)$ denotes an
analyst-specified reference treatment distribution depending only on the
moderator summary $M_{i,t}=m$, satisfying 
$\sum_{a}\widetilde p_t(a\mid m)=1$; it is used for treatment centering and as the
numerator of the stabilized weight $W_{i,t}$, and is not the full-history
propensity $p_t(a\mid H_{i,t})$. Setting
$\widehat\mu_t(H_{i,t})=\sum_{a}\widetilde p_t(a\mid M_{i,t})\,\widehat\mu_t(H_{i,t},a)$ and
$W_{i,t}:=\widetilde p_t(A_{i,t}\mid M_{i,t})/p_t(A_{i,t}\mid H_{i,t})$, the causal
estimate is obtained on the remaining data by
\begin{equation}
\label{wcls}
\widehat{\beta}
=
\underset{b}{\argmin}
\sum_{i=1}^{n}\sum_{t=1}^{T}
\nbracket{Y_{i,t}-\widehat{\mu}_{t}(H_{i,t})
-\nbracket{A_{i,t}-\widetilde{p}_t(1\mid M_{i,t})}f_{t}(M_{i,t})^{\top}b}^{2}W_{i,t}.
\end{equation}
The roles of the two parts may be reversed so that~\eqref{wcls} uses all data, or
extended to $K$ folds; we use the two-stage version for simplicity.


\paragraph{Why a linear model?} With a large number of moderators $f_t(m)\in\mathbb{R}^p$, flexible machine-learning models may improve prediction but offer limited interpretability and uncertainty quantification. High-dimensional linear models are more transparent but can still mask important moderators with noise covariates, while the popular alternative, one-at-a-time marginal analyses can produce false positives when moderators are strongly correlated.

\section{Two-step method for excursion effect estimation}
\label{sec:twostep}

We now consider a two-step method: first use lasso-penalized regression \citep{tibshirani1996regression} to select relevant moderators, $\widehat E\left(\{ \bar X_{i,T}, \bar A_{i,T}, \bar Y_{i,T} \}_{i=1}^n\right) \subseteq [p]$, by solving
\[\underset{\beta}{\text{minimize}} \left\{\frac{1}{\sqrt{n}}\sum_{i=1}^{n} \sum_{t=1}^{T}\nbracket{Y_{i,t} - \widehat{\mu}_{t}(H_{i,t}) - \nbracket{A_{i,t}-\widetilde{p}(1|M_{i,t}) }f_{t}(M_{i,t})^{\top}\beta}^{2}W_{i,t} + \lambda\|\beta\|_1 \right\}.\]
We select the moderators in $f_{t}(M_{i,t})\in \mathbb{R}^{p}$ with nonzero lasso coefficients. We denote the identified set as $\{\widehat{E}\left(\{ \bar X_{i,T}, \bar A_{i,T}, \bar Y_{i,T} \}_{i=1}^n\right)=E\}$, and $f_t^E(M_{i,t}) \in \mathbb{R}^{|E|}$ the restriction of the moderators to that set. We then consider the working model 
$\beta^{E}(t;m) = \left(f^{E}_t(m)\right)^{\top}  \beta^{E}.$
This approach yields a simpler, interpretable low-dimensional model, alleviating limitations of both high-dimensional and marginal analyses. Inference for $\beta^E$ requires adjustments for the data-driven selection procedure.
For example, to estimate the CATE, \cite{zhao2021selective} applied the lasso for feature selection and a polyhedral approach~\citep{Lee_2016} to adjust for selection.
 

\subsection{Nuisance parameters}
\label{sub:nuisance}

A challenge in causal inference is that estimation error in outcome and treatment models complicates both the model selection event and the pivotal statistics for selective inference~\citep{zhao2021selective}. In an MRT, the treatment probabilities are known by design; so no treatment-model estimation is needed. To extend the framework to observational studies where propensity scores are unknown and must be estimated, we now state the assumptions required when the randomization probabilities are not known.

\begin{assumption} 
\label{assumption:consistency}
Let~$\widehat{p}(a|h)$ denote the estimated randomization probability and $
    \|g_t(\cdot,a)\|_{2,n}^2
    =
    \mathbb P_n\!\left[g_t(H_t,a)^2\right]
    =
    \frac1n\sum_{i=1}^n g_t(H_{i,t},a)^2$ denote the empirical \(L^2\)-norm of a scalar-valued function \(g_t(H_t,a)\). We require the following two assumptions:
\ 1. Accuracy of treatment model: $\sum_{t=1}^T \sum_{a \in \{ 0,1 \}} \| \widehat{p}_t (a | H_t) - p_t (a | H_t) \|_{2,n}^2 = o_p (n^{-1/2}).$  \ 2. Boundedness: each term in~$f_t (M_t)$ and the treatment effect~$\beta(t;m)$ are uniformly bounded.
\end{assumption}
    
These assumptions are analogous to Assumptions 2 and 3 in~\cite{zhao2021selective}. Standard causal assumptions stated in Section~\ref{sub:notations} and Assumption~\ref{assumption:consistency} ensure the projection with estimated propensity weights will converge to the projection with the known treatment propensity weights at~$o_p (n^{-1/2})$ rate.  To establish the asymptotic distribution under fixed set~$E$, as discussed in~\cite{shi2023metalearning} we require the following assumption.
\begin{assumption}[Accuracy of outcome model] 
\label{eq:outcomeaccuracy}
Let~$\mu_t(H_t, A_t) := \mathbb{E}[Y_{t+1} \mid H_t, A_t]$, then we assume the outcome model~$\widehat{\mu}_t$ satisfies
$\sum_{t=1}^T \sum_{a \in \{ 0,1 \}} \| \widehat{\mu}_t (H_t, a) - \mu_t (H_t, a)  \|_{2,n} = o_p (1)$ and 
    $
    \widehat{B} = \sum_{t=1}^T \sum_{a \in \{ 0,1 \}} \| \widehat{p}_t (a | H_t) - p_t (a | H_t) \|_{2,n} \| \widehat{\mu}_t (H_t, a) - \mu_t (H_t, a)  \|_{2,n} = o_p (n^{-1/2}).
    $
\end{assumption}

The condition on~$\widehat{B}$ is doubly robust~\citep{chernozhukov2018double}: causal effect estimates remain consistent if at least one nuisance parameter is consistent.  
This holds under correctly specified outcome or treatment models and can also hold for structured machine-learning nuisance estimators, such as the lasso or random forests~\citep{chernozhukov2018double}. In a randomized experiment, if the randomization probabilities are known, setting $\widehat{p}_t=p_t$ makes treatment-model error zero; if a parametric treatment mechanism is estimated at $O_p(n^{-1/2})$ rate, the outcome model need only be consistent. Under Assumptions~\ref{assumption:consistency} and~\ref{eq:outcomeaccuracy}, \eqref{wcls} leverages Robinson's transformation and Neyman orthogonality to ensure consistency.  
If the treatment model is misspecified, however, the criteria will not have the correct projection interpretation.
\cite{shi2023metalearning} provide an alternative criteria (the DR-WCLS) that guarantees the projection has the desired weights~$\widetilde{p}_t(1|M_t) (1- \widetilde{p}_t(1|M_t))$ under Assumption~\ref{eq:outcomeaccuracy}.  
Our framework also accommodates DR-WCLS and cross-fitting \citep{newey2018crossfitting}: cross-fitted nuisances enter as fixed plug-ins, with moderator selection and selective inference still performed once on the full sample. For clarity, we use a single split and hold the estimated nuisances fixed during selection.


\subsection{A general M-estimation framework for selective inference}
\label{sub:generalnotation}

We recast the two-step procedure of Section~\ref{sec:twostep} in a
generic notational framework, so that our method applies more generally to M-estimation problems beyond WCLS. 
While WCLS serves as the running example in our paper, extensions to more general M-estimators, including those for binary outcomes based on relative-risk models, as well as to other penalized selection procedures are provided in Supplement~\ref{app:extension}.

With a slight abuse of notation, let $\{Y_i,X_i\}_{i\in[n]}$ be $n$ i.i.d.\ observations
from an unknown distribution $\mathbb{F}_n$. We assume
that the target causal effect is consistently estimated by the $M$-estimator obtained by minimizing  a user-specified loss $\psi(\cdot\,;\cdot)$; the conditions on $\psi$ are collected in Assumption~\ref{assump:loss}.
For example, in the MRT setting,  $Y_i\in\mathbb{R}^{T}$ is the (stacked)
response and $X_i\in\mathbb{R}^{T\times p}$ the stacked temporal covariates for
individual $i$, both obtained through a suitable transformation of the raw data, as highlighted in Remark \ref{rem:transformation}. 

\begin{remark}
  The WCLS criterion in \eqref{wcls} can be seen as a special case of this general M-estimation formulation. Setting
$\widetilde{Y}_{i,t}=\sqrt{W_{i,t}}\,(Y_{i,t}-\widehat\mu_t(H_{i,t}))$,
$\widetilde{X}_{i,t}=\sqrt{W_{i,t}}\,(A_{i,t}-\widetilde{p}_t(1\mid M_{i,t}))\,f_t(M_{i,t})$, and 
stacking the temporal observations of the $i$-th individual $X_i=[\widetilde X_{i,1}\cdots\widetilde X_{i,T}]^{\top},
\ Y_i=[\widetilde Y_{i,1}\cdots\widetilde Y_{i,T}]^{\top}$, the WCLS estimator is exactly the M-estimator minimizing the quadratic loss $\psi(X_ib;Y_i)=\tfrac12\|Y_i-X_ib\|_2^2$.
  \label{rem:transformation}
\end{remark}

\subsubsection{Selection via randomized lasso}
\label{sec:randlasso}
We propose selecting a smaller subset of important moderators via the randomized lasso.
Specifically, we add an independent $p$-dimensional Gaussian randomization vector $\sqrt{n}\omega_n\sim\mathcal{N}(0_p,\Omega)$, with a prespecified covariance matrix $\Omega$, to the penalized objective as follows:
\begin{equation}
\label{rand:lasso}
\underset{b \in \mathbb{R}^p}{\text{minimize}} \left\{\frac{1}{\sqrt{n}}\sum_{i=1}^{n}\psi(X_{i}b; Y_{i}) + \lambda\|b\|_1-\sqrt{n}\omega_{n}^{\top}b \right\},
\end{equation}
whose solution is the randomized lasso estimator $\lasso\in\mathbb{R}^{p}$.  A simple randomization scheme, also used later in our simulations, adds $p$ i.i.d. Gaussian noise variables to the lasso objective, i.e., $\Omega = \tau\cdot \widehat{\sigma}^2 I_{p}$. 
Here, the parameter $\tau$ plays a role similar to the split ratio in data splitting, determining how much information is used for selection versus how much is exclusively reserved for inference. Unlike data splitting, however, this additive Gaussian randomization uses all available samples for selection and yields a pivot for the selected target based on the full dataset, rather than only on the held-out samples. As illustrated in Figure~\ref{plot1} earlier, a small amount of randomization (with $\tau=0.5$, corresponding to `SI (low)') retains the selection performance of standard (non-randomized) lasso while also ensuring short and bounded interval estimates based on the full set of available samples. 

The selected set
\[\widehat{E}=\{j\in [p]: |\sign(\lassoj)|=1\}\] 
collects the indices of variables with nonzero lasso coefficients. 
On the observed samples, we observe $\{\widehat{E}=E\}$, where $E$ is the observed value of the selected set $\widehat{E}$.
Let $E'$ denote the complement set of $E$, and let $|E|=q$ and $|E'|=q'=p-q$.
For WCLS, \eqref{rand:lasso} is the
criterion \eqref{wcls} with an added $\ell_1$-penalty and Gaussian randomization, with $E$ indexing the selected effect moderators included in the working model $\beta^{E}(t;m)=(f^{E}_t(m))^{\top}\beta^{E}$.

\subsubsection{Post-selection target and refitted M-estimator}
\label{sec:refit and target}

Before proceeding further, we state the assumptions on the loss function in \eqref{rand:lasso}, which are required for the inferential results that follow in Section \ref{sec:SI}.
 
\begin{assumption}[Regularity conditions on the loss function]
\label{assump:loss}
The loss function  $\psi(\,\cdot\,;y)$ in \eqref{rand:lasso} depends
on the coefficient $b$ only through the linear predictor $X_i b$ and is twice continuously differentiable with respect to that argument. 
\end{assumption} 
Let $\estE$ denote the estimator refitted on the selected features using $\{Y_i,X_{i,E}\}_{i\in[n]}$, and let $\targetE$ denote its population analog:
\begin{equation}
\label{refit}
\estE = \underset{b\in \mathbb{R}^q}{\text{argmin}} \sum_{i=1}^{n}\psi(X_{i,E}b; Y_{i}); \quad \targetE = \underset{b\in \mathbb{R}^q}{\text{argmin}}\ \Ee_{\mathbb{F}_{n}}\rbracket{\psi(X_{1,E}b; Y_{1})},
\end{equation}
assuming that, for the selected set $E$, the population risk $b\mapsto \mathbb{E}_{\mathbb{F}_n}\psi(X_{1,E}b;Y_1)$ has a unique minimizer.
We write $\estj,\targetj\in\mathbb{R}$ for their respective $j$th components.
Throughout, for each $j\in E$, $\targetj$ is our target of inference, and we construct a pivot for it using the appropriate conditional distribution of $\estj$.

Additionally, we define two population quantities on which the distribution of  the refitted M-estimator $\estE$ depends, both marginally for any arbitrary fixed set $E$ and conditionally on the selection event $\{\widehat{E}=E\}$:
\begin{equation}
\label{eq:HK}
H=\mathbb{E}_{\mathbb{F}_n}\!\left[\frac{1}{n}\sum_{i=1}^{n}X_i^{\top}\nabla^{2}\psi\!\left(X_{i,E}\targetE; Y_i\right)X_i\right],
\quad
K=\Cov_{\mathbb{F}_n}\!\left(\frac{1}{\sqrt{n}}\sum_{i=1}^{n}X_i^{\top}\nabla\psi\!\left(X_{i,E}\targetE; Y_i\right)\right).
\end{equation}
Here, $H$ is the expected Hessian of the loss and $K$ is the covariance matrix of the estimating score when the loss is viewed as being derived from a log-likelihood function.
Ignoring selection, the standard sandwich covariance matrix of the refitted M-estimator
$\estE$, defined in \eqref{refit}, is $\Sigma_{E,E}=H_{E,E}^{-1}K_{E,E}H_{E,E}^{-1}$, with $H$ and $K$ as in \eqref{eq:HK}.
\begin{remark}
For WCLS, $\targetj$ represents the causal moderation effect associated with the $j$th selected moderator within the selected working model. Specifically, it is the $j$th coefficient in the $L_2$-projection of the causal excursion effect onto the subspace spanned by the selected features, and $\estj$ denotes the corresponding coefficient in the refitted WCLS estimator, using only the selected moderators, $\{Y_i, X_{i,E}\}_{i\in [n]}$. 
Following up on Remark \ref{rem:transformation}, note that the loss function in our general M-estimation formulation for the problem satisfies Assumption \ref{assump:loss}; in particular, $\nabla^{2}\psi=I_n$,  and $H=\mathbb{E}_{\mathbb{F}_n}[n^{-1}\sum_i X_i^{\top}X_i]$ is the population Gram matrix of the design, while $K$ is the covariance matrix of the centered residual score.
\end{remark}

\section{Selective inference for excursion effect}
\label{sec:SI}

This section develops the methodology for valid inference on $\targetj$ by
conditioning on the selection event $\{\widehat{E}=E\}$, leading to the inferential guarantee in \eqref{cond:guarantee}. Proofs of all results presented in this section are provided in Supplement~\ref{app:SIproofs}.

\subsection{Main steps toward valid conditional inference}
\label{sub:mainsteps}

Our goal is to conduct inference using the conditional distribution
$\estj \mid \{\widehat{E}=E\}$.
However, this distribution does not directly yield valid inference because it depends on a high-dimensional nuisance parameter, and it does not readily yield feasible inference because the conditioning event is difficult to characterize in terms of the observed data.
Our method circumvents both challenges through three steps, as outlined below.

\begin{enumerate}[leftmargin=*]
\item In \textbf{Step 1}, we identify the master statistics
$\sqrt{n}\nbracket{(\estj)^\top \ (\gammaj)^\top}^\top$, whose conditional distribution we
characterize for valid selective inference. This distribution involves our target
parameter $\targetj$ as well as a $(p-1)$-dimensional nuisance parameter $\gammaej$. 
The exact definitions of $\gammaj$ and $\gammaej$ are provided in
Section~\ref{step1}. 
Conditioning on the observed value of $\gammaj$ yields a univariate
conditional distribution $\estj \mid \gammaj, \{\widehat{E}=E\}$, free of the
nuisance vector $\gammaej$.
\item In \textbf{Step 2}, we simplify the conditioning event $\{\widehat{E}=E\}$ by
conditioning on additional information from the randomized lasso selection. Specifically,
we identify a proper subset of this event, which is equivalent to truncating a real-valued randomized lasso variable to a simple interval. This yields the simplified univariate conditional distribution
\begin{equation}
\estj \; \big\lvert \; \gammaj, \ \cbracket{\unicond\in [I_{-}^{E\cdot j}, I_{+}^{E\cdot j}] ,\ \widehat{S}_{n, E'}= S_{E'}, \ \widehat{V}_n^{E\cdot j} = V^{E\cdot j}},
\label{preview:cond}
\end{equation}
where the randomized lasso variables $\unicond, \widehat{V}_n^{E\cdot j}, \widehat{S}_{n, E'}$ and the interval
$[I_{-}^{E\cdot j}, I_{+}^{E\cdot j}]$ are defined in Section~\ref{step2}; here $S_{E'}$ and $V^{E\cdot j}$ denote the observed values of $\widehat{S}_{n,E'}$ and $\widehat{V}_n^{E\cdot j}$, respectively, and conditioning on the statistic $\gammaj$ is understood to mean conditioning on its observed value.
\item Finally, in \textbf{Step 3} described in Section~\ref{step3}, a change-of-variables map derived from the Karush--Kuhn--Tucker (K.K.T.) stationarity conditions of the randomized lasso \eqref{rand:lasso} yields the joint density of the master statistics and the randomized lasso variables, from which we obtain a closed-form pivot; Section~\ref{sec:theorypivot} establishes the asymptotic coverage of confidence intervals obtained by inverting this pivot.
\end{enumerate}

\subsubsection{Step 1: Master statistics and their distribution}
\label{step1}
 
Fixing notation, for a matrix (or vector) $A$ whose rows are indexed by a set
$\mathcal{I}$ and a subset $\mathcal{E}\subseteq\mathcal{I}$,
let $R_{\mathcal{E}} A$ denote the submatrix of $A$ consisting of
the rows indexed by $\mathcal{E}$; for a singleton set
$\mathcal{E} = \{ j \}$, we simply write $R_j A$. 
In particular, the rows and coordinates of objects indexed by the selected set, such as $\Sigma_{E,E}$, $\estE$, or $Q^{E\cdot j}$, are labelled by the elements of $E$; for example, $R_j\Sigma_{E,E}$ denotes the row of $\Sigma_{E,E}$ corresponding to $j \in E$. 
For a $p \times p$ matrix $A$ and $\mathcal{E}\subseteq[p]$, we use $A = \rbracket{A_{\mathcal{E}} \ A_{\mathcal{E}'}}$, where
$A_{\mathcal{E}}=\rbracket{A_{\mathcal{E},\mathcal{E}}^{\top}  \ A_{\mathcal{E}',\mathcal{E}}^{\top}}^{\top}$,
$A_{\mathcal{E}'} = \rbracket{A_{\mathcal{E},\mathcal{E}'}^{\top}  \ A_{\mathcal{E}',\mathcal{E}'}^{\top}}^{\top}$,
to denote submatrices that partition the columns and rows of $A$ based on the set
$\mathcal{E}$ and its complement $\mathcal{E}' = \mathcal{E}^c$.

Recall from Section~\ref{sec:refit and target} that, ignoring selection, $\Sigma_{E,E}$ is the covariance matrix of the refitted M-estimator.
The $j$th column of this matrix is $\Sigma_{E,j}=\Sigma_{E,E}R_j^{\top}$, and the variance
corresponding to the $j$th component is
$(\sigma^{E\cdot j})^2=R_j\Sigma_{E,E}R_j^{\top}$.
Next, we define the statistic
\[\gammaj  = \begin{bmatrix}
R_{E\setminus j}\nbracket{\estE - \frac{1}{(\sigma^{E\cdot j})^{2}}\Sigma_{E,j}\estj} \\ \frac{1}{n}\sum_{i=1}^{n}X_{i,E'}^{\top}\nabla\psi\nbracket{X_{i,E}\estE ;Y_{i}}  - (H_{E',E}-K_{E',E}K_{E,E}^{-1}H_{E,E})\estE  \end{bmatrix} \in \mathbb{R}^{p-1},\]
where the first $q-1$ components are obtained from $\estE$ and the last $p-q$ components
involve moderators in $E'$ as well as the residuals from the refitted model. 
When inferring for $\targetj$, $\gammaej =\mathbb{E}_{\mathbb{F}_n}[\gammaj]$
plays the role of nuisance parameters, while $\gammaj$ captures the remaining information used in selection and is asymptotically uncorrelated with $\estj$, as shown in Proposition~\ref{prop:asymptotic}.
For fixed $E \subseteq [p]$ and $j\in E$, i.e., ignoring selection, Proposition \ref{prop:asymptotic} establishes the asymptotic normality of the master statistics $\sqrt{n} \rbracket{(\estj)^\top  \ (\gammaj)^\top}^\top \in \mathbb{R}^{p}$.

\begin{proposition}
\label{prop:asymptotic}
Fix $E \subseteq [p]$ and $j \in E$, let $\zeta_n=\frac{1}{\sqrt{n}}\sum_{i=1}^{n}K^{-1/2}(b_{i,n}-\mathbb{E}_{\mathbb{F}_n}[b_{i,n}])$, where $b_{i,n} =  X_{i}^{\top}\nabla\psi\nbracket{X_{i,E}\targetE ; Y_{i}}$, and let $A_{1}$, $A_{2}$ be as in Table~\ref{Table:matrices}. Then:
\begin{enumerate}[leftmargin=*, itemsep=2pt]
\item $\displaystyle \sqrt{n}\begin{bmatrix} \estj- \targetj \\ \gammaj -  \gammaej \end{bmatrix} = \begin{bmatrix} A_1 \\ A_2 \end{bmatrix} \zeta_n+ R_{n,1}$, \ where $R_{n,1}=o_{p}(1)$;
\item $\displaystyle \sqrt{n} \begin{bmatrix}\estj - \targetj \\  \gammaj - \gammaej \end{bmatrix} \indist \mathcal{N}_{p} \nbracket{0_{p},\ \begin{bmatrix}\sigmaj & 0 \\ 0 & \sigmaE \end{bmatrix}}$, \ where $\sigmaE= A_{2}A_{2}^{\top}$.
\end{enumerate}
\end{proposition}

\subsubsection{Step 2: Conditioning event}
\label{step2}
Without loss of generality, we rearrange the columns of $X$ and write
$X_{i}= \rbracket{X_{i,E} \ X_{i,E'}}$ and $\lasso = \rbracket{(\lassoE)^{\top} \ 0_{q'}^\top}^\top$, so that $X_{i}\lasso = X_{i,E}\lassoE$.
The K.K.T. (Karush--Kuhn--Tucker) stationarity conditions of the randomized lasso are given as:
\begin{align}
\label{KKT}
    \frac{1}{\sqrt{n}}\sum_{i=1}^{n} X_{i}^{\top}\nabla\psi(X_{i,E}\lassoE; Y_{i}) + \lambda  S = \sqrt{n}\omega_{n},
\end{align}
where $S$ is the observed value of the subgradient of the $\ell_1$-penalty at the solution
$\widehat{S}_n= \left\{\partial \|b\|_1\right\}_{b= \lasso}$; the sub-vectors $S_{E}$ and $S_{E'}$ are the observed values of $\widehat{S}_{n,E}$ and $\widehat{S}_{n,E'}$.
 
\begin{proposition}
\label{prop:sel:event}
It holds that
$\left\{\widehat{E}=E\right\} = \underset{S_{E}\in \{-1,1\}^q}{\bigcup}\left\{ \widehat{S}_{n,E}= S_E, \;\| \widehat{S}_{n,E'} \|_{\infty} \leq 1\right\}.$
\end{proposition} Proposition~\ref{prop:sel:event} describes the selection event $\{\widehat{E}=E\}$ in terms of the subgradient variables but does not directly yield a closed-form pivot. 
We therefore condition on additional information from the randomized lasso solution, which produces a much simpler truncation region and, consequently, a simplified conditional density, while retaining the inferential guarantee in \eqref{cond:guarantee} by the tower property of expectations.
 
\noindent{\textbf{Additional conditioning steps.}} \quad Here, we describe the additional conditioning steps which reduce the conditioning event, or, equivalently, the truncation region of the conditional distribution, from a union of polyhedrons to an orthant and then to a single interval: 
\begin{equation}
  \underbrace{\bigl\{\widehat{S}_n = S,\ \widehat{V}_n^{\eta} = V^{\eta}\bigr\}}
    _{\text{\footnotesize interval}}
  \ \subset\
  \underbrace{\bigl\{\widehat{S}_n = S\bigr\}}
    _{\text{\footnotesize orthant}}
  \ \subset\
  \underbrace{\bigl\{\widehat{E} = E\bigr\}}
    _{\text{\footnotesize union of polyhedra}}
    \label{prog:conditioning}
\end{equation}
Moving from right to left, we condition on progressively more information, where $\widehat{V}_n^{\eta}$ is as defined below. 

First, conditioning on the observed subgradient selects the polyhedron in the union of Proposition~\ref{prop:sel:event} that is compatible with the observed signs and turns the inequalities on the inactive subgradient variables into equalities:
\begin{equation}
\label{orthant}
\cbracket{\widehat{S}_{n}=S}
\,=\, \cbracket{ \widehat{S}_{n,E'}=S_{E'},\  \sqrt{n}|\lassoE| \in \mathbb{R}_{+}^{q}}
\,\subset \cbracket{\widehat{E}=E},
\end{equation} 
This conditioning set is equivalent to truncating the magnitudes of the active lasso coefficients $\sqrt{n}|\lassoE| = \Dg(S_{E})\sqrt{n}\lassoE$ to the positive orthant in $\mathbb{R}^{q}$.

Let $\widetilde{H}_{E} = H_{E}\,\Dg(S_{E}) \in \mathbb{R}^{p\times q}$, so that $H_{E}\sqrt{n}\lassoE = \widetilde{H}_{E}\sqrt{n}|\lassoE|$. 
For any $\eta\in\mathbb{R}^{q}\setminus\{0\}$, define $Q^{\eta}=\Lambda\eta/(\eta^{\top}\Lambda\eta)$, where $\Lambda = (\widetilde{H}_E^{\top}\Omega^{-1}\widetilde{H}_E)^{-1}$. 
Then, by straightforward algebra, the magnitudes of the active lasso coefficients can be written as:
\begin{equation}
\label{decomp:eta}
\sqrt{n}|\lassoE|
\;=\; Q^{\eta}\,\eta^{\top}\sqrt{n}|\lassoE|
\;+\; \nbracket{I_{q}-Q^{\eta}\eta^{\top}}\sqrt{n}|\lassoE|,
\end{equation}
decomposing this vector into the scalar-valued component $\widehat{U}^{\eta}_n=\eta^{\top}\sqrt{n}|\lassoE| \in \mathbb{R}$ along the direction $\eta$ and the vector-valued residual component $\widehat{V}^{\eta}_n=\nbracket{I_{q}-Q^{\eta}\eta^{\top}}\sqrt{n}|\lassoE|$.
If we further condition on the residual component $\widehat{V}^{\eta}_n$, as in the left-hand-side event in \eqref{prog:conditioning}, the real-valued randomized lasso variable $\widehat{U}^{\eta}_n$ becomes the sole remaining source of randomness; this reduces the truncation region from an orthant to a single interval.

\noindent{\textbf{Choice of conditioning event.}} It remains only to choose the direction $\eta$, which in turn determines the conditioning event or truncating interval, as formalized in Proposition \ref{prop:condevent}.

Let $P_1^{E\cdot j}=-(\sigma^{E\cdot j})^{-2}K_E H_{E,E}^{-1}R_j^{\top}$, which, as seen in Proposition~\ref{prop:KKTmaster}, is the coefficient of $\sqrt{n}\estj$ in the K.K.T. stationarity conditions  of the lasso.
Set
\begin{equation*}
\eta^{E\cdot j} = \widetilde{H}_E^{\top}\Omega^{-1}P_1^{E\cdot j},\qquad
Q^{E\cdot j} = Q^{\eta^{E\cdot j}} = \frac{\Lambda\,\eta^{E\cdot j}}{(\eta^{E\cdot j})^{\top}\Lambda\,\eta^{E\cdot j}},
\end{equation*}
and the corresponding randomized lasso variables
\begin{equation}
\label{opt:vars}
\widehat{U}_n^{E\cdot j} = \sqrt{n}(\eta^{E\cdot j})^{\top}|\lassoE| \in \mathbb{R} , \;
\widehat{V}_n^{E\cdot j}=\sqrt{n}\nbracket{I_{q}-  Q^{E\cdot j} (\eta^{E\cdot j}) ^{\top}} |\lassoE| \in \mathbb{R}^{q}.
\end{equation}
Because $(\eta^{E\cdot j})^{\top}Q^{E\cdot j}=1$ by construction, we have $(\eta^{E\cdot j})^{\top}\widehat{V}_n^{E\cdot j}=0$, i.e., $\widehat{V}_n^{E\cdot j}$ lies in the $(q-1)$-dimensional subspace $\{v\in\mathbb{R}^{q}: (\eta^{E\cdot j})^{\top}v=0\}$. 
As discussed in \cite[Proposition~3.2]{panigrahi2024exact}, this specific choice of direction $\eta= \eta^{E\cdot j}$ ensures that the conditioned residual carries as little information about $\targetj$ as possible, allowing the resulting intervals approach the naive intervals as the effect of selection diminishes.
 
\begin{proposition}
\label{prop:condevent}
    Let the estimators $\widehat{U}_n^{E\cdot j}$ and $\widehat{V}_n^{E\cdot j}$ be as defined in \eqref{opt:vars}. For $j \in E$, we have that
    \[
\cbracket{\widehat{S}_n= S,\ \widehat{V}_n^{E\cdot j} = V^{E\cdot j}}
= \left\{\unicond\in [I_{-}^{E\cdot j}, I_{+}^{E\cdot j}] ,\ \widehat{S}_{n, E'}= S_{E'}, \ \widehat{V}_n^{E\cdot j} = V^{E\cdot j}\right\},\]
where
$I_{-}^{E\cdot j} = \max _{k\in E: R_{k}Q^{E\cdot j}>0}\left\{\frac{-R_{k} V^{E\cdot j}}{R_{k}Q^{E\cdot j}}\right\}$, and $I_{+}^{E\cdot j}=\min _{k\in E: R_{k}Q^{E\cdot j}<0}\left\{\frac{-R_{k} V^{E\cdot j}}{R_{k} Q^{E\cdot j}}\right\}$, with the conventions $\max\emptyset = -\infty$ and $\min\emptyset = +\infty$ when the corresponding index set is empty.
\end{proposition}
 

\subsubsection{Step 3: Pivot via conditioning} 
\label{step3} 
We now characterize the conditional distribution in \eqref{preview:cond}, which yields a closed-form pivot, by first relating the master statistics to the randomized lasso solution.
\begin{proposition} 
\label{prop:KKTmaster}
The K.K.T. stationarity conditions at the randomized lasso can be rewritten as 
\[P_{1}^{E\cdot j} \sqrt{n}\estj  + P_{2}^{E\cdot j}\sqrt{n}\gammaj + H_{E}\sqrt{n}\lassoE + 
         \lambda \rbracket{ S_{E}^{\top} \ 
        S_{E'}^{\top} }^{\top}= \sqrt{n}\omega_{n} + R_{n,2}\]
where $R_{n,2}= o_{p}(1)$, and the expressions for matrices $P_{1}^{E\cdot j}$, $P_{2}^{E\cdot j}$ are provided in Table~\ref{Table:matrices}. 
\end{proposition} 
Hereafter, we define 
\begin{equation}
\label{tilde:omega}
\sqrt{n} \widetilde{\omega}_n = \sqrt{n}\omega_{n} + R_{n,2},
\end{equation}
which is asymptotically distributed as a $\mathcal{N}\left(0_p, \Omega\right)$ random variable.
We connect the result of Proposition \ref{prop:KKTmaster} to Steps 1 and 2 in Sections \ref{step1} and \ref{step2}, respectively.
\begin{enumerate}[leftmargin=*]
\item[(i)] By Proposition~\ref{prop:KKTmaster}, the active and inactive subgradient events $\{\widehat{S}_{n,E}=S_E\}$ and $\{\widehat{S}_{n,E'}=S_{E'}\}$ can be written explicitly in terms of $\widetilde{\omega}_n$ and the master statistics
$\sqrt{n}\estj$,$\sqrt{n}\gammaj$ (see Corollary~\ref{cor:signevents} in Supplement~\ref{app:SIproofs}).
Together with Proposition~\ref{prop:sel:event}, this result shows the selection event $\{\widehat{E}=E\}$ depends on the observed data only through the master statistics from Section \ref{step1}.
\item[(ii)]Given the conditioning event from Step 2, characterizing the conditional distribution in \eqref{preview:cond} requires the joint distribution of the master statistics and the randomized lasso variables  before truncation, i.e., the marginal distribution of $(\sqrt{n}\estj,\sqrt{n}\gammaj,\unicond,\extracond,\widehat S_{n,E'})$. This follows by applying a change of variables to the known density of the randomization, through the map
\begin{equation}
\label{CoV:map}
\Pi_{\sqrt{n}\estj, \sqrt{n}\gammaj}(u, v, z) =  L^{E\cdot j} u +  \widetilde{H}_{E} v + \lambda \rbracket{S_E^{\top} \ z^{\top}}^{\top} + P_{1}^{E\cdot j} \sqrt{n}\estj  + P_{2}^{E\cdot j}\sqrt{n}\gammaj
\end{equation}
whose value, at the observed randomized lasso solution, coincides with the left-hand side of the K.K.T. stationarity conditions in Proposition~\ref{prop:KKTmaster} after further decomposing the active lasso coefficients in the direction of $\eta=\eta^{E\cdot j}$, as shown in \eqref{decomp:eta}. 
In particular, the linear term involving the active lasso coefficients can be written as 
$H_{E}\sqrt{n}\lassoE =\widetilde{H}_{E}\sqrt{n}|\lassoE| = L^{E\cdot j}\widehat{U}_n^{E\cdot j} + \widetilde{H}_{E}\widehat{V}_n^{E\cdot j}$.
Theorem \ref{thm:CoV}, stated next, shows that applying the change-of-variables map in \eqref{CoV:map} yields a joint density for the master statistics and the randomized lasso variables.
\end{enumerate}

\begin{theorem}
\label{thm:CoV}
Let $E$ and $S_{E}$ be a given set of indices and vector of signs, respectively. 
Let $\Pi_{\sqrt{n}\estj, \sqrt{n}\gammaj}(\cdot)$ be defined according to \eqref{CoV:map}.
Then, the asymptotic marginal density of 
$\left(\sqrt{n}\estj, \sqrt{n}\gammaj, \unicond, \extracond,  \widehat{S}_{n, E'}\right)$ is equal to 
\begin{align*}
& \phi(\sqrt{n}\estj;\sqrt{n}\targetj, (\sigma^{E\cdot j})^{2})\times \phi\nbracket{\sqrt{n}\gammaj;\sqrt{n}\gammaej, \Sigma^{E\cdot j} }\\
&\qquad\qquad \times \phi\left(\Pi_{\sqrt{n}\estj, \sqrt{n}\gammaj}(\unicond, \widehat{V}_n^{E\cdot j}, \widehat{S}_{n, E'});0_{p},\Omega\right)\times |\det(H_{E,E})|\,\lambda^{p-q}.
\end{align*}
\end{theorem}

A proof of Theorem~\ref{thm:CoV} is given in Supplement~\ref{app:SIproofs}.

When $\{p(Y_{i}|X_{i})\}_{i\in [n]}$ is Gaussian and we are in a fixed-$X$ regime, the densities in Proposition \ref{prop:asymptotic} and Theorem \ref{thm:CoV} are exact, i.e., $R_{n,1}=R_{n,2}=0$, and therefore, the master statistics are exactly normal and linearly related to the lasso variables through the K.K.T. map. 
This case was addressed in \cite{panigrahi2024exact}, where conditioning on the event in Proposition \ref{prop:condevent} yields an exact pivot for $\targetj$, constructed as follows.

\paragraph{Constructing an exact pivot for normal data.} Assume for now that the density in Theorem \ref{thm:CoV} is exact.
It follows directly that the conditional density 
\[\nbracket{ \sqrt{n}(\estj)^\top \ \sqrt{n}(\gammaj)^\top \ \unicond}^\top\Big\lvert \cbracket{\unicond\in [I_{-}^{E\cdot j}, I_{+}^{E\cdot j}] , \widehat{V}_n^{E\cdot j} = V^{E\cdot j}, \widehat{S}_{n, E'}= S_{E'}}\]
is proportional to
\begin{align*}
& \phi(\sqrt{n}\estj;\sqrt{n}\targetj, (\sigma^{E\cdot j})^{2})\times \phi\nbracket{\sqrt{n}\gammaj;\sqrt{n}\gammaej, \Sigma^{E\cdot j} }\\
&\qquad\qquad\times \phi\left(\Pi_{\sqrt{n}\estj, \sqrt{n}\gammaj}(\unicond, V^{E\cdot j}, S_{ E'});0_{p},\Omega\right)\times \mathbbm{1}_{[I_{-}^{E\cdot j}, I_{+}^{E\cdot j}]}(\unicond),
\end{align*} 
ignoring constants. 
Conditioning further on the observed $\sqrt{n}\gammaj$, in order to remove the nuisance parameter vector $\gammaej$, and marginalizing over $\unicond$ gives a conditional density for $\sqrt{n}\estj$ involving only $\targetj$: 
\[
C^{-1}\cdot \phi(\sqrt{n}\estj;\sqrt{n}\targetj, (\sigma^{E\cdot j})^{2}) \cdot F(\sqrt{n}\estj, \sqrt{n}\gammaj)
\]
where 
\begin{align}
F(\sqrt{n}\estj, \sqrt{n}\gammaj) = \int_{ I_{-}^{E\cdot j}}^{I_{+}^{E\cdot j}} \phi\nbracket{L^{E\cdot j} t +P_{1}^{E\cdot j} \sqrt{n} \estj + P_{2}^{E\cdot j} \sqrt{n} \gammaj+T^{E\cdot j} ; 0_{p},\Omega}dt,
\label{def:F}
\end{align}
 with
\[T^{E\cdot j}=\widetilde{H}_{E}V^{E\cdot j} + \lambda \rbracket{S_E^{\top} \ S_{E'}^{\top}}^{\top} \text{ and } 
C= \int_{-\infty}^{\infty}\phi(x; \sqrt{n}\targetj, (\sigma^{E\cdot j})^{2})F(x, \sqrt{n}\gammaj)dx.\]
Applying a probability integral transform to this conditional density function yields an exact pivot
\begin{align}
\label{exact:pivot}
    P^{E\cdot j}(\estj, \gammaj; \targetj) = \frac{\int_{-\infty}^{\sqrt{n}\estj}\phi(x; \sqrt{n}\targetj, (\sigma^{E\cdot j})^{2})F(x, \sqrt{n}\gammaj)dx}{\int_{-\infty}^{\infty}\phi(x; \sqrt{n}\targetj, (\sigma^{E\cdot j})^{2})F(x, \sqrt{n}\gammaj)dx},
\end{align}
which is distributed as a $\text{Unif}(0,1)$ random variable. 

In the following section, our theory shows how we can utilize the same pivot in \eqref{exact:pivot}, as though the data were normally distributed, to conduct asymptotic selective inference.

\subsection{Uniformly asymptotic selective inference}
\label{sec:theorypivot}

To obtain confidence intervals for $\targetj$, we invert our proposed pivot. 
For example, two-sided confidence intervals at level $\alpha$ are constructed as
\[
\left(L_{\alpha,n}^j, U_{\alpha,n}^j\right)=\left\{\targetj: P^{E\cdot j}(\estj, \gammaj; \targetj) \in\left[\frac{\alpha}{2}, 1-\frac{\alpha}{2}\right]\right\}.
\]
In this section, we demonstrate that these intervals achieve the desired coverage probability as $n\to \infty$ when
$(X_{i},Y_{i})_{i\in[n]}$ are i.i.d. variables from $\mathbb{F}_{n}$.
This leads to our main result, Theorem~\ref{main}, which offers asymptotic conditional guarantees for these selective confidence intervals, uniformly across all distributions in a large collection $\mathcal{F}_n=\{\mathbb{F}_{n}\}$, for $n\in\mathbb{N}$ that meet the following assumptions. To simplify notation, we drop the superscript $E\cdot j$ from the pivot and its associated matrices from this point onward.

We consider a sequence of parameters that depend on $n$ as
$\sqrt{n}\rbracket{
    (\targetj)^\top (\gammaej)^\top}^\top  = O(r_n),$ 
where $r_n$ is a non-negative sequence of real numbers growing strictly slower than $n^{1/6}$, i.e. $r_n = o(n^{1/6})$.
Notably, under the assumptions of this section, $\log\mathbb{P}_{\mathbb{F}_n}[\widehat E=E]\asymp -r_n^{2}$.
Put in words, the selection probability decays exponentially at the rate $r_n^2$, as shown in \cite{panigrahi2023carving}. 
When $r_n = O(1)$, the selection probability remains bounded away from zero; when $r_n \to\infty$, it vanishes to zero exponentially fast.
By allowing $r_n$ to grow, our results substantially extend existing theory developed under conditions corresponding to selection probabilities bounded away from zero, including the asymptotic theory for effect modification in \cite{zhao2021selective}.

\begin{assumption}
\label{assump:1}
Let $b_{i,n}= X_{i}^{\top}\nabla\psi\nbracket{X_{i,E}\targetE ; Y_{i}}$ denote the score when evaluated at the population target.
We assume that there exists a constant $B_0>0$ such that, for any $t\in \mathbb{R}$ and $i \in [n]$, we have
$\lim_n \sup_{\mathbb{F}_n\in \mathcal{F}_n} \mathbb{E}_{\mathbb{F}_n}\rbracket{\exp \nbracket{t\norm{b_{i,n}-\mathbb{E}_{\mathbb{F}_n}\rbracket{b_{i,n}}}}} \leq \exp(B_0t^{2})$.
\end{assumption}

\begin{assumption}
\label{assump:2}
Let $\widetilde{\omega}_{n}$ be the random variable defined in \eqref{tilde:omega}, with a Lebesgue density $q_{n}$ for $n\ge n_{0}$. 
Define 
$G_{n}(x,y) = \int_{I_{-}^{E\cdot j}}^{I_{+}^{E\cdot j}} q_{n}\nbracket{L^{E\cdot j} t +P_{1}^{E\cdot j} x + P_{2}^{E\cdot j}y+T^{E\cdot j}}dt,$ with $L^{E\cdot j}$, $P_{1}^{E\cdot j}$, $P_{2}^{E\cdot j}$ and $T^{E\cdot j}$ as in \eqref{def:F}. 
Then, assume that $\lim_{n \to\infty}\sup_{(x,y)}\left \lvert \frac{G_{n}(x,y)}{F(x,y)} -1 \right \rvert =0.$  
\end{assumption}

\begin{assumption}
\label{assump:3}
When $r_n\to \infty$ as $n\to \infty$, 
$\lim _n \sup _{\mathbb{F}_n \in \mathcal{F}_n} \frac{1}{r_{n}^{2}}\log \mathbb{P}_{\mathbb{F}_n}\rbracket{\frac{\norm{R_{n,1}}}{r_{n}}>\epsilon} = -\infty$, for any $\epsilon > 0$.
\end{assumption}

\begin{remark}
All the expectations and probabilities in this section are taken over the generating distribution,  $\mathbb{F}_n \in \mathcal{F}_n$.
\end{remark}

The condition in Assumption \ref{assump:1} states that the centered score $X_{i}^{\top}\nabla\psi\nbracket{X_{i,E}\targetE ; Y_{i}}-\mathbb{E}_{\mathbb{F}_n}\rbracket{X_{i}^{\top}\nabla\psi\nbracket{X_{i,E}\targetE ; Y_{i}}}$ must be sub-Gaussian.
This condition is satisfied if the variables $X_{i} \in \mathbb{R}^{p}$ are uniformly bounded, i.e., $\left \| X_{i} \right \|_{\infty} \leq C$ for some $C>0$, and if the score variable derived from the loss function is sub-Gaussian.
The condition in Assumption \ref{assump:2} ensures that $\sqrt{n}\widetilde{\omega}_n$ in \eqref{tilde:omega} behaves similarly to its limiting normal counterpart $\sqrt{n}\omega_n$.
Since $\sqrt{n}(\widetilde{\omega}_n - \omega_n) = o_p(1)$, this condition is automatically satisfied when the sequence $r_n$ does not grow with $n$.
Finally, Assumption \ref{assump:3} imposes conditions on the tail behavior of the error variable $R_{n,1}$. This condition is only needed when $r_n \to \infty$, a setting not covered by \cite{zhao2021selective}, and one in which the polyhedral method is typically numerically unstable and unreliable.
For $r_n\to \infty$, this condition requires the approximation error $R_{n,1}$ to be negligible relative to $r_n$, in the sense that the probability of an error of order $r_n$ decays faster than the probability of the selection event.


\begin{theorem} \label{main}
  For each $n \in \mathbb{N}$, let  $\mathcal{F}_n$ be a collection of data-generating distributions, satisfying Assumptions~\ref{assump:1}, \ref{assump:2}, \ref{assump:3}. Then, we have that
\[
\lim _n \sup _{\mathbb{F}_n \in \mathcal{F}_n}\left|\mathbb{P}_{\mathbb{F}_n}\left[\targetj \in\left(L_{\alpha,n}^j, U_{\alpha,n}^j\right)\Big\lvert\cbracket{\widehat{S}_n= S, \widehat{V}_n^{E\cdot j} = V^{E\cdot j}}\right]-(1-\alpha)\right|=0.
\]  
\end{theorem}

Put in words, our main result establishes that for any fixed $\epsilon>0$, there exists $N(\epsilon)\in\mathbb{N}$ such that the interval $\left(L_{\alpha,n}^j, U_{\alpha,n}^j\right)$ has coverage probability at least $1-\alpha-\epsilon$ for any $n\geq N(\epsilon)$ and any data-generating distribution in $\mathcal{F}_n$. This guarantee is stated conditional on the the event in Proposition \ref{prop:condevent}, which is a proper subset of ${\widehat{E}=E}$. 
Integrating over the additional variables on which we condition then yields the corresponding asymptotic guarantee conditional on the selection event ${\widehat{E}=E}$.
Supporting results for the theory developed in this section, along with detailed proofs, are collected in Supplement~\ref{app:asymp}.

\section{Simulation Study} 
\label{sec:sim}

In this section, we assess the performance of our method compared to existing approaches
using synthetic MRT data. Our data generating model is motivated from the experimental
design in \cite{MRT}. We simulate $n$ i.i.d.\ participants over $T=30$ decision points with $p=50$ potential moderators, of which a sparse subset $E^* \subset [p]$ with $|E^*|=5$ represents the true effect moderators. The moderator process follows a stationary AR(1) model with autocorrelation $\rho_{\mathrm{state}}$, and outcomes are generated as
\[
Y_{i,t} = \theta_1^\top \widetilde{M}_{i,t} + \bigl(A_{i,t} - p_t(1 \mid H_{i,t})\bigr)\,\beta^{*\top} \widetilde{M}_{i,t} + \epsilon_{i,t},
\]
where $\widetilde{M}_{i,t}$ denotes the centered moderators (innovations of the AR process), $\theta_1 \in \mathbb{R}^p$ is a dense nuisance coefficient, $\beta^* = (\beta_{11}\cdot\mathbf{1}_{|E^*|}^\top, \mathbf{0}_{p-|E^*|}^\top)^\top$ is the sparse treatment effect of interest, and $\{\epsilon_{i,t}\}$ are longitudinally correlated errors with $\mathrm{Corr}(\epsilon_{i,t},\epsilon_{i,s}) = \rho_{\mathrm{error}}^{|t-s|}$. The nuisance component $\theta_1$ is estimated via OLS of $Y$ on $\widetilde{M}$ on a held-out subset of $n/3$ individuals, and the residualized outcome is used for both lasso selection and inference. More details of the data-generating mechanism, parameter values, and evaluation metrics are given in Supplement~\ref{app:addsim}.

\paragraph{Methods \& findings.}
We compare the proposed selective-inference procedure, denoted as `SI', with three alternatives: (i) the polyhedral method \citep{zhao2021selective}; (ii) the data splitting approach, using $70\%$ of individuals for selection and $30\%$ for inference; and (iii) the naive approach, which ignores the selection step.  
In each setting, all competing methods produce confidence intervals for the selected working model, targeting FCR$=0.10$, equivalently $90\%$ coverage.
SI uses the randomized lasso in~\eqref{rand:lasso} with
\(\Omega=\tau\,\widehat\sigma^2 I_p\), whereas all other procedures use the standard lasso. In all cases, the lasso penalty is set to $\lambda = \mathrm{median}\{\|X^\top\epsilon\|_\infty : \epsilon \sim \mathcal{N}(0,\widehat{\sigma}^2 I_n)\}$  where the median is approximated using 2000 Monte Carlo draws, following \citet{Negahban}. 
The methods are evaluated over $500$ Monte Carlo replications.

We first investigate how the level of randomization affects selection quality, measured by recall, and inferential performance, measured by coverage, as shown in Figure~\ref{plot1}. We fix $n=60$ and $\beta_{11}=0.2$ and set the randomization covariance to $\Omega=\tau\times\widehat\sigma^2 I_p$, with $\tau\in\{0.5,1,2\}$ corresponding to low, moderate, and high levels of randomization, respectively.
SI attains the nominal coverage level under all three randomization settings. 
Selection quality is largely preserved under low and moderate randomization but deteriorates under high randomization. 
Excessive randomization can therefore compromise selection: although a noise variable selected at the first stage may subsequently be ruled out through inference, a true signal omitted during selection cannot be recovered; 
this asymmetry motivates using a small amount of randomization. 
Accordingly, all subsequent experiments use the low-randomization setting $\tau=0.5$, corresponding to $\Omega=\frac{1}{2}\widehat\sigma^2 I_p$.

\begin{figure}[ht] 
\centering
\includegraphics[width=1\textwidth]{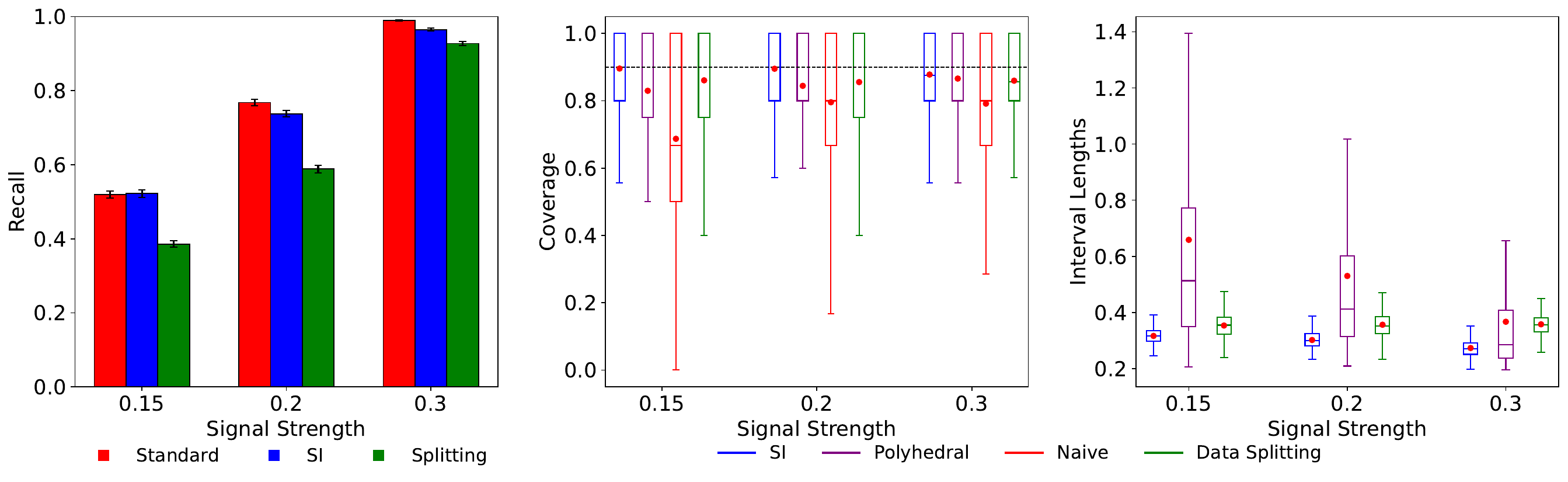}
 \caption{Recall (left), coverage (center), and bounded CI length (right), as in Figure~\ref{plot1}, for varying signal strength $\beta_{11}\in\{0.15,0.20,0.30\}$ with $n=60$ fixed. The polyhedral method yields infinitely long intervals in \textbf{\textcolor{red}{14.2\%}}, \textbf{\textcolor{red}{13\%}}, \textbf{\textcolor{red}{8.2\%}} of replications for $\beta_{11}=0.15,0.20,0.30$, respectively.}
\label{nsplot}
\end{figure}

\begin{figure}[ht] 
\centering
\includegraphics[width=1\textwidth]{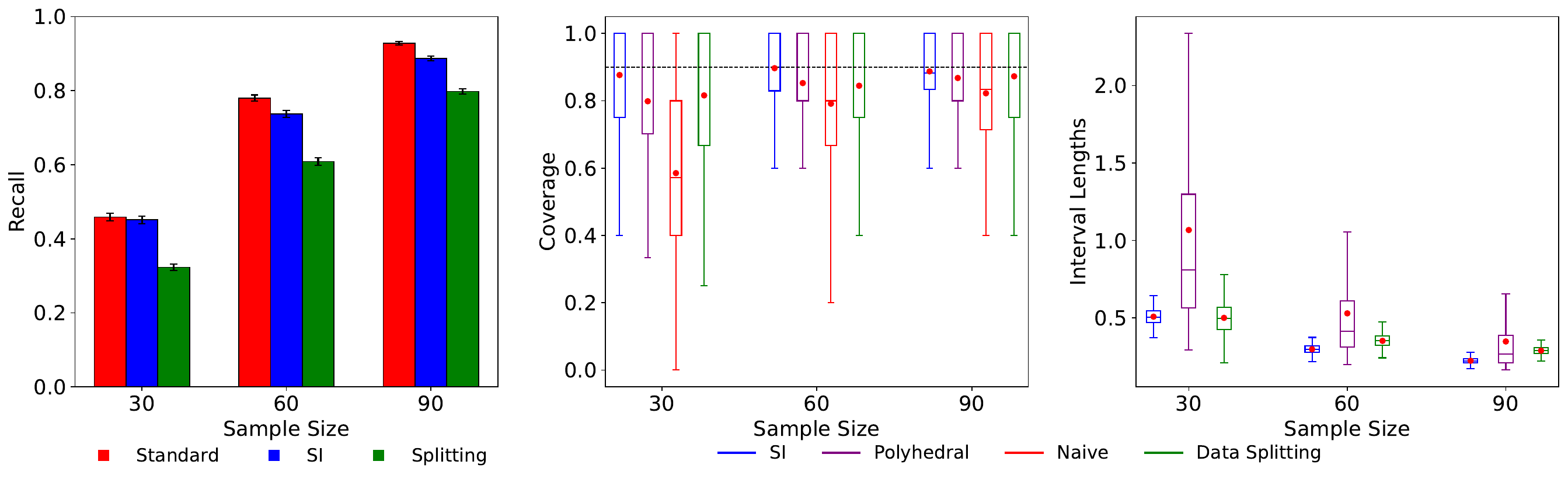}
\caption{Recall (left), coverage (center), and bounded CI length (right), as in Figure~\ref{plot1}, for varying sample size $n\in\{30,60,90\}$ with $\beta_{11}=0.2$ fixed. The polyhedral method yields infinitely long intervals in \textbf{\textcolor{red}{17.2\%}}, \textbf{\textcolor{red}{13.2\%}}, \textbf{\textcolor{red}{9\%}} of replications for $n=30,60,90$, respectively.}
\label{esplot}
\end{figure}

Figure~\ref{nsplot} displays results as signal strength varies over $\beta_{11} \in \{ 0.15, 0.2, 0.3\}$ with $n=60$ fixed, and Figure~\ref{esplot} displays results as the number of individuals varies over $n \in \{30, 60, 90\}$ with $\beta_{11}=0.2$ fixed. 
The key takeaways from these results are as follows:
\begin{enumerate}[leftmargin=*]
    \item The naive approach consistently under-covers, confirming the unreliability of inference that ignores selection. The polyhedral method and data splitting also under-cover at low signal strengths or small sample sizes, attaining nominal coverage only as the signal or sample size grows.

    \item Our method achieves nominal coverage across all signal regimes and sample sizes without sacrificing selection quality: the randomized lasso retains nearly the same recall as the full-data ordinary lasso and outperforms data splitting, which uses only part of the data for selection. This is practically important because valid inference is most informative when the selected model retains the relevant signals. 

   \item Interval lengths show a similarly sharp contrast. Polyhedral intervals are substantially wider than SI intervals, particularly under weak signals or small sample sizes, and have infinite length with non-negligible frequency even in moderate- and high-signal regimes. We therefore report this frequency for the polyhedral method separately in the captions and compute their average interval lengths only across replications in which all intervals are finite. SI also produces shorter intervals than data splitting as the signal strength or sample size increases. Moreover, SI yields shorter intervals as selection becomes easier, whereas intervals produced by  data splitting remain insensitive to selection difficulty because they are based on a fixed set of held-out samples.
\end{enumerate}

The instability of the polyhedral method arises when the selection event has low probability, which is not uncommon in high-dimensional regression. Conditioning on such an event leaves little information for inference, producing a hard-truncated distribution with low power and potential numerical instability. The resulting nonsmooth density can also yield poor asymptotic approximations in finite samples, as demonstrated in prior work; see, for example, \cite{wang2025asymptotically}. By contrast, even a small amount of randomization yields a smooth, soft-truncated density and preserves a nonzero amount of information after selection through the added noise, thereby improving stability, power, and the accuracy of asymptotic approximations made post selection.

Additional simulations in Supplement~\ref{app:addsim} assess sensitivity to correlation among moderators, model misspecification through the addition of a nonlinear baseline component to the generative model, and non-Gaussian outcome errors in the misspecified model, including Laplace and centered exponential errors. The conclusions are consistent across these settings: our proposed selective inference method maintains nominal coverage, while the naive method undercovers and the polyhedral method produces a substantial fraction of infinitely long intervals.

Python code to reproduce the experiments is available at our
\href{https://github.com/BakshiSoham/SI-MRT.git}{GitHub repository}.

\section{Data Application}
\label{sec:real}
We apply our method to the Virtual Application-supported Environment To Increase
Exercise (VALENTINE) Study \citep{VALENTINE}, a micro-randomized
cardiac-rehabilitation trial that delivers contextually tailored mobile-health
notifications to promote physical activity. We analyze Apple Watch data
($23{,}462$ observations from $62$ participants) with the log post-notification
step count as the proximal outcome. Effect modification---the contexts in
which notifications are most effective---is of primary interest. Study details, data
processing, and moderator construction are in Supplement~\ref{app:DA}.

\paragraph{Methods.} Selection requires a nuisance estimate of the outcome model. We split
participants by \texttt{id}, using $30\%$ to fit an OLS working model
\citep{dempsey2020} that supplies both the plug-in nuisance and the noise scale
$\widehat{\sigma}$, and the remaining $70\%$ for penalized selection and inference. As
in Section~\ref{sub:nuisance}, only a consistent nuisance estimator is
needed---not a correctly specified one---because the known randomization
probabilities protect the excursion-effect target. Our method uses randomized
lasso (low randomization, $\tau=0.5$); for comparison we run the
polyhedral and data splitting with
standard lasso, all sharing a common $\lambda$ chosen following \cite{Negahban}.

\begin{figure}[h]
    \centering
    \includegraphics[width=0.65\textwidth]{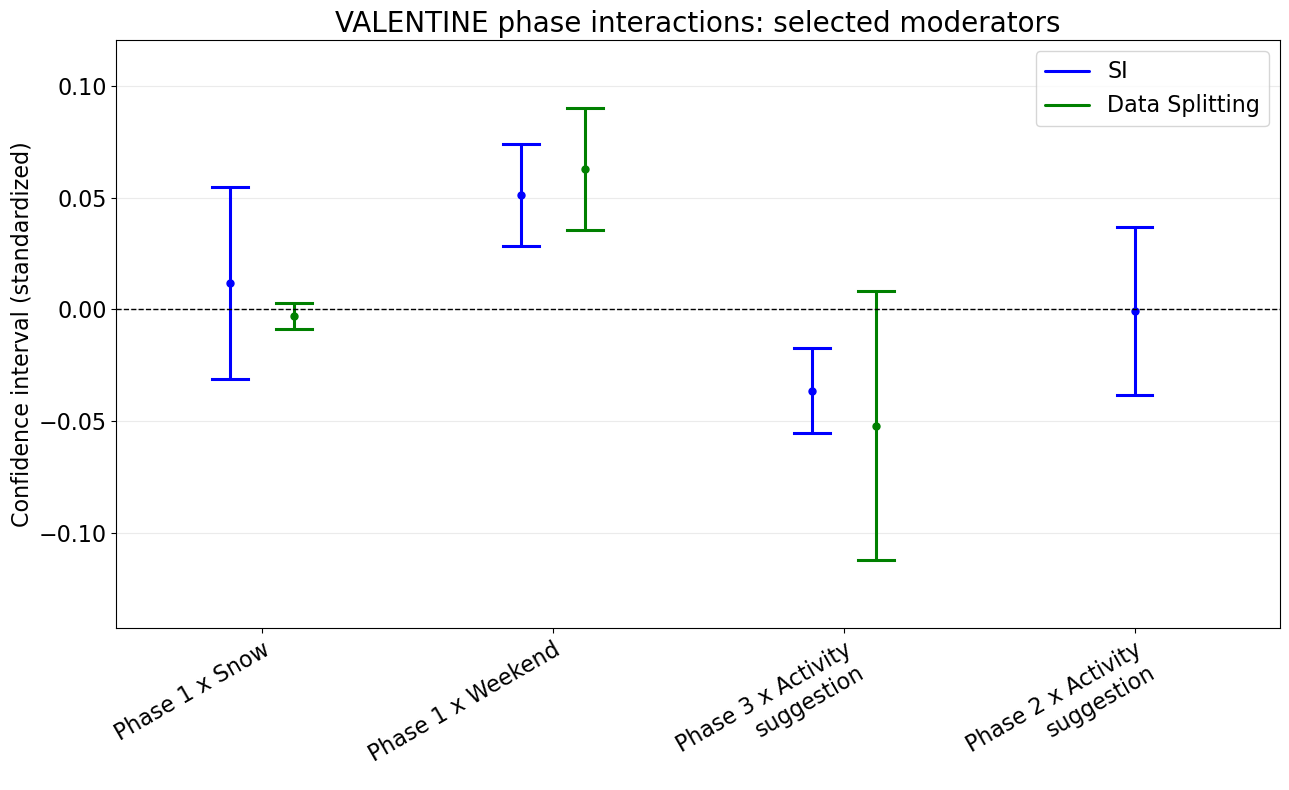}
    \caption{VALENTINE data application, phase--time interaction setting ($58$ candidate moderators): $90\%$ selective confidence intervals for the selected moderators, by method. SI selects $4$ moderators with average interval length $0.061$; data splitting selects $5$ moderators with average length $0.081$; the polyhedral method produces infinitely long intervals here and is omitted.}
    \label{fig:phase}
\end{figure}

We highlight the phase--time interaction setting, where $58$ candidate moderators
make selection necessary (\Figref{fig:phase}). The polyhedral intervals are infinite,
so we compare SI with data splitting. Both find a positive
Phase\,1\,$\times$\,weekend effect; SI additionally resolves a negative
Phase\,3\,$\times$\,activity-suggestion effect that splitting, limited to its
$30\%$ inference fold, cannot distinguish from zero. SI intervals are shorter on
average across selected moderators.

\paragraph{Results \& Interpretations.} Substantively, moderation concentrates early in rehabilitation and wanes over time: notifications are most effective on weekends and in Phase\,1, while generic activity-suggestion notifications become counterproductive by Phase\,3. Because \texttt{IsActivity} contrasts activity-suggestion messages (e.g.\ prompting a walk) against anti-sedentary messages (e.g.\ prompting a brief stretch or stand-up break) as the reference content, the negative Phase\,3\,$\times$\,activity-suggestion effect indicates that by the later phase, activity-suggestion notifications are comparatively less effective than, and can even undercut the benefit of, anti-sedentary notifications, while the latter continue to help. A plausible explanation is participant burden: anti-sedentary messages request a brief, low-effort action (standing up, a short stretch), whereas activity-suggestion messages ask for a more demanding action (going for a walk), which may become harder to sustain as rehabilitation progresses. Unlike one-at-a-time marginal screening, which flags a long, false-positive-prone list, our procedure returns a short, jointly valid set in which the time-in-study (phase) terms are retained, indicating that modification is driven by rehabilitation stage and recent-activity context rather than many weakly associated features. This is directly useful for tailoring: prioritize weekend and early-phase notifications and avoid generic activity prompts late in the program; a formal comparison of method-specific notification policies is a natural next step. Supplement~\ref{app:DA} reports additional analyses in the baseline ($25$-moderator) and extended ($94$-moderator) settings: in the baseline setting, our method identifies four significant moderators with valid, shorter intervals while splitting distinguishes none, and the Phase\,3\,$\times$\,activity effect recurs under our method across settings, confirming the power gain and its stability as the candidate set grows.

\section{Discussion}
\label{dis}
This paper develops a selective inference approach for causal effect moderation that allows researchers to first identify a simpler, interpretable model using penalized regression and then conduct uniformly valid semiparametric inference on the effects of the selected moderators. Unlike traditional approaches such as data splitting, our two-step method uses Gaussian randomization to leverage all observed samples for both selection and inference, while allowing analysts to control the trade-off between predictive accuracy and inferential precision. With a small amount of randomization, our method delivers valid selective inference, even in limited samples, while preserving the quality of model selection.

Our asymptotic framework can be extended in several directions. Penalties such as the group lasso, which aggregate variables measured across multiple time points within a specified window, can be applied to identify important moderators. Prior work by \cite{panigrahi2023approximate} and \cite{huang2023selective} introduced an approximate likelihood-based approach for such penalties; a pivot-based approach, as developed here, would eliminate the need for such approximations and is a natural extension. Our current work evaluates the reliability and power of the proposed pivot for a fixed $\lambda$; an important extension for practical data analysis is to accommodate $\lambda$ selected via cross-validation. Recent work by \cite{liu2025flexibleselectiveinferenceflowbased} accounts for cross-validated choices of $\lambda$ and could be readily combined with our pivot to yield valid selective inference. A comprehensive investigation of such a selective inference procedure under adaptive tuning is left for future work.

\section*{Supplementary Materials}
The online supplementary materials, organized into Sections~\ref{app:FCR}--\ref{app:DA} and referenced as Supplement~ throughout, contain the proofs of all theoretical results in the paper, extensions of our method to other penalized selection procedures and to the linear- and relative-risk contrasts, additional simulation results, and further details of the VALENTINE data application. Python code to reproduce all experiments is available at our \href{https://github.com/BakshiSoham/SI-MRT.git}{GitHub repository}.

\section*{Acknowledgments \& Funding}
Data used in the preparation of this article were obtained from the VALENTINE study (\url{https://www.nature.com/articles/s41746-023-00921-9}). The authors gratefully acknowledge Dr. Jessica Golbus for providing us access to this dataset. (uncomment for the final, unblinded version) W. Dempsey's research is supported in part by NIH grants 1R01GM152549-01 and 1P50DA054039-01. 

\section*{Disclosure Statement}
The authors report there are no competing interests to declare.

\section*{}

\bibliographystyle{plainnat}
\bibliography{references}

\newpage
\section*{Appendix: Table of notations and key matrices}
{\small
\renewcommand{\arraystretch}{1.1}\setlength{\tabcolsep}{3pt}
\begin{table}[h!]
\centering
\begin{tabular}{lcc}
\toprule
 Variable & Definition & Dimension \\
\midrule
\multicolumn{3}{l}{\emph{Estimators based on the randomized lasso solution}}\\[2pt]
$\lasso$ & Randomized lasso solution & $p$\\
$\widehat{E}\subseteq [p]$ & $\left\{j\in [p]: |\sign(\lassoj)|= 1\right\}$ & $q$ \\
$\lassoE$ & $(\lassoj: j\in E)$ & $q$ \\
$\widehat{S}_{n} =[\widehat{S}_{n,E}^{\top} \ \widehat{S}_{n,E'}^{\top}]^{\top}$ & $\left\{\partial \|b\|_1\right\}_{b= \lasso}$ & $p$ \\
$\widehat{U}_n^{E\cdot j}$ & $\sqrt{n}(\eta^{E\cdot j})^{\top}|\lassoE|$  & $1$\\
$\widehat{V}_n^{E\cdot j}$ & $\sqrt{n}\nbracket{I_{q}-  Q^{E\cdot j} (\eta^{E\cdot j}) ^{\top}} |\lassoE|$   & $q-1$\,$^{\dagger}$\\
\midrule
\multicolumn{3}{l}{\emph{Key matrices and vectors}}\\[2pt]
$A_{1}$ & $\begin{bmatrix}-R_j H_{E,E}^{-1} & 0_{p-q}^{\top}\end{bmatrix} K^{1/2}$ & $1\times p$ \\[4pt]
$A_{2}$ & $\begin{bmatrix} 
R_{E\setminus j}\big(\tfrac{1}{(\sigma^{E\cdot j})^{2}}\Sigma_{E,j}R_j H_{E,E}^{-1}-H_{E,E}^{-1}\big) & 0_{q-1,p-q} \\
- K_{E',E}K_{E,E}^{-1} & I_{p-q,p-q} 
\end{bmatrix} K^{1/2}$ & $(p-1)\times p$ \\[4pt]
$P_{1}^{E\cdot j}$ & $-\tfrac{1}{(\sigma^{E\cdot j})^{2}}K_{E}H_{E,E}^{-1}R_j^{\top}$ & $p$ \\[4pt]
$P_{2}^{E\cdot j}$ & $\begin{bmatrix} 
- H_{E,E}R_{E\setminus j}^{\top} & 0_{q,\,p-q}\\
- K_{E',E}K_{E,E}^{-1}H_{E,E}R_{E\setminus j}^{\top} & I_{p-q,\,p-q}
\end{bmatrix}$ & $p\times (p-1)$ \\[4pt]
$\widetilde{H}_{E}$ & $H_{E}\,\Dg(S_{E})$ & $p\times q$ \\
$\Lambda$ & $(\widetilde{H}_E^\top \Omega^{-1} \widetilde{H}_E)^{-1}$ & $q\times q$ \\
$\eta^{E\cdot j}$ & $\widetilde{H}_E^\top\Omega^{-1}P_{1}^{E\cdot j}$ & $q$ \\
$Q^{E\cdot j}$ & $\Lambda \eta^{E\cdot j}/\{(\eta^{E\cdot j})^\top \Lambda \eta^{E\cdot j}\}$ & $q$ \\
$L^{E\cdot j}$ & $\widetilde{H}_E Q^{E\cdot j}$ & $p$ \\[2pt]
$J_{\mathrm{block}}$ & $\begin{bmatrix} H_{E,E}\Dg(S_E) & 0_{q,p-q} \\ H_{E',E}\Dg(S_E) & \lambda I_{p-q,p-q} \end{bmatrix}$, \ $|\det(J_{\mathrm{block}})| = |\det(H_{E,E})|\,\lambda^{p-q}$ & $p\times p$ \\
\bottomrule
\end{tabular}
\caption{Notation for estimators based on the randomized lasso solution (top) and key matrices and vectors used in our results (bottom).
$^{\dagger}$ denotes effective dimension rather than dimension of the vector; since $(\eta^{E\cdot j})^{\top}\widehat{V}_n^{E\cdot j}=0$, and therefore, its effective dimension equals $q-1$.}
\label{Table:matrices}
\end{table}}

\clearpage
\renewcommand{\thesection}{\Alph{section}}
\renewcommand{\thetable}{\Alph{table}}
\renewcommand{\theHsection}{supp.\Alph{section}}
\renewcommand{\theHsubsection}{\theHsection.\arabic{subsection}}
\renewcommand{\theHequation}{\theHsection.\arabic{equation}}
\renewcommand{\theHtable}{supp.\Alph{table}}
\renewcommand{\theHtheorem}{\theHsection.\arabic{theorem}}
\renewcommand{\theHproposition}{\theHsection.\arabic{proposition}}
\renewcommand{\theHcorollary}{\theHsection.\arabic{corollary}}
\renewcommand{\theHremark}{\theHsection.\arabic{remark}}
\renewcommand{\theHassumption}{\theHsection.\arabic{assumption}}
\setcounter{section}{0}
\setcounter{table}{0}
\numberwithin{equation}{section}

\let\bibliography\arxivbibliography
\let\bibliographystyle\arxivbibliographystyle

\spacingset{1.9}

\begin{center}
{\large\bf Supplementary Material for:\\  Selective Inference for Time-Varying Moderated Effects}
\end{center}
\section{False Coverage Rate Control}
\label{app:FCR}
We now show that the conditional coverage guarantee in~\eqref{cond:guarantee} implies control of the false coverage rate (FCR). 
Our proof closely follows the arguments in \cite{benjamini2005false} and the treatment of post-selection inference in \cite{Lee_2016}, but we adapt the notation to our present setting.

\begin{lemma}[FCR Control]\label{lemma:fcr}
Suppose that for every model $E$ and coordinate $j\in E$,
\[
\mathbb{P}\!\left[\beta^{E\cdot j}\in \widehat{C}^{E\cdot j} \,\Big|\, \widehat{E}=E \right] \geq 1-\alpha.
\]
Then the false coverage rate
\[
\mathrm{FCR} := \mathbb{E}\!\left[\dfrac{\Big|\{j\in \widehat{E}: \beta^{\widehat{E}\cdot j} \notin \widehat{C}^{\widehat{E}\cdot j}\}\Big|}{\max(|\widehat{E}|,1)} \right]
\]
is controlled at level $\alpha$, i.e.
\[
\mathrm{FCR} \leq \alpha.
\]
\end{lemma}

\begin{proof}
By assumption, for any fixed $E$ and $j\in E$,
\[
\mathbb{P}\!\left[\beta^{E\cdot j}\notin \widehat{C}^{E\cdot j} \,\Big|\, \widehat{E}=E\right] \leq \alpha.
\]
Summing over $j\in E$ yields
\[
\mathbb{E}\!\left[\Big|\{ j\in E : \beta^{E\cdot j}\notin \widehat{C}^{E\cdot j}\}\Big| \,\Big|\, \widehat{E}=E \right] \leq \alpha |E|.
\]
Dividing both sides by $\max(|E|,1)$ gives
\[
\mathbb{E}\!\left[\frac{\Big|\{ j\in E: \beta^{E\cdot j}\notin \widehat{C}^{E\cdot j}\}\Big|}{\max(|E|,1)} \,\Big|\, \widehat{E}=E \right] \leq \alpha.
\]
Taking expectation with respect to the randomness in $\widehat{E}$ and applying the tower property, we obtain
\[
\mathrm{FCR} = \mathbb{E}\!\left[\mathbb{E}\!\left[\frac{\Big|\{ j\in \widehat{E}: \beta^{\widehat{E}\cdot j}\notin \widehat{C}^{\widehat{E}\cdot j}\}\Big|}{\max(|\widehat{E}|,1)} \,\Big|\, \widehat{E}\right]\right] \leq \alpha.
\]
\end{proof}

\begin{corollary}[Average Coverage]\label{cor:avg}
Under the conditions of Lemma~\ref{lemma:fcr}, the average coverage across selected parameters is at least $1-\alpha$, i.e.
\[
\mathbb{E}\!\left[\frac{1}{\max(|\widehat{E}|,1)} \sum_{j\in \widehat{E}} 
\mathbbm{1}\!\left\{\beta^{\widehat{E}\cdot j}\in \widehat{C}^{\widehat{E}\cdot j}\right\}\right] \geq 1-\alpha.
\]
\end{corollary}

\begin{proof}
Note that
\[
\frac{1}{\max(|\widehat{E}|,1)} \sum_{j\in \widehat{E}} 
\mathbbm{1}\!\left\{\beta^{\widehat{E}\cdot j}\in \widehat{C}^{\widehat{E}\cdot j}\right\}
= 1 - \frac{\Big|\{ j\in \widehat{E} : \beta^{\widehat{E}\cdot j}\notin \widehat{C}^{\widehat{E}\cdot j}\}\Big|}{\max(|\widehat{E}|,1)}.
\]
Taking expectations and invoking Lemma~\ref{lemma:fcr} gives the stated bound.
\end{proof}

\section{Causal Assumptions}
\label{app:causalassumptions}
We state the three fundamental assumptions used in Section~\ref{sec:intro} to reformulate the causal excursion effect in terms of the observed data \cite{robins97}:
\begin{assumption}
\label{assmp:causal}
     We assume consistency, positivity, and sequential ignorability. For $t=1, \ldots, T$:
     \begin{enumerate}
         \item {\bf Consistency:}
         $\{X_{i,t}\left(\bar{A}_{i,t-1}\right), A_{i,t}\left(\bar{A}_{i,t-1}\right), Y_{i,t}\left(\bar{A}_{i,t}\right)\}=\left\{X_{i,t}, A_{i,t}, Y_{i,t}\right\},$ i.e., observed values equal the corresponding potential outcomes;
         \item {\bf Positivity:} $\mathbb{P}\left[A_{i,t}=a\mid H_{i,t}=h\right]>0$ given that the joint density for $\left\{A_{i,t}, H_{i,t}\right\}$ when evaluated at $(a, h)$, is greater than zero;
         \item {\bf Sequential ignorability:} Conditional on the observed history $H_{i,t}$, \[\left\{Y_{i,t}\left(\bar{a}_{i,t}\right), X_{i,t+1}\left(\bar{a}_{i,t}\right), A_{i,t+1}\left(\bar{a}_{i,t}\right), \ldots, Y_{i,T}\left(\bar{a}_{i,T-1}\right)\right\} \indep A_{i,t},\] i.e., the potential outcomes are independent of the current treatment given the past history.
     \end{enumerate}
\end{assumption}

\section{Proofs of results in Section~\ref{sec:SI}}
\label{app:SIproofs}

\begin{corollary}
\label{cor:signevents}
As a direct consequence of Proposition~\ref{prop:KKTmaster}, it holds that \begin{align*}\{\widehat{S}_{n,E}= S_E\} =  \left\{\sign\left(H_{E,E}^{-1}\left\{\sqrt{n}\widetilde{\omega}_{n,E} - R_{E} \nbracket{P_{1}^{E\cdot j} \sqrt{n}\estj  + P_{2}^{E\cdot j}\sqrt{n}\gammaj} - \lambda S_E \right\}\right) =S_E \right\} 
\\
\left\{ \widehat{S}_{n,E'}  =S_{E'}\right\}= \Big\{ \frac{1}{\lambda} \Big[\sqrt{n}\widetilde{\omega}_{n,E'} - R_{E'} \nbracket{P_{1}^{E\cdot j} \sqrt{n}\estj  + P_{2}^{E\cdot j}\sqrt{n}\gammaj} -H_{E',E}H_{E,E}^{-1}\Big\{\sqrt{n}\widetilde{\omega}_{n,E}\\
\qquad\qquad- R_{E} \nbracket{P_{1}^{E\cdot j} \sqrt{n}\estj  + P_{2}^{E\cdot j}\sqrt{n}\gammaj} - \lambda S_{E} \Big\}\Big] =S_{E'}\Big\}.
\end{align*}
\end{corollary}
\begin{proof}
Both identities follow by partitioning the stationarity relation of
Proposition~\ref{prop:KKTmaster} into its $E$ and $E'$ blocks and solving for the
subgradients $\widehat{S}_{n,E}$ and $\widehat{S}_{n,E'}$, respectively.
\end{proof}

\begin{proof}[Proof of Proposition \ref{prop:asymptotic}]
Throughout this proof and the rest of the supplementary materials, $A_{1} \in \mathbb{R}^{1\times p}$ and $A_{2}\in \mathbb{R}^{(p-1)\times p}$ denote the matrices defined in Table~\ref{Table:matrices} of the main paper, namely
\begin{equation*}
A_{1} = \begin{bmatrix}-R_j H_{E,E}^{-1} & 0_{p-q}^{\top}\end{bmatrix} K^{1/2}, \qquad
A_{2} = \begin{bmatrix} 
R_{E\setminus j}\big(\tfrac{1}{(\sigma^{E\cdot j})^{2}}\Sigma_{E,j}R_j H_{E,E}^{-1}-H_{E,E}^{-1}\big) & 0_{q-1,p-q} \\
- K_{E',E}K_{E,E}^{-1} & I_{p-q,p-q} 
\end{bmatrix} K^{1/2},
\end{equation*}
and $b_{i,n}$ and $\zeta_{n}$ are as in the statement of Proposition~\ref{prop:asymptotic}, with the shorthand $a_{i,n} = K^{-1/2}\nbracket{b_{i,n}-\mathbb{E}_{\mathbb{F}_n}\rbracket{b_{i,n}}}$, so that $\zeta_{n}=\frac{1}{\sqrt{n}}\sum_{i=1}^{n}a_{i,n}$.

To start, note that $\frac{1}{\sqrt{n}}\sum_{i=1}^{n} X_{i,E}^{\top}\nabla\psi(X_{i,E}\estE;Y_{i})= 0$, and Taylor expanding the expression on the left-hand side display around $\targetE$, we have that:
 \begin{align}
     & \frac{1}{\sqrt{n}}\sum_{i=1}^{n}X_{i,E}^{\top}\nabla\psi\nbracket{X_{i,E}\targetE ;Y_{i}} - \frac{1}{n} \sum_{i=1}^{n} X_{i,E}^{\top}
\nabla^{2}\psi\nbracket{X_{i,E}\targetE ; Y_{i}}X_{i,E}\sqrt{n}(\estE - \targetE) = 0 \\
& \implies \sqrt{n}(\estE - \targetE) = -H_{E,E}^{-1}\frac{1}{\sqrt{n}} \sum_{i=1}^{n} X_{i,E}^{\top}\nabla\psi\nbracket{X_{i,E}\targetE ; Y_{i}}+ o_{p}(1). 
\label{TS:1}
 \end{align}Multiplying both sides of \eqref{TS:1} with the matrix $R_{j}$, we obtain
\begin{align*}
     \sqrt{n}(\estj - \targetj) &= -R_{j} H_{E,E}^{-1}\frac{1}{\sqrt{n}} \sum_{i=1}^{n} X_{i,E}^{\top}\nabla\psi\nbracket{X_{i,E}\targetE ; Y_{i}}+ o_{p}(1) = A_{1}\zeta_{n} + o_{p}(1). 
\end{align*} Next, observe that $\gammaj  = \begin{bmatrix} T_{1}
 \\ T_{2} \end{bmatrix} \in \mathbb{R}^{p-1},$ where 
\begin{align}
     & T_{1}= R_{E\setminus j}\nbracket{\estE - \frac{1}{(\sigma^{E\cdot j})^{2}}\Sigma_{E,j}\estj} \in \mathbb{R}^{q-1}  \text{ and } \\
     & T_{2} = \frac{1}{n}\sum_{i=1}^{n}X_{i,E'}^{\top}\nabla\psi\nbracket{X_{i,E}\estE ; Y_{i}}  + (K_{E',E}K_{E,E}^{-1}H_{E,E}-H_{E',E})\estE \in \mathbb{R}^{p-q}. 
\label{T1:T2}
\end{align} 
Multiplying both sides of \eqref{TS:1} with $R_{E\setminus j}$, we obtain that
\begin{align*}
&\sqrt{n}\nbracket{T_{1} - \mathbb{E}[T_{1}]} =R_{E\setminus j}\sqrt{n}(\estE - \targetE) - \frac{1}{(\sigma^{E\cdot j})^{2}}R_{E\setminus j}\Sigma_{E,j}\sqrt{n}(\estj - \targetj) \\
&= R_{E\setminus j} \nbracket{-H_{E,E}^{-1} + \frac{1}{(\sigma^{E\cdot j})^{2}}\Sigma_{E,j}R_{j}H_{E,E}^{-1}}\frac{1}{\sqrt{n}} \sum_{i=1}^{n} X_{i,E}^{\top}\nabla\psi\nbracket{X_{i,E}\targetE ; Y_{i}}+ o_{p}(1).
\end{align*}
Next, observe that Taylor expanding $\frac{1}{\sqrt{n}}\sum_{i=1}^{n}X_{i,E'}^{\top}\nabla\psi(X_{i,E}\estE ; Y_{i}) \in \mathbb{R}^{p-q}$ around $\targetE$ yields: \begin{align*}
&\frac{1}{\sqrt{n}}\sum_{i=1}^{n}X_{i,E'}^{\top}\nabla\psi(X_{i,E}\estE ; Y_{i})\\
=& \frac{1}{\sqrt{n}}\sum_{i=1}^{n}X_{i,E'}^{\top}\nabla\psi(X_{i,E}\targetE;Y_{i}) - \frac{1}{\sqrt{n}}\sum_{i=1}^{n}X_{i,E'}^{\top}\nabla^{2}\psi(X_{i,E}\targetE ;Y_{i})X_{i,E}\nbracket{\estE - \targetE} \\
=& \frac{1}{\sqrt{n}}\sum_{i=1}^{n}X_{i,E'}^{\top}\nabla\psi(X_{i,E}\targetE;Y_{i}) -  H_{E',E}\sqrt{n}\nbracket{\estE - \targetE} + o_{p}(1).
\end{align*}
Adding $\sqrt{n}(K_{E',E}K_{E,E}^{-1}H_{E,E}-H_{E',E})\estE$ to both sides of the last display, we obtain that:
\[
\sqrt{n} T_{2} = \frac{1}{\sqrt{n}}\sum_{i=1}^{n}X_{i,E'}^{\top}\nabla\psi(X_{i,E}\targetE ;Y_{i}) +  K_{E',E}K_{E,E}^{-1}H_{E,E}\sqrt{n}\estE - H_{E',E} \sqrt{n}\targetE.
\]
This leads us to note that
\begin{align*}
\sqrt{n} \nbracket{T_{2} - \mathbb{E}[T_{2}]} &= \frac{1}{\sqrt{n}}\nbracket{\sum_{i=1}^{n}X_{i,E'}^{\top}\nabla\psi(X_{i,E}\targetE ;Y_{i}) - \mathbb{E}\rbracket{\sum_{i=1}^{n}X_{i,E'}^{\top}\nabla\psi(X_{i,E}\targetE ;Y_{i})}} \\
& \;\;\;\; +  K_{E',E}K_{E,E}^{-1}H_{E,E}\sqrt{n}(\estE-\targetE) \\
&= \frac{1}{\sqrt{n}}\nbracket{\sum_{i=1}^{n}X_{i,E'}^{\top}\nabla\psi(X_{i,E}\targetE ; Y_{i}) - \mathbb{E}\rbracket{\sum_{i=1}^{n}X_{i,E'}^{\top}\nabla\psi(X_{i,E}\targetE ;Y_{i})}} \\
& \;\;\;\;-  K_{E',E}K_{E,E}^{-1}\frac{1}{\sqrt{n}} \sum_{i=1}^{n} X_{i,E}^{\top}\nabla\psi\nbracket{X_{i,E}\targetE ; Y_{i}}+ o_{p}(1).
\end{align*}The claims now follow upon expressing the displays above in terms of the centered score vector $\zeta_{n}$; note that the $E$-block of the score is automatically centered, since \[\mathbb{E}_{\mathbb{F}_n}\rbracket{\sum_{i=1}^{n}X_{i,E}^{\top}\nabla\psi(X_{i,E}\targetE ;Y_{i})}=0\] by the definition of $\targetE$ in \eqref{refit}.
The second part of the claim is a direct consequence of the asymptotic representation in the first part.
\end{proof}

\begin{proof}[Proof of Proposition \ref{prop:KKTmaster}]
A Taylor expansion of K.K.T. stationarity conditions around $\estE$ yields:
\begin{align*}
\sqrt{n}\omega_{n}+ R_{n,2} &=   \frac{1}{\sqrt{n}}\sum_{i=1}^{n}X_{i}^{\top}\nabla\psi(X_{i,E}\estE ;Y_{i}) +\frac{1}{\sqrt{n}}\sum_{i=1}^{n}X_{i}^{\top}\nabla^{2}\psi(X_{i,E}\estE ; Y_{i})X_{i,E}\nbracket{\lassoE-\estE} + \lambda \rbracket{ S_{E}^{\top} \ S_{E'}^{\top}}^{\top}
\end{align*}where $R_{n,2}$ is an $o_{p}(1)$ term with the Taylor series remainder.
Using the fact that \[\sum_{i=1}^{n}X_{i,E}^{\top}\nabla\psi(X_{i,E}\estE ;Y_{i})=0\] and replacing the sample average $\frac{1}{n}\sum_{i=1}^{n}X_{i}^{\top}\nabla^{2}\psi(X_{i,E}\estE ; Y_{i})X_{i,E}$ with its expected value $H_E$, the above equation can be expressed as
\[ \begin{bmatrix}
         0_{E} \\
         \frac{1}{\sqrt{n}}\sum_{i=1}^{n}X_{i,E'}^{\top}\nabla\psi\nbracket{X_{i,E}\estE ; Y_{i}}\\
     \end{bmatrix} +  \begin{bmatrix}
          H_{E,E} \\ H_{E',E}\end{bmatrix} \sqrt{n}\lassoE -  \begin{bmatrix}
          H_{E,E} \\ H_{E',E}\end{bmatrix} \sqrt{n}\estE  + \lambda \begin{bmatrix}
          S_{E} \\  S_{E'} \end{bmatrix}  = \sqrt{n}\omega_{n}+ R_{n,2},\]
where $R_{n,2}$ is an $o_{p}(1)$ term. Recall the definition  of $\gammaj  = \begin{bmatrix} T_{1}
 \\ T_{2} \end{bmatrix} \in \mathbb{R}^{p-1}$ in \eqref{T1:T2}.
Since, 
\[\frac{1}{\sqrt{n}}\sum_{i=1}^{n}X_{i,E'}^{\top}\nabla\psi\nbracket{X_{i,E}\estE ; Y_{i}} = \sqrt{n}T_{2}+\nbracket{H_{E',E} - K_{E',E}K_{E,E}^{-1}H_{E,E}} \sqrt{n}\estE,\]
we further obtain that
\[ \begin{bmatrix}
         0_{q} \\
         \sqrt{n}T_{2}\\
     \end{bmatrix} - \begin{bmatrix}
          H_{E,E} \\ K_{E',E}K_{E,E}^{-1}H_{E,E}\end{bmatrix} \sqrt{n} \estE+ H_{E}\sqrt{n}\lassoE   + \lambda \begin{bmatrix}
         S_{E} \\  S_{E'} \end{bmatrix}= \sqrt{n}\omega_{n} + R_{n,2}.\] 
For the second term on the left-hand side display, note that  
\begin{align*}
\begin{bmatrix}H_{E,E} \\ K_{E',E}K_{E,E}^{-1}H_{E,E}\end{bmatrix}
\sqrt{n} \estE =& \begin{bmatrix}H_{E,E} \\ K_{E',E}K_{E,E}^{-1}H_{E,E}\end{bmatrix}\Big(\sqrt{n} \estE - \frac{1}{(\sigma^{E\cdot j})^{2}}\Sigma_{E,j}\sqrt{n}\estj \\
&\qquad\qquad+ \frac{1}{(\sigma^{E\cdot j})^{2}}\Sigma_{E,j}\sqrt{n}\estj\Big)\\
          =& \begin{bmatrix}H_{E,E}R_{E\setminus j}^{\top} \\ K_{E',E}K_{E,E}^{-1}H_{E,E}R_{E\setminus j}^{\top}\end{bmatrix} R_{E\setminus j} \sqrt{n}\nbracket{\estE - \frac{1}{(\sigma^{E\cdot j})^{2}}\Sigma_{E,j}\estj}\\
          & \;\;+ \begin{bmatrix}H_{E,E} \\ K_{E',E}K_{E,E}^{-1}H_{E,E}\end{bmatrix} \frac{1}{(\sigma^{E\cdot j})^{2}}\Sigma_{E,j}\sqrt{n}\estj \\
          =& \begin{bmatrix}H_{E,E}R_{E\setminus j}^{\top} \\ K_{E',E}K_{E,E}^{-1}H_{E,E}R_{E\setminus j}^{\top}\end{bmatrix} \sqrt{n}T_{1}  + \frac{1}{(\sigma^{E\cdot j})^{2}}K_{E}H_{E,E}^{-1}R_{j}^{\top}\sqrt{n}\estj.
\end{align*}
Therefore, we have
\[\begin{bmatrix}
         0_{q} \\
         \sqrt{n}T_{2}\\
     \end{bmatrix} - \begin{bmatrix}
          H_{E,E} \\ K_{E',E}K_{E,E}^{-1}H_{E,E}\end{bmatrix} \sqrt{n} \estE = P_{1}^{E\cdot j} \sqrt{n}\estj  + P_{2}^{E\cdot j}\sqrt{n}\gammaj,\]
which leads to our assertion. 
\end{proof}

\begin{proof}[Proof of Proposition \ref{prop:condevent}]
Firstly observe \[\cbracket{\widehat{S}_n= S, \widehat{V}_n^{E\cdot j} = V^{E\cdot j}}= \cbracket{\Dg(S_{E})\lassoE > 0, \widehat{S}_{n, E'}= S_{E'}, \widehat{V}_n^{E\cdot j} = V^{E\cdot j}}.\] Now, basic algebra leads to
\begin{align*}
\sqrt{n}\Dg(S_E)\lassoE = \sqrt{n} |\lassoE| &= \nbracket{I_{q}-  Q^{E\cdot j} (\eta^{E\cdot j}) ^{\top}} \sqrt{n}|\lassoE| + Q^{E\cdot j} (\eta^{E\cdot j})^{\top}\sqrt{n}|\lassoE|\\
&= \widehat{V}_n^{E\cdot j} + Q^{E\cdot j} \unicond.
\end{align*} 
Thus when $\widehat{V}_n^{E\cdot j} = V^{E\cdot j}$ the constraint $\Dg(S_{E})\lassoE > 0$ becomes  \[Q^{E\cdot j} \unicond > - V^{E\cdot j} \iff \max _{k\in E: R_{k}Q^{E\cdot j}>0}\left\{\frac{-R_{k} V^{E\cdot j}}{R_{k}Q^{E\cdot j}}\right\} < \unicond < \min _{k\in E: R_{k}Q^{E\cdot j}<0}\left\{\frac{-R_{k} V^{E\cdot j}}{R_{k} Q^{E\cdot j}}\right\}.\]
\end{proof}

\begin{proof}[Proof of Theorem \ref{thm:CoV}]
First note for a fixed set $E$ and signs $S_{E}$, $\sqrt{n}\estj$, $\sqrt{n}\gammaj$, and $\sqrt{n}\omega_{n}$ are independent variables and have joint normal density
\[\phi(\sqrt{n}\estj;\sqrt{n}\targetj, \sigmaj)\phi\nbracket{\sqrt{n}\gammaj;\sqrt{n}\gammaej, \sigmaE}\phi(\sqrt{n}\omega_{n} ; 0, \Omega).\] Using the mapping in \eqref{CoV:map} in the main paper and Proposition~\ref{prop:condevent}, we know that \[\Pi_{\sqrt{n}\estj, \sqrt{n}\gammaj}(\unicond, \widehat{V}_n^{E\cdot j}, \widehat{S}_{n, E'}) = \sqrt{n} \widetilde{\omega}_n = \sqrt{n}\omega_n + R_{n,2}.\]

Also, as $\widetilde{H}_{E}\sqrt{n}|\lassoE| = L^{E\cdot j}\unicond +  \widetilde{H}_{E}\extracond$, the function can be seen as 
\begin{align*}
 \widetilde{\Pi}_{\sqrt{n}\estj, \sqrt{n}\gammaj}(\sqrt{n}|\lassoE|, \widehat{S}_{n, E'})&=  \widetilde{H}_{E}\sqrt{n}|\lassoE| + \lambda \begin{bmatrix}
             S_{E} \\ \widehat{S}_{n,E'}
         \end{bmatrix} + P_{1}^{E\cdot j} \sqrt{n}\estj  + P_{2}^{E\cdot j}\sqrt{n}\gammaj \\
&= \Pi_{\sqrt{n}\estj, \sqrt{n}\gammaj}(\unicond, \widehat{V}_n^{E\cdot j}, \widehat{S}_{n, E'}).
\end{align*}To derive the density for the optimization variables, consider the following change of variables $f : \mathbb{R}^{2p} \to \mathbb{R}^{2p}$, for $(\sqrt{n}\estj, \sqrt{n}\gammaj, \sqrt{n}|\lassoE|, \widehat{S}_{n, E'})\in \mathbb{R} \times \mathbb{R}^{p-1} \times\mathbb{R}^{q} \times\mathbb{R}^{p-q}$

\begin{align*}
f(\sqrt{n}\estj, \sqrt{n}\gammaj, \sqrt{n}|\lassoE|, \widehat{S}_{n, E'}) &= \begin{bmatrix}
  \sqrt{n}\estj \\ \sqrt{n}\gammaj \\ \widetilde{\Pi}_{\sqrt{n}\estj, \sqrt{n}\gammaj}(\sqrt{n}|\lassoE|, \widehat{S}_{n, E'}) \end{bmatrix} \text{ so } \\
f : &  \begin{bmatrix}
  \sqrt{n}\estj \\ \sqrt{n}\gammaj \\ \sqrt{n}|\lassoE| \\ \widehat{S}_{n, E'}
\end{bmatrix}
 \mapsto \begin{bmatrix}
  \sqrt{n}\estj \\ \sqrt{n}\gammaj \\ \sqrt{n}\widetilde{\omega}_{n}.
\end{bmatrix}
\end{align*}Hence, asymptotically the joint density of the variables $\sqrt{n}\estj, \sqrt{n}\gammaj,  \sqrt{n}|\lassoE|, \widehat{S}_{n, E'}$ (with $\pi$ denoting the joint normal density displayed at the start of this proof) is equal to 
\begin{align*}
    &= |\det(J)|\,\pi\nbracket{f(\sqrt{n}\estj, \sqrt{n}\gammaj,  \sqrt{n}|\lassoE|, \widehat{S}_{n, E'})} \\
    &= |\det(J)| \phi(\sqrt{n}\estj;\sqrt{n}\targetj, (\sigma^{E\cdot j})^{2})\phi\nbracket{\sqrt{n}\gammaj;\sqrt{n}\gammaej, \sigmaE} \\
   & \times \phi\nbracket{P_{1}^{E\cdot j} \sqrt{n}\estj  + P_{2}^{E\cdot j}\sqrt{n}\gammaj + \widetilde{H}_{E}\sqrt{n}|\lassoE| + 
         \lambda \begin{bmatrix} S_{E} \\ 
        \widehat{S}_{n,E'} \end{bmatrix};0_{p},\Omega},
\end{align*}
where 
\begin{align*}
 J&= \begin{bmatrix}
    1 & 0 & 0 & 0 \\
    0 & I & 0 & 0 \\
    R_{E}P_{1}^{E\cdot j} & R_{E}P_{2}^{E\cdot j} & H_{E,E}\Dg(S_E) & 0 \\
    R_{E'}P_{1}^{E\cdot j} & R_{E'}P_{2}^{E\cdot j} & H_{E',E}\Dg(S_E) & \lambda I \\
\end{bmatrix} \ \in \mathbb{R}^{2p\times 2p}.
\end{align*}
Since $J$ is block lower-triangular, its determinant equals that of its lower-right block, which is the matrix $J_{\mathrm{block}}$ of Table~\ref{Table:matrices}:
\begin{align*}
|\det(J)| = \left|\det(J_{\mathrm{block}})\right| = \left|\det \begin{bmatrix}
    H_{E,E}\Dg(S_E) & 0 \\
    H_{E',E}\Dg(S_E) & \lambda I \end{bmatrix}\right| = |\det(H_{E,E})|\,\lambda^{p-q},
\end{align*}
where the last equality uses $|\det\Dg(S_E)|=1$, since $S_E\in\{\pm1\}^{q}$.
\end{proof}
         
\section{Asymptotic theory: proofs of results in Section~\ref{sec:theorypivot}}
\label{app:asymp}
\subsection{Main Results}
\begin{remark}
The exact pivot in the main paper is written for the finite truncation interval $[I_{-}^{E\cdot j},I_{+}^{E\cdot j}]$. The asymptotic bounds below are stated for the one-sided lower-tail weight $[c,\infty)$, and we write $c=I_{-}^{E\cdot j}$ to keep notation light. For a finite interval, the selection weight is the difference of two such lower-tail weights, evaluated at $c=I_{-}^{E\cdot j}$ and $c=I_{+}^{E\cdot j}$; the derivative and Stein bounds below apply to the two endpoint weights term by term. The lower-bound condition needed for the resulting finite-interval denominator is encoded in Assumption~\ref{assump:2} of the main paper.
\end{remark}
Before developing the proofs for the asymptotic theory in this paper, we establish a few more notations. 
Recall from Proposition~\ref{prop:asymptotic} that our master statistics satisfy the following representation:
\[
\sqrt{n}\begin{bmatrix}
    \estj- \targetj \\
\gammaj -  \gammaej
\end{bmatrix} = \begin{bmatrix}
   A_1 \\
A_2 
\end{bmatrix}
\zeta_n+ R_{n,1},
\]
 where $R_{n,1}=o_p(1)$.
 
For the proofs in this section, we will express the pivot in \eqref{exact:pivot} in terms of the standardized variable, $\Upsilon_{n}$, which is defined as:
\begin{align}
  \Upsilon_{n} &= \sqrt{n} \begin{bmatrix}
   A_1 \\
A_2 
\end{bmatrix}^{-1} \begin{bmatrix}
    \estj- \targetj\\
\gammaj- \gammaej
\end{bmatrix} = \zeta_{n} + \Delta_{n,1}, \text{ with } \Delta_{n,1} := \begin{bmatrix} A_1 \\ A_2 \end{bmatrix}^{-1} R_{n,1}.
\label{Upsilon:defn}
\end{align}
Here the matrix $\rbracket{A_{1}^{\top} \ A_{2}^{\top}}^{\top}$ is nonsingular whenever $K$ is positive definite and $H_{E,E}$ is invertible, so that $\Upsilon_{n}$ in \eqref{Upsilon:defn} is well defined.

Throughout the proofs we also use the shorthand
\begin{equation}
\Exp(x, \Sigma) := \exp\left(-\frac{1}{2} x^{\top} \Sigma x\right),
\label{defn:Exp}
\end{equation}
defined for a vector $x \in \mathbb{R}^p$ and a positive semidefinite matrix $\Sigma \in \mathbb{R}^{p \times p}$.
We record the Gaussian marginalization identity used repeatedly below: for $Z \sim \mathcal{N}(0_p, I_{p,p})$, any $m \in \mathbb{R}^p$, $s\in[0,1]$ and any positive semidefinite $\Xi$,
\begin{equation}
\mathbb{E}_{\mathcal{N}}\left[\Exp\left(\sqrt{s}\,\widetilde{P} Z+m, \Xi\right)\right]
= \frac{\Exp\left(m,\left(\Xi^{-1}+s\,\widetilde{P}\widetilde{P}^{\top}\right)^{-1}\right)}{\left|I_{p,p}+s\,\widetilde{P}^{\top}\Xi\widetilde{P}\right|^{1/2}},
\label{defn:marginal}
\end{equation}
where $\left(\Xi^{-1}+s\,\widetilde{P}\widetilde{P}^{\top}\right)^{-1}$ is read as $\lim_{c\to\infty}\left(\left(\Xi+c^{-1}I_{p,p}\right)^{-1}+s\,\widetilde{P}\widetilde{P}^{\top}\right)^{-1}$ when $\Xi$ is singular. In particular, taking $\Xi=\Omega^{-1}$ and $s=1-t$ gives $\left(\Omega+(1-t)\widetilde{P}\widetilde{P}^{\top}\right)^{-1}$, and taking $\Xi=\Theta$ and $s=1$ gives $\Theta\left(\Omega+\widetilde{P}\widetilde{P}^{\top}\right)$, with $\Theta(\cdot)$ as follows.
For a positive definite $\Sigma \in \mathbb{R}^{p\times p}$ we further write
\begin{equation}
\Theta(\Sigma) := \Sigma^{-1}-\frac{\Sigma^{-1} L L^{\top} \Sigma^{-1}}{L^{\top} \Sigma^{-1} L}
= \lim_{c \to \infty}\left(\Sigma + c\,L L^{\top}\right)^{-1},
\label{defn:Theta}
\end{equation}
the precision matrix of a $\mathcal{N}(0_p,\Sigma)$ variable after the direction $L$ has been marginalized out; note $\Theta(\Sigma) L = 0_p$ and $0 \preceq \Theta(\Sigma) \preceq \Sigma^{-1}$. We abbreviate $\Theta := \Theta(\Omega)$.

Then,  since $\sqrt{n}\estj = A_{1}\Upsilon_{n} + \sqrt{n}\targetj$, $\sqrt{n}\gammaj = A_{2}\Upsilon_{n} + \sqrt{n}\gammaej$, our pivot in terms of the standardized variables $\Upsilon$ is equal to:
\begin{align*}
    P^{E\cdot j}(\Upsilon_{n}) = \frac{\int_{-\infty}^{A_{1}\Upsilon_{n} + \sqrt{n}\targetj}\phi(x; \sqrt{n}\targetj, \sigmaj)F(x, A_{2}\Upsilon_{n} + \sqrt{n}\gammaej)dx}{\int_{-\infty}^{\infty}\phi(x; \sqrt{n}\targetj, \sigmaj)F(x, A_{2}\Upsilon_{n} + \sqrt{n}\gammaej)dx}.
\end{align*} Here $F$ is the function that adjusts for the selection process,
\begin{equation}
F(x_{1},x_{2}) = \int_{I_{-}^{E\cdot j}}^{\infty} \phi\nbracket{Lt +P_{1}^{E\cdot j} x_{1} + P_{2}^{E\cdot j} x_{2} + T; 0_{p},\Omega}dt,
\label{StandardizedPivot}
\end{equation}
which we also write in terms of the standardized variable $\Upsilon_{n}$, using $\sqrt{n}\estj = A_{1}\Upsilon_{n} + \sqrt{n}\targetj$ and $\sqrt{n}\gammaj = A_{2}\Upsilon_{n} + \sqrt{n}\gammaej$, as
\begin{equation}
F(\Upsilon_{n}) := F\nbracket{A_{1}\Upsilon_{n} + \sqrt{n}\targetj, \ A_{2}\Upsilon_{n} + \sqrt{n}\gammaej} = \int_{I_{-}^{E\cdot j}}^{\infty} \phi\nbracket{Lt +\widetilde{P} \Upsilon_{n} + T_{n}; 0_{p},\Omega}dt,
\label{StandardizedPivotUpsilon}
\end{equation} where
\begin{equation}
\widetilde{P}= (P_{1}^{E\cdot j}A_{1} + P_{2}^{E\cdot j}A_{2}),  \ T_{n}=  P_{1}^{E\cdot j} \sqrt{n} \targetj + P_{2}^{E\cdot j} \sqrt{n} \gammaej + T.
\label{StandardizedNotations}
\end{equation}
Throughout, whether $F$ is evaluated at a pair of arguments or at a single standardized argument is indicated by the argument itself. To prove our main result in Theorem \ref{main}, we start with a supporting result that provides sufficient conditions for the main result to hold.


\begin{theorem}
 For $\mathcal{H} \in \mathbb{C}^3(\mathbb{R}, \mathbb{R})$ an arbitrary bounded function with bounded derivatives up to the third order and for an increasing sequence of sets $\{D_{n}: n\in \mathbb{N}\}$ in $\mathbb{R}^p$ such that
\[
\lim _n \sup _{\mathbb{F}_n \in \mathcal{F}_n} \mathbb{P}_{\mathbb{F}_n}\left[\zeta_n \in D_n^c\right]=0,
\]
and
\[
\lim _n \sup _{\mathbb{F}_n \in \mathcal{F}_n} \frac{\mathbb{E}_{\mathbb{F}_n}\left[F\left(\zeta_{n}\right) \mathbbm{1}_{D_{n}^c}\left(\zeta_{n}\right)\right]}{\mathbb{E}_{\mathbb{F}_n}\left[F\left(\zeta_{n}\right)\right]}=0,
\]
define the relative difference terms
\begin{align*}
\mathrm{RD}_n^{(1)}&=\frac{\left|\mathbb{E}_{\mathbb{F}_n}\left[F\left(\zeta_{n}\right) \mathbbm{1}_{D_n}\left(\zeta_{n}\right)\right]-\mathbb{E}_{\mathcal{N}}\left[F(Z) \mathbbm{1}_{D_n}(Z)\right]\right|}{\mathbb{E}_{\mathcal{N}}\left[F(Z)\right]} \\
\mathrm{RD}_n^{(2)}&=\frac{\left|\mathbb{E}_{\mathbb{F}_n}\left[\mathcal{H} \circ P^{E\cdot j}\left(\zeta_{n}\right) \times F\left(\zeta_{n}\right) \mathbbm{1}_{D_n}\left(\zeta_{n}\right)\right]-\mathbb{E}_{\mathcal{N}}\left[\mathcal{H} \circ P^{E\cdot j}(Z) \times F(Z) \mathbbm{1}_{D_n}(Z)\right]\right|}{\mathbb{E}_{\mathcal{N}}\left[F(Z)\right]}.
\end{align*} Suppose that
\[\lim _n \sup _{\mathbb{F}_n \in \mathcal{F}_n} \mathrm{RD}_n^{(1)}=0, \quad \lim _n \sup _{\mathbb{F}_n \in \mathcal{F}_n} \mathrm{RD}_n^{(2)}=0.\] 
Then under Assumption \ref{assump:2}, Theorem \ref{main} holds.
\label{thm:sufficient}
\end{theorem}

\begin{proof}
To prove the claim in Theorem \ref{main}, it is sufficient to show that for any arbitrary bounded function $\mathcal{H} \in \mathbb{C}^3(\mathbb{R}, \mathbb{R})$ with bounded derivatives up to the third order, the following statement holds: 
\begin{align*}
\lim _n \sup _{\mathbb{F}_n \in \mathcal{F}_n} \Bigl\lvert \mathbb{E}_{\mathbb{F}_n} & \left[\mathcal{H} \circ P^{E\cdot j}\left(\zeta_{n}\right) \Bigm\vert \cbracket{\widehat{S}_n= S, \widehat{V}_n^{E\cdot j} = V^{E\cdot j}} \right] \\
& -\mathbb{E}_{\mathcal{N}}\left[\mathcal{H} \circ P^{E\cdot j}(Z) \Bigm\vert \cbracket{\widehat{S}_n= S, \widehat{V}_n^{E\cdot j} = V^{E\cdot j}}\right] \Bigr\rvert=0.
\end{align*}
By applying Lemma~\ref{supporting3}, we observe that a sufficient condition for the convergence mentioned above is to establish a bound on
\begin{equation}
\left|\frac{\mathbb{E}_{\mathbb{F}_n}\left[\mathcal{H} \circ P^{E\cdot j}\left(\zeta_{n}\right) \times F\left(\zeta_{n}\right)\right]}{\mathbb{E}_{\mathbb{F}_n}\left[F\left(\zeta_{n}\right)\right]}-\frac{\mathbb{E}_{\mathcal{N}}\left[\mathcal{H} \circ P^{E\cdot j}(Z) \times F(Z)\right]}{\mathbb{E}_{\mathcal{N}}\left[F(Z)\right]}\right|.
\label{inter:bound}
\end{equation}
In the rest of the proof, we obtain a bound for \eqref{inter:bound} in terms of the relative difference terms in our assertion. Note that the difference in \eqref{inter:bound} is bounded as 
\[
\mathrm{Bd}_0+\mathrm{Bd}_1+\mathrm{Bd}_2+\mathrm{Bd}_3
\]
where
\begin{align*}
& \mathrm{Bd}_0=\left|\frac{\mathbb{E}_{\mathbb{F}_n}\left[\mathcal{H} \circ P^{E\cdot j}\left(\Upsilon_{n}\right) \times F\left(\Upsilon_{n}\right)\right]}{\mathbb{E}_{\mathbb{F}_n}\left[F\left(\Upsilon_{n}\right)\right]}-\frac{\mathbb{E}_{\mathbb{F}_n}\left[\mathcal{H} \circ P^{E\cdot j}(\zeta_{n}) \times F(\zeta_{n})\right]}{\mathbb{E}_{\mathbb{F}_n}\left[F(\zeta_{n})\right]}\right|\\
&
\mathrm{Bd}_1=\left|\frac{\mathbb{E}_{\mathbb{F}_n}\left[\mathcal{H} \circ P^{E\cdot j}\left(\zeta_{n}\right) \times F\left(\zeta_{n}\right) \mathbbm{1}_{D_{n}}\left(\zeta_{n}\right)\right]}{\mathbb{E}_{\mathbb{F}_n}\left[F\left(\zeta_{n}\right)\right]}-\frac{\mathbb{E}_{\mathbb{F}_n}\left[\mathcal{H} \circ P^{E\cdot j}\left(\zeta_{n}\right) \times F\left(\zeta_{n}\right) \mathbbm{1}_{D_{n}}\left(\zeta_{n}\right)\right]}{\mathbb{E}_{\mathcal{N}}\left[F(Z)\right]}\right|, \\
& \mathrm{Bd}_2=\left|\frac{\mathbb{E}_{\mathbb{F}_n}\left[\mathcal{H} \circ P^{E\cdot j}\left(\zeta_{n}\right) \times F\left(\zeta_{n}\right) \mathbbm{1}_{D_{n}}\left(\zeta_{n}\right)\right]}{\mathbb{E}_{\mathcal{N}}\left[F(Z)\right]}-\frac{\mathbb{E}_{\mathcal{N}}\left[\mathcal{H} \circ P^{E\cdot j}(Z) \times F(Z) \mathbbm{1}_{D_{n}}(Z)\right]}{\mathbb{E}_{\mathcal{N}}\left[F(Z)\right]}\right|, \\
& \mathrm{Bd}_3=\left|\frac{\mathbb{E}_{\mathbb{F}_n}\left[\mathcal{H} \circ P^{E\cdot j}\left(\zeta_{n}\right) \times F\left(\zeta_{n}\right) \mathbbm{1}_{D_{n}^c}\left(\zeta_{n}\right)\right]}{\mathbb{E}_{\mathbb{F}_n}\left[F\left(\zeta_{n}\right)\right]}-\frac{\mathbb{E}_{\mathcal{N}}\left[\mathcal{H} \circ P^{E\cdot j}(Z) \times F(Z) \mathbbm{1}_{D_{n}^c}(Z)\right]}{\mathbb{E}_{\mathcal{N}}\left[F(Z)\right]}\right|.
\end{align*}

To begin with, according to Lemma~\ref{supporting2}, we can conclude that: \[\lim _n \sup _{\mathbb{F}_n \in \mathcal{F}_n} \mathrm{Bd}_0 = 0.\]
Moving to the next term, we observe that
\begin{align*}
\mathrm{Bd}_1 & \leq\left|\mathbb{E}_{\mathbb{F}_n}\left[\mathcal{H} \circ P^{E\cdot j}\left(\zeta_{n}\right) \times F\left(\zeta_{n}\right) \mathbbm{1}_{D_{n}}\left(\zeta_{n}\right)\right]\right| \times\left|\frac{1}{\mathbb{E}_{\mathbb{F}_n}\left[F\left(\zeta_{n}\right)\right]}-\frac{1}{\mathbb{E}_{\mathcal{N}}\left[F\left(\zeta_{n}\right)\right]}\right| \\
& \leq \|\mathcal{H}\|_{\infty} \times\left|\frac{\mathbb{E}_{\mathbb{F}_n}\left[F\left(\zeta_{n}\right) \mathbbm{1}_{D_{n}}\left(\zeta_{n}\right)\right]}{\mathbb{E}_{\mathbb{F}_n}\left[F\left(\zeta_{n}\right)\right]}-\frac{\mathbb{E}_{\mathbb{F}_n}\left[F\left(\zeta_{n}\right) \mathbbm{1}_{D_{n}}\left(\zeta_{n}\right)\right]}{\mathbb{E}_{\mathcal{N}}\left[F\left(\zeta_{n}\right)\right]}\right|.
\end{align*}
By applying the triangle inequality once more, we obtain the following bound for $\mathrm{Bd}_1$:
\begin{align*}
& \mathrm{Bd}_1 \leq \|\mathcal{H}\|_{\infty} \times\left\{\left|\frac{\mathbb{E}_{\mathbb{F}_n}\left[F\left(\zeta_{n}\right) \mathbbm{1}_{D_{n}}\left(\zeta_{n}\right)\right]}{\mathbb{E}_{\mathbb{F}_n}\left[F\left(\zeta_{n}\right)\right]}-\frac{\mathbb{E}_{\mathcal{N}}\left[F(Z) \mathbbm{1}_{D_{n}}(Z)\right]}{\mathbb{E}_{\mathcal{N}}\left[F(Z)\right]}\right|\right. \\
&\left.+\left|\frac{\mathbb{E}_{\mathcal{N}}\left[F(Z) \mathbbm{1}_{D_{n}}(Z)\right]}{\mathbb{E}_{\mathcal{N}}\left[F(Z)\right]}-\frac{\mathbb{E}_{\mathbb{F}_n}\left[F\left(\zeta_{n}\right) \mathbbm{1}_{D_{n}}\left(\zeta_{n}\right)\right]}{\mathbb{E}_{\mathcal{N}}\left[F\left(\zeta_{n}\right)\right]}\right|\right\} \\
&= \|\mathcal{H}\|_{\infty} \times\left\{\left|\frac{\mathbb{E}_{\mathbb{F}_n}\left[F\left(\zeta_{n}\right) \mathbbm{1}_{D_{n}^c}\left(\zeta_{n}\right)\right]}{\mathbb{E}_{\mathbb{F}_n}\left[F\left(\zeta_{n}\right)\right]}-\frac{\mathbb{E}_{\mathcal{N}}\left[F(Z) \mathbbm{1}_{D_{n}^c}(Z)\right]}{\mathbb{E}_{\mathcal{N}}\left[F(Z)\right]}\right|\right. \\
&\left.+\left|\frac{\mathbb{E}_{\mathcal{N}}\left[F(Z) \mathbbm{1}_{D_{n}}(Z)\right]}{\mathbb{E}_{\mathcal{N}}\left[F(Z)\right]}-\frac{\mathbb{E}_{\mathbb{F}_n}\left[F\left(\zeta_{n}\right) \mathbbm{1}_{D_{n}}\left(\zeta_{n}\right)\right]}{\mathbb{E}_{\mathcal{N}}\left[F\left(\zeta_{n}\right)\right]}\right|\right\} \\
& \leq \|\mathcal{H}\|_{\infty} \times\left\{2 \sup _{\mathbb{F}_n \in \mathcal{F}_n} \frac{\mathbb{E}_{\mathbb{F}_n}\left[F\left(\zeta_{n}\right) \mathbbm{1}_{D_{n}^c}\left(\zeta_{n}\right)\right]}{\mathbb{E}_{\mathbb{F}_n}\left[F\left(\zeta_{n}\right)\right]}+\mathrm{RD}_n^{(1)}\right\}.
\end{align*}

It is easy to see that $\mathrm{Bd}_2$ is equal to $\mathrm{RD}_n^{(2)}$, and that

\[
\mathrm{Bd}_3 \leq \|\mathcal{H}\|_{\infty} \times 2 \sup _{\mathbb{F}_n \in \mathcal{F}_n} \frac{\mathbb{E}_{\mathbb{F}_n}\left[F\left(\zeta_{n}\right) \mathbbm{1}_{D_{n}^c}\left(\zeta_{n}\right)\right]}{\mathbb{E}_{\mathbb{F}_n}\left[F\left(\zeta_{n}\right)\right]}
\]

Thus, we conclude that
\begin{align*}
&  \ \  \lim _n \sup _{\mathbb{F}_n \in \mathcal{F}_n}\left|\frac{\mathbb{E}_{\mathbb{F}_n}\left[\mathcal{H} \circ P^{E\cdot j}\left(\zeta_{n}\right) \times F\left(\zeta_{n}\right)\right]}{\mathbb{E}_{\mathbb{F}_n}\left[F\left(\zeta_{n}\right)\right]}-\frac{\mathbb{E}_{\mathcal{N}}\left[\mathcal{H} \circ P^{E\cdot j}(Z) \times F(Z)\right]}{\mathbb{E}_{\mathcal{N}}\left[F(Z)\right]}\right| \\
 \leq  & \|\mathcal{H}\|_{\infty} \times \lim _n \sup _{\mathbb{F}_n \in \mathcal{F}_n} \mathrm{RD}_n^{(1)}+\lim _n \sup _{\mathbb{F}_n \in \mathcal{F}_n} \mathrm{RD}_n^{(2)}\\
&\qquad +\|\mathcal{H}\|_{\infty} \times 4 \lim _n \sup _{\mathbb{F}_n \in \mathcal{F}_n} \frac{\mathbb{E}_{\mathbb{F}_n}\left[F\left(\zeta_{n}\right) \mathbbm{1}_{D_{n}^c}\left(\zeta_{n}\right)\right]}{\mathbb{E}_{\mathbb{F}_n}\left[F\left(\zeta_{n}\right)\right]}.
\end{align*} Using the weighted-tail condition in the statement of the theorem,
we conclude that the difference in \eqref{inter:bound} uniformly converges to $0$.
This completes our proof. 
\end{proof}

To prove our main result on the asymptotic validity of the pivot, we establish the convergence of the two relative difference terms identified in Theorem \ref{thm:sufficient}.
We divide the proof of our result into two cases based on how the sequence of parameters $\sqrt{n}\begin{bmatrix}
  (\targetj)^\top &  (\gammaej)^\top
\end{bmatrix}^\top$  grows with increasing $n$.
First, \Thmref{RD1} demonstrates the convergence of both relative differences when $r_n$ does not grow with $n$. Then, \Thmref{RD2} extends this proof to the case where we allow $r_n$ to grow to $\infty$.

\begin{theorem} 
\label{RD1}
Suppose that $r_{n} \leq C$ for some constant $C \in \mathbb{R}$. 
Under Assumptions \ref{assump:1} and \ref{assump:2}, we have that
\[\lim _n \sup _{\mathbb{F}_n \in \mathcal{F}_n} \mathrm{RD}_n^{(1)}=0, \quad \lim _n \sup _{\mathbb{F}_n \in \mathcal{F}_n} \mathrm{RD}_n^{(2)}=0,\] 
where we let $D_n = \mathbb{R}^p$ for all $n \in \mathbb{N}$.
\end{theorem}

\begin{proof}
We start by analyzing the denominator of the two relative difference terms and note that
\[
\mathbb{E}_{\mathcal{N}}\left[F(Z)\right] \propto \mathbb{E}_{\mathcal{N}}\left[\int_{I_{-}^{E\cdot j}}^{\infty} \exp \left\{-\frac{1}{2}\left(L t+\widetilde{P} Z+T_{n} \right)^{\top} \Omega^{-1}\left(L t+\widetilde{P} Z+T_{n} \right)\right\} d t\right] .
\]
Let $\mathcal{D}_{C_1}=\left[-C_1 \cdot \mathbf{1}_p, C_1 \cdot \mathbf{1}_p\right] \subseteq \mathbb{R}^p$ for a large enough $C_1>0$ such that 
\[\mathbb{P}_{\mathcal{N}}\left[Z\in \mathcal{D}_{C_1}\right] \geq\frac{1}{2}.\]
Writing $T_{n} + LI_{-}^{E\cdot j} = -r_{n}b$ for a fixed vector $b \in \mathbb{R}^{p}$, as in the parameterization introduced before Theorem~\ref{RD2}, and noting that $\|r_{n}b\| = O(1)$ in the present case, for $Z\in \mathcal{D}_{C_1}$ we have that
\[
\exp \left\{-\frac{1}{2}\left(L t+\widetilde{P} Z-r_n b- L I_{-}^{E\cdot j}\right)^{\top} \Omega^{-1}\left(L t+\widetilde{P} Z - r_n b-L I_{-}^{E\cdot j}\right)\right\} \geq d_2,
\]
where $d_2$ is some positive constant. 
This leads us to observe that
\[ \mathbb{E}_{\mathcal{N}}\rbracket{F(Z)} \geq C_{2} \times  \mathbb{P}_{\mathcal{N}}\rbracket{Z \in \mathcal{D}_{C_1}} \geq C_{2} \times \frac{1}{2}.\] Our assertion holds if we show that the numerators in the two relative difference terms converge to $0$.
In the remaining proof, we show that:
\begin{enumerate}
\item $\sup _{\mathbb{F}_n \in \mathcal{F}_n}\left|\mathbb{E}_{\mathbb{F}_n}\left[F\left(\zeta_{n}\right)\right]-\mathbb{E}_{\mathcal{N}}\left[F(Z)\right]\right| \lesssim \dfrac{1}{\sqrt{n}}$
\item $\sup _{\mathbb{F}_n \in \mathcal{F}_n}\left|\mathbb{E}_{\mathbb{F}_n}\left[\mathcal{H} \circ P^{E\cdot j}\left(\zeta_{n}\right) \times F\left(\zeta_{n}\right)\right]-\mathbb{E}_{\mathcal{N}}\left[\mathcal{H} \circ P^{E\cdot j}(Z) \times F(Z)\right]\right| \lesssim \dfrac{1}{\sqrt{n}}$.
\end{enumerate} We then apply the Stein bound from Lemma~\ref{lem:stein} to the real-valued functions $G^{(1)}(Z)=F(Z)$,  $G^{(2)}(Z)=\mathcal{H} \circ P^{E\cdot j}(Z) \times F(Z)$, using $\mathcal{W}_{\alpha, \kappa}, \zeta_{n}[-1], a_{1, n}$, and $a_{1, n}^*$ as defined in Lemma~\ref{lem:stein}. 
This yields:
\begin{equation}
\begin{split}
& \left|\mathbb{E}_{\mathbb{F}_n}\left[G(\zeta_{n})\right]-\mathbb{E}_{\mathcal{N}}[G(Z)]\right| \\
& \lesssim \frac{1}{\sqrt{n}} \sum_{\substack{\lambda, \gamma \in \mathbb{N}: \\ \lambda+\gamma \leq 3}} \sum_{i_1, i_2, i_3 \in[p]} \mathbb{E}_{\mathbb{F}_n}\Bigg[\left\|a_{1, n}\right\|^\lambda\left\|a_{1, n}^*\right\|^\gamma \\
&\qquad\qquad\times \sup _{\alpha, \kappa \in[0,1]} \int_0^1 \frac{\sqrt{t}}{2} \mathbb{E}_{\mathcal{N}}\left[\left|\partial_{i_1, i_2, i_3}^3 G\left(\sqrt{t} \mathcal{W}_{\alpha, \kappa}+\sqrt{1-t} Z\right)\right|\right] d t\Bigg].
\end{split}
\label{Stn:bdd}
\end{equation}
Using the behavior of the pivot and weight function in Propositions~\ref{pivotderivative} and \ref{weightderivative}, we have that
\begin{align*}
\left|\partial_{i_1, i_2, i_3}^3 G^{(l)}\left(\sqrt{t} \mathcal{W}_{\alpha, \kappa}+\sqrt{1-t} Z\right)\right| & \lesssim \sum_{l=0}^3\left\|\widetilde{P} \sqrt{t} \mathcal{W}_{\alpha, \kappa}+\widetilde{P} \sqrt{1-t} Z+T_{n}\right\|^l \\
& \lesssim \sum_{l=0}^3 \sqrt{t}\left\|\mathcal{W}_{\alpha, \kappa}\right\|^l+\sqrt{1-t}\|Z\|^l,
\end{align*}
since $\|T_n\|= O(1)$ in this case.

Revisiting the Stein's bound in \eqref{Stn:bdd}, we conclude that:
\begin{align*}
& \frac{1}{\sqrt{n}} \sum_{\substack{\lambda, \gamma \in \mathbb{N}: \\
\lambda+\gamma \leq 3}} \sum_{i_1, i_2, i_3} \mathbb{E}_{\mathbb{F}_n}\left[\left\|a_{1, n}\right\|^\lambda\left\|a_{1, n}^*\right\|^\gamma \sup _{\alpha, \kappa \in[0,1]} \int_0^1 \frac{\sqrt{t}}{2} \mathbb{E}_{\mathcal{N}}\left[\sum_{l=0}^3 \sqrt{t}\left\|\mathcal{W}_{\alpha, \kappa}\right\|^l+\sqrt{1-t}\|Z\|^l\right] d t\right] \\
& \lesssim \frac{1}{\sqrt{n}} \sum_{\lambda, \gamma \in \mathbb{N}: \lambda+\gamma \leq 3} \mathbb{E}_{\mathbb{F}_n}\left[\left\|a_{1, n}\right\|^\lambda\left\|a_{1, n}^*\right\|^\gamma \sup _{\alpha, \kappa \in[0,1]}\left\|\mathcal{W}_{\alpha, \kappa}\right\|^3\right] \\
& \lesssim \frac{1}{\sqrt{n}} \sup _n \sup _{\mathbb{F}_n \in \mathcal{F}_n} \mathbb{E}_{\mathbb{F}_n}\left[\left\|a_{1, n}\right\|^6\right] \lesssim \frac{1}{\sqrt{n}},
\end{align*}
where the last inequality is Proposition~\ref{prop:sixthmoment}. This proves our claim.
\end{proof} Now we consider the set-up, where the parameters grow as $O(r_{n})$, where $r_n \to \infty$, i.e., \[\sqrt{n}\begin{bmatrix}
  (\targetj)^\top &  (\gammaej)^\top
\end{bmatrix}^\top  = r_n\beta,\] for some given constant vector $\beta \in \mathbb{R}^{p}$. 
Observe that $T_n= O(r_{n})$. 
To simplify the notations in our proof, we will hereafter simply write
\[T_{n} + LI_{-}^{E\cdot j} =  P_{1} \sqrt{n} \targetj + P_{2} \sqrt{n} \gammaej + T + LI_{-}^{E\cdot j} = -r_{n}b \in \mathbb{R}^{p},\]
where $b\in \mathbb{R}^{p}$ is a fixed vector.

\begin{theorem} 
\label{RD2}
Let $r_n \to \infty$ as $n \to \infty$, with a rate  of $ o(n^{\frac{1}{6}})$.
Throughout this proof we take the increasing sequence of sets in Theorem~\ref{thm:sufficient} to be $D_n = \mathcal{D}_n$, the box $\left[-d_0 r_n\cdot \mathbf{1}_p,\, d_0 r_n\cdot \mathbf{1}_p\right]$ of Proposition~\ref{prop:A21}; the ordinary tail condition follows from the construction of this box, and the weighted-tail condition is verified by the numerator and denominator bounds in Propositions~\ref{prop:A23} and~\ref{prop:A24}. 
Then under Assumptions \ref{assump:1}, \ref{assump:2}, \ref{assump:3}, it holds that
\[\lim _n \sup _{\mathbb{F}_n \in \mathcal{F}_n} \mathrm{RD}_n^{(1)}=0, \quad \lim _n \sup _{\mathbb{F}_n \in \mathcal{F}_n} \mathrm{RD}_n^{(2)}=0.\] 
\end{theorem}

\begin{proof}
For this proof, we consider the lower-tail regime in which the normalizing denominator has the small-denominator rate in Proposition~\ref{prop:A23}. The complementary regime is handled by the same Stein bound with the larger denominator lower bound from Proposition~\ref{prop:A23}.

For the numerators of the relative differences, we will show in the rest of the proof that:
\noindent \textbf{(B1)}. $\sup _{\mathbb{F}_n \in \mathcal{F}_n}\left|\mathbb{E}_{\mathbb{F}_n}\left[F\left(\zeta_n\right) \mathbbm{1}_{\mathcal{D}_n}\left(\zeta_n\right)\right]-\mathbb{E}_{\mathcal{N}}\left[F(Z) \mathbbm{1}_{\mathcal{D}_n}(Z)\right]\right| \lesssim \dfrac{r_n^2}{\sqrt{n}} \Exp\left(-r_n b,\left[\Omega + \widetilde{P}\widetilde{P}^{\top}\right]^{-1}\right);$\\
\noindent \textbf{(B2)}.
\begin{align*}
&\sup _{\mathbb{F}_n \in \mathcal{F}_n} \left|\mathbb{E}_{\mathbb{F}_n}\left[\mathcal{H} \circ P^{E\cdot j}\left(\zeta_n\right) \times F\left(\zeta_n\right) \mathbbm{1}_{\mathcal{D}_n}\left(\zeta_n\right)\right]  -\mathbb{E}_{\mathcal{N}}\left[\mathcal{H} \circ P^{E\cdot j}(Z) \times F(Z) \mathbbm{1}_{\mathcal{D}_n}(Z)\right]\right| \\
&\qquad\qquad\lesssim  \frac{r_n^2}{\sqrt{n}} \Exp\left(-r_n b,\left[\Omega + \widetilde{P}\widetilde{P}^{\top}\right]^{-1}\right),
\end{align*}
Define the real-valued functions
\begin{align*}
& \bar{G}^{(1)}(Z)=\Exp\left(\widetilde{P}Z+T_n+L I_{-}^{E\cdot j}, \Omega^{-1}\right) \times \mathcal{K}^{1}_n(Z) \mathbbm{1}_{\mathcal{D}_n}(Z) \\
& \bar{G}^{(2)}(Z)=\mathcal{H} \circ P^{E\cdot j}(Z) \times \Exp\left(\widetilde{P}Z+T_n+L I_{-}^{E\cdot j}, \Omega^{-1}\right) \times \mathcal{K}^{1}_n(Z) \mathbbm{1}_{\mathcal{D}_n}(Z)
\end{align*}
where $\mathcal{K}^{1}_n$ and the increasing sequence of sets $\mathcal{D}_n$ are as defined in Proposition~\ref{prop:A21}. 
Note that, $\bar{G}^{(1)}(Z)$ and $\bar{G}^{(2)}(Z)$ are equal to $F(Z)$ and $\mathcal{H} \circ P^{E\cdot j}(Z) \times F(Z)$ respectively on the set $\mathcal{D}_n$. Using the Stein bound in Lemma~\ref{lem:stein}, for $G=\bar{G}^{\mathcal{P}, \lambda}$ for $l \in\{1,2\}$, and  the properties of our pivot, $F$ and $\mathcal{K}^{1}_n$ in Propositions \ref{pivotderivative} and \ref{weightderivative}, we note that
\begin{align*}
\left|\partial_{i_1, i_2, i_3}^3 \bar{G}^{(l)}\left(\sqrt{t} \mathcal{W}_{\alpha, \kappa}+\sqrt{1-t} Z\right)\right| \lesssim & \sum_{l=0}^3 r_n^{-1}\left\|\widetilde{P} \sqrt{t} \mathcal{W}_{\alpha, \kappa}+\widetilde{P} \sqrt{1-t} Z+T_n\right\|^l \\
& \times \Exp\left(\widetilde{P} \sqrt{t} \mathcal{W}_{\alpha, \kappa}+\widetilde{P} \sqrt{1-t} Z+T_n, \Omega^{-1}\right).
\end{align*}
Therefore, the expected value of the above-stated partial derivative satisfies:
\begin{align*}
& \quad \mathbb{E}_{\mathcal{N}}\left[\left|\partial_{i_1, i_2, i_3}^3 \bar{G}^{(l)}\left(\sqrt{t} \mathcal{W}_{\alpha, \kappa}+\sqrt{1-t} Z\right)\right|\right] \\
& \lesssim \sum_{\substack{\lambda, \gamma \in \mathbb{N}: \\
\lambda+\gamma \leq 3}} r_n^{-1}\left\|\mathcal{W}_{\alpha, \kappa}\right\|^\lambda\left\|T_n\right\|^\gamma \Exp\left(\widetilde{P} \sqrt{t} \mathcal{W}_{\alpha, \kappa}+T_n,\left[\Omega + (1-t)\widetilde{P}\widetilde{P}^{\top}\right]^{-1}\right) \\
& \lesssim \sum_{\substack{\bar{\lambda}, \bar{\kappa}, \breve{\lambda}, \breve{\kappa} \in \mathbb{N}: \\
\bar{\lambda}+\bar{\kappa}+\breve{\lambda}+\breve{\kappa} \leq 3}} r_n^{\breve{\lambda}-1}\left\|\zeta_{n}[-1]\right\|^{\breve{\kappa}}\left\|\frac{a_{1, n}}{\sqrt{n}}\right\|^{\bar{\lambda}}\left\|\frac{a_{1, n}^*}{\sqrt{n}}\right\|^{\bar{\kappa}} \Exp\left(\widetilde{P} \sqrt{t} \mathcal{W}_{\alpha, \kappa}-r_n b,\left[\Omega + (1-t)\widetilde{P}\widetilde{P}^{\top}\right]^{-1}\right).
\end{align*}
where the second display integrates $Z$ out using \eqref{defn:marginal} with $\Xi=\Omega^{-1}$ and $s=1-t$, and the polynomial factors are bounded by absorbing them into the Gaussian tail.
Plugging this into the Stein bound obtained in Lemma~\ref{lem:stein}, we have that
\begin{align*}
& \quad \left|\mathbb{E}_{\mathbb{F}_n}\left[\bar{G}^{(l)}\left(\zeta_n\right)\right]-\mathbb{E}_{\mathcal{N}}\left[\bar{G}^{(l)}(Z)\right]\right| \\
& \lesssim \frac{1}{\sqrt{n}} \sum_{\substack{\lambda, \gamma \in \mathbb{N}: i_1, i_2, i_3 \in[p] \\
\lambda+\gamma \leq 3}} \mathbb{E}_{\mathbb{F}_n}\left[\left\|a_{1, n}\right\|^\lambda\left\|a_{1, n}^*\right\|^\gamma \sup _{\alpha, \kappa \in[0,1]} \int_0^1 \frac{\sqrt{t}}{2}\right. \\
& \left.\times \sum_{\substack{\bar{\lambda}, \bar{\kappa}, \breve{\lambda}, \breve{\kappa} \in \mathbb{N}: \\
\bar{\lambda}+\bar{\kappa}+\breve{\lambda}+\breve{\kappa} \leq 3}} r_n^{\breve{\lambda}-1}\left\|\zeta_{n}[-1]\right\|^{\breve{\kappa}}\left\|\frac{a_{1, n}}{\sqrt{n}}\right\|^{\bar{\lambda}}\left\|\frac{a_{1, n}^*}{\sqrt{n}}\right\|^{\bar{\kappa}} \Exp\left(\widetilde{P} \sqrt{t} \mathcal{W}_{\alpha, \kappa}-r_n b,\left[\Omega + (1-t)\widetilde{P}\widetilde{P}^{\top}\right]^{-1}\right) d t\right].
\end{align*} Next, we simplify the bound on the right-hand side display.
In particular, we show the simplification for $\bar{\lambda}=\bar{\kappa}=\breve{\kappa}=0, \breve{\lambda}=3$, noting that a similar approach applies to other values of $\bar{\lambda}, \bar{\kappa}, \breve{\lambda}, \breve{\kappa}$. 
In this case, applying the result from Proposition~\ref{prop:A24} gives us:
\[\sup _{\mathbb{F}_n \in \mathcal{F}_n}\left|\mathbb{E}_{\mathbb{F}_n}\left[\bar{G}^{(l)}\left(\zeta_n\right)\right]-\mathbb{E}_{\mathcal{N}}\left[\bar{G}^{(l)}(Z)\right]\right| \lesssim \frac{r_n^2}{\sqrt{n}} \Exp\left(-r_n b,\left[\Omega + \widetilde{P}\widetilde{P}^{\top}\right]^{-1}\right).\] Finally, it is immediate from Proposition~\ref{prop:A23} that
\[
\left(\mathbb{E}_{\mathcal{N}}\left[F(Z)\right]\right)^{-1} \lesssim r_n \left[\Exp\left(-r_n b,\left[\Omega +\widetilde{P}\widetilde{P}^{\top}\right]^{-1}\right)\right]^{-1}.\] Combining this bound with \textbf{(B1)} and \textbf{(B2)} gives $\mathrm{RD}_n^{(1)},\, \mathrm{RD}_n^{(2)} \lesssim r_n^{3}/\sqrt{n} \to 0$, since $r_n = o(n^{1/6})$, which completes the proof of our claims.
\end{proof}

\subsection{Supporting Results}
\begin{lemma}
\label{supporting1}
    Let $D_{n}$ be an increasing sequence of sets in $\mathbb{R}^{p}$ such that \[\lim _n \sup _{\mathbb{F}_n \in \mathcal{F}_n} \mathbb{P}_{\mathbb{F}_n}\rbracket{\zeta_{n} \in D_{n}^{c}}=0.\] Suppose also that $\sup_z F(z)<\infty$ and $\inf_{n,\mathbb{F}_n\in\mathcal{F}_n}\mathbb{E}_{\mathbb{F}_n}[F(\zeta_n)]>0$. Then we have  

    \[
\lim _n \sup _{\mathbb{F}_n \in \mathcal{F}_n} \frac{\mathbb{E}_{\mathbb{F}_n}\left[F\left(\zeta_{n}\right) \mathbbm{1}_{D_{n}^{c}}\left(\zeta_{n}\right)\right]}{\mathbb{E}_{\mathbb{F}_n}\left[F(\zeta_{n})\right]}=0 .
\]
\end{lemma}

\begin{proof}
Let $M=\sup_z F(z)$ and $m=\inf_{n,\mathbb{F}_n\in\mathcal{F}_n}\mathbb{E}_{\mathbb{F}_n}[F(\zeta_n)]$. Then
\[
\sup_{\mathbb{F}_n\in\mathcal{F}_n}
\frac{\mathbb{E}_{\mathbb{F}_n}\left[F(\zeta_n)\mathbbm{1}_{D_n^c}(\zeta_n)\right]}
{\mathbb{E}_{\mathbb{F}_n}\left[F(\zeta_n)\right]}
\leq \frac{M}{m}\sup_{\mathbb{F}_n\in\mathcal{F}_n}\mathbb{P}_{\mathbb{F}_n}\left[\zeta_n\in D_n^c\right],
\]
which converges to zero by assumption.
\end{proof}

\begin{lemma}
\label{supporting2}
Under Assumption~\ref{assump:3}, it holds that
\[\lim _n \sup _{\mathbb{F}_n \in \mathcal{F}_n} 
\left|\frac{\mathbb{E}_{\mathbb{F}_n}\left[\mathcal{H} \circ P^{E\cdot j}\left(\Upsilon_{n}\right) \times F\left(\Upsilon_{n}\right)\right]}{\mathbb{E}_{\mathbb{F}_n}\left[F\left(\Upsilon_{n}\right)\right]}-\frac{\mathbb{E}_{\mathbb{F}_n}\left[\mathcal{H} \circ P^{E\cdot j}(\zeta_{n}) \times F(\zeta_{n})\right]}{\mathbb{E}_{\mathbb{F}_n}\left[F(\zeta_{n})\right]}\right|=0.
\]
\end{lemma}

\begin{proof}
To prove this, we apply the triangle inequality: 
\begin{align*}
&\left|\frac{\mathbb{E}_{\mathbb{F}_n}\left[\mathcal{H} \circ P^{E\cdot j}\left(\Upsilon_{n}\right) \times F\left(\Upsilon_{n}\right)\right]}{\mathbb{E}_{\mathbb{F}_n}\left[F\left(\Upsilon_{n}\right)\right]}-\frac{\mathbb{E}_{\mathbb{F}_n}\left[\mathcal{H} \circ P^{E\cdot j}(\zeta_{n}) \times F(\zeta_{n})\right]}{\mathbb{E}_{\mathbb{F}_n}\left[F(\zeta_{n})\right]}\right|   \\
\leq &\left|\frac{\mathbb{E}_{\mathbb{F}_n}\left[\mathcal{H} \circ P^{E\cdot j}\left(\Upsilon_{n}\right) \times F\left(\Upsilon_{n}\right)\right]}{\mathbb{E}_{\mathbb{F}_n}\left[F\left(\Upsilon_{n}\right)\right]}-\frac{\mathbb{E}_{\mathbb{F}_n}\left[\mathcal{H} \circ P^{E\cdot j}\left(\Upsilon_{n}\right) \times F\left(\Upsilon_{n}\right)\right]}{\mathbb{E}_{\mathbb{F}_n}\left[F(\zeta_{n})\right]}\right| \\
+&\left|\frac{\mathbb{E}_{\mathbb{F}_n}\left[\mathcal{H} \circ P^{E\cdot j}\left(\Upsilon_{n}\right) \times F\left(\Upsilon_{n}\right)\right]}{\mathbb{E}_{\mathbb{F}_n}\left[F(\zeta_{n})\right]}-\frac{\mathbb{E}_{\mathbb{F}_n}\left[\mathcal{H} \circ P^{E\cdot j}(\zeta_{n}) \times F(\zeta_{n})\right]}{\mathbb{E}_{\mathbb{F}_n}\left[F(\zeta_{n})\right]}\right|.
\end{align*}
For the first term on the right-hand side of the display, note that
\begin{align*}
& \left|\frac{\mathbb{E}_{\mathbb{F}_n}\left[\mathcal{H} \circ P^{E\cdot j}\left(\Upsilon_{n}\right) \times F\left(\Upsilon_{n}\right)\right]}{\mathbb{E}_{\mathbb{F}_n}\left[F\left(\Upsilon_{n}\right)\right]}-\frac{\mathbb{E}_{\mathbb{F}_n}\left[\mathcal{H} \circ P^{E\cdot j}\left(\Upsilon_{n}\right) \times F\left(\Upsilon_{n}\right)\right]}{\mathbb{E}_{\mathbb{F}_n}\left[F(\zeta_{n})\right]}\right|\\
&\leq\left|\mathbb{E}_{\mathbb{F}_n}\left[\mathcal{H} \circ P^{E\cdot j}\left(\Upsilon_{n}\right) \times F\left(\Upsilon_{n}\right)\right]\right| \times\left|\frac{1}{\mathbb{E}_{\mathbb{F}_n}\left[F\left(\Upsilon_{n}\right)\right]}-\frac{1}{\mathbb{E}_{\mathbb{F}_n}\left[F\left(\zeta_{n}\right)\right]}\right| \\
& \leq \|\mathcal{H}\|_{\infty} \times\left|\frac{\mathbb{E}_{\mathbb{F}_n}\left[F\left(\Upsilon_{n}\right)\right]}{\mathbb{E}_{\mathbb{F}_n}\left[F\left(\Upsilon_{n}\right)\right]}-\frac{\mathbb{E}_{\mathbb{F}_n}\left[F\left(\Upsilon_{n}\right)\right]}{\mathbb{E}_{\mathbb{F}_n}\left[F\left(\zeta_{n}\right)\right]}\right| \\
& =\|\mathcal{H}\|_{\infty} \times\left|\frac{\mathbb{E}_{\mathbb{F}_n}\left[F\left(\zeta_{n}\right)\right]-\mathbb{E}_{\mathbb{F}_n}\left[F\left(\Upsilon_{n}\right)\right]}{\mathbb{E}_{\mathbb{F}_n}\left[F\left(\zeta_{n}\right)\right]}\right|.
\end{align*}
We will now show that
\begin{equation}
\label{to:show}
\lim _n \sup _{\mathbb{F}_n \in \mathcal{F}_n}\left|\frac{\mathbb{E}_{\mathbb{F}_n}\left[F\left(\zeta_{n}+\Delta_{n,1}\right)-F\left(\zeta_{n}\right)\right]}{\mathbb{E}_{\mathbb{F}_n}\left[F\left(\zeta_{n}\right)\right]}\right|=0,
\end{equation}
using the fact that $\Upsilon_{n}=\zeta_{n}+\Delta_{n,1}$. 

Note that, for any $\varepsilon>0$, there exists $n_0$ such that for all $n \geq n_0$, it holds that
\[
\sup_{\mathbb{F}_n \in \mathcal{F}_n}\mathbb{E}_{\mathbb{F}_n}\left[\left|\frac{F\left(\zeta_{n}+\Delta_{n,1}\right)}{F\left(\zeta_{n}\right)}-1\right|\right]<\varepsilon .
\]
This bound is a consequence of Assumption~\ref{assump:3}: since $\Delta_{n,1}=\rbracket{A_{1}^{\top} \ A_{2}^{\top}}^{-\top} R_{n,1}$, the tail condition on $R_{n,1}$ ensures that the perturbation $\Delta_{n,1}$ is negligible on the scale $r_{n}$ at which the log-weight $\log F$ varies, following the argument in \cite{snigdhacarving}.
Then by a simple rearrangement, we obtain that
\begin{align*}
& \sup_{\mathbb{F}_n \in \mathcal{F}_n}\mathbb{E}\left[F\left(\zeta_{n}\right) \times\left\{\left|\frac{F\left(\zeta_{n}+\Delta_{n,1}\right)}{F\left(\zeta_{n}\right)}-1\right|-\varepsilon\right\}\right] \\
& \quad \leq \sup _z F(z) \times \sup_{\mathbb{F}_n \in \mathcal{F}_n} \mathbb{E}_{\mathbb{F}_n}\left[\left|\frac{F\left(\zeta_{n}+\Delta_{n,1}\right)}{F\left(\zeta_{n}\right)}-1\right|-\varepsilon\right]<0
\end{align*}
for all $n \geq n_0$. Hence, for any given $\varepsilon>0$, there exists a $n_0$ such that for all $n \geq n_0$ and $\mathbb{F}_n \in \mathcal{F}_n$,
\[
\lim _n \sup _{\mathbb{F}_n \in \mathcal{F}_n}\mathbb{E}_{\mathbb{F}_n}\left[F\left(\zeta_{n}\right) \times\left|\frac{F\left(\zeta_{n}+\Delta_{n,1}\right)}{F\left(\zeta_{n}\right)}-1\right|\right]<\lim _n \sup _{\mathbb{F}_n \in \mathcal{F}_n} \varepsilon \mathbb{E}_{\mathbb{F}_n}\left[F\left(\zeta_{n}\right)\right],
\]
i.e., the claim in \eqref{to:show} holds.

For the second term in this display, note that it is equal to: 
\[
\left|\frac{\mathbb{E}_{\mathbb{F}_n}\left[\mathcal{H} \circ P^{E\cdot j}\left(\Upsilon_{n}\right) \times F\left(\Upsilon_{n}\right)\right]-\mathbb{E}_{\mathbb{F}_n}\left[\mathcal{H} \circ P^{E\cdot j}\left(\zeta_{n}\right) \times F\left(\zeta_{n}\right)\right]}{\mathbb{E}_{\mathbb{F}_n}\left[F\left(\zeta_{n}\right)\right]}\right| .
\]
A similar proof strategy can be applied to conclude that
\[
\lim _n \sup _{\mathbb{F}_n \in \mathcal{F}_n}\left|\frac{\mathbb{E}_{\mathbb{F}_n}\left[\mathcal{H} \circ P^{E\cdot j}\left(\zeta_{n}+\Delta_{n,1}\right) \times F\left(\zeta_{n}+\Delta_{n,1}\right)\right]-\mathbb{E}_{\mathbb{F}_n}\left[\mathcal{H} \circ P^{E\cdot j}\left(\zeta_{n}\right) \times F\left(\zeta_{n}\right)\right]}{\mathbb{E}_{\mathbb{F}_n}\left[F\left(\zeta_{n}\right)\right]}\right|=0 .
\]

\end{proof}

\begin{lemma}
\label{supporting3}
Suppose that \[\lim _n \sup _{\mathbb{F}_n \in \mathcal{F}_n} 
\left|\frac{\mathbb{E}_{\mathbb{F}_n}\left[\mathcal{H} \circ P^{E\cdot j}\left(\Upsilon_{n}\right) \times F\left(\Upsilon_{n}\right)\right]}{\mathbb{E}_{\mathbb{F}_n}\left[F\left(\Upsilon_{n}\right)\right]}-\frac{\mathbb{E}_{\mathcal{N}}\left[\mathcal{H} \circ P^{E\cdot j}(Z) \times F(Z)\right]}{\mathbb{E}_{\mathcal{N}}\left[F(Z)\right]}\right|=0,
\]
Then the following holds: 
\begin{align*}
\lim _n \sup _{\mathbb{F}_n \in \mathcal{F}_n} \Bigl\lvert \mathbb{E}_{\mathbb{F}_n} & \left[\mathcal{H} \circ P^{E\cdot j}\left(\Upsilon_{n}\right) \Bigm\vert \cbracket{\widehat{S}_n= S, \widehat{V}_n^{E\cdot j} = V^{E\cdot j}} \right] \\
& -\mathbb{E}_{\mathcal{N}}\left[\mathcal{H} \circ P^{E\cdot j}(Z) \Bigm\vert \cbracket{\widehat{S}_n= S, \widehat{V}_n^{E\cdot j} = V^{E\cdot j}}\right] \Bigr\rvert=0.
\end{align*}
\end{lemma}

\begin{proof}
Suppose that $Z$ follows $\mathcal{N}_{p}(0_p, I_{p, p})$, then conditional density of $Z$ at the point $z$ is equal to
\[
\frac{\phi\left(z; 0_p, I_{p, p}\right) F(z)}{\int \phi\left(z^{\top} ; 0_p, I_{p, p}\right) F\left(z^{\top}\right) d z^{\top}}=\left(\mathbb{E}_{\mathcal{N}}\left[F(Z)\right]\right)^{-1} \phi\left(z; 0_p, I_{p, p}\right) F(z).
\] This means that the ratio between the conditional and unconditional densities of $\Upsilon_{n}$, at the point $Z$, is to
\[
R(Z)=\left(\mathbb{E}_{\mathcal{N}}\left[F(Z)\right]\right)^{-1} F(Z), 
\]thus
\begin{align*}
\mathbb{E}_{\mathcal{N}}\left[\mathcal{H} \circ P^{E\cdot j}(Z) \big \lvert \cbracket{\widehat{S}_n= S, \widehat{V}_n^{E\cdot j} = V^{E\cdot j}} \right] & =\mathbb{E}_{\mathcal{N}}\left[\mathcal{H} \circ P^{E\cdot j}(Z) R(Z)\right] \\
& =\frac{\mathbb{E}_{\mathcal{N}}\left[\mathcal{H} \circ P^{E\cdot j}(Z) \times F(Z)\right]}{\mathbb{E}_{\mathcal{N}}\left[F(Z)\right]}
\end{align*}
Now for general distributions, i.e., $\Upsilon_{n}$ follows $\mathbb{F}_{n}$, the weight $F(Z)$ might not be a Gaussian integral, as the change of variables is not exact. However, from Assumption~\ref{assump:2}, for sufficiently large $n$, we can use the Lebesgue density $q_{n}$ and it's corresponding weight $G_{n}(Z)$ to get:  \[\mathbb{E}_{\mathbb{F}_{n}}\rbracket{\mathcal{H} \circ P^{E\cdot j}(\Upsilon_{n}) | \cbracket{\widehat{S}_n= S, \widehat{V}_n^{E\cdot j} = V^{E\cdot j}}} = \frac{ \mathbb{E}_{\mathbb{F}_{n}}\rbracket{ \mathcal{H} \circ P^{E\cdot j}(\Upsilon_{n}) G_{n}(\Upsilon_{n}) }}{\mathbb{E}_{\mathbb{F}_{n}}  [G_{n}(\Upsilon_{n})]}.\] So in order to get \[\lim _n \sup _{\mathbb{F}_n \in \mathcal{F}_n} \left| \mathbb{E}_{\mathbb{F}_{n}}\rbracket{\mathcal{H} \circ P^{E\cdot j}(\Upsilon_{n}) | \cbracket{\widehat{S}_n= S, \widehat{V}_n^{E\cdot j} = V^{E\cdot j}}} - \frac{\mathbb{E}_{\mathbb{F}_n}\left[\mathcal{H} \circ P^{E\cdot j}(\Upsilon_{n}) \times F(\Upsilon_{n})\right]}{\mathbb{E}_{\mathbb{F}_n}\left[F(\Upsilon_{n})\right]} \right| = 0\]
it is sufficient to prove
\begin{align*}
& \lim _n \sup _{\mathbb{F}_n \in \mathcal{F}_n}\left|\frac{\mathbb{E}_{\mathbb{F}_n}\left[\mathcal{H} \circ P^{E\cdot j}\left(\Upsilon_{n}\right) \times G_{n}\left(\Upsilon_{n}\right)\right]}{\mathbb{E}_{\mathbb{F}_n}\left[G_{n}\left(\Upsilon_{n}\right)\right]}-\frac{\mathbb{E}_{\mathbb{F}_n}\left[\mathcal{H} \circ P^{E\cdot j}(\Upsilon_{n}) \times F(\Upsilon_{n})\right]}{\mathbb{E}_{\mathbb{F}_n}\left[F(\Upsilon_{n})\right]}\right|=0.
\end{align*}

Note that the limit on the left-hand side is further bounded by
\begin{align*}
& \underbrace{\lim _n \sup _{\mathbb{F}_n}\left|\frac{\mathbb{E}_{\mathbb{F}_n}\left[\mathcal{H} \circ P^{E\cdot j}\left(\Upsilon_{n}\right) \times G_{n}\left(\Upsilon_{n}\right)\right]}{\mathbb{E}_{\mathbb{F}_n}\left[G_{n}\left(\Upsilon_{n}\right)\right]}-\frac{\mathbb{E}_{\mathbb{F}_n}\left[\mathcal{H} \circ  P^{E\cdot j}\left(\Upsilon_{n}\right) \times G_{n}\left(\Upsilon_{n}\right)\right]} {\mathbb{E}_{\mathbb{F}_n}\left[F\left(\Upsilon_{n}\right)\right]}\right|}_{\text{(F1)}}\\
&+ \underbrace{\lim _n \sup _{\mathbb{F}_n}\left|\frac{\mathbb{E}_{\mathbb{F}_n}\left[\mathcal{H} \circ  P^{E\cdot j}\left(\Upsilon_{n}\right) \times G_{n}\left(\Upsilon_{n}\right)\right]}{\mathbb{E}_{\mathbb{F}_n}\left[F\left(\Upsilon_{n}\right)\right]}-\frac{\mathbb{E}_{\mathbb{F}_n}\left[\mathcal{H} \circ  P^{E\cdot j}\left(\Upsilon_{n}\right) \times F\left(\Upsilon_{n}\right)\right]}{\mathbb{E}_{\mathbb{F}_n}\left[F\left(\Upsilon_{n}\right)\right]}\right|}_{\text{(F2)}}.
\end{align*}
Thus, our proof is complete by showing the limits of the two terms in the sum is $0$. Observe that
\begin{align*}
\text{(F1)}&\leq \lim _n \sup _{\mathbb{F}_n} \int\left|\mathcal{H} \circ  P^{E\cdot j}\left(\Upsilon_{n}\right)\right|\left|\frac{G_{n}\left(\Upsilon_{n}\right)}{\mathbb{E}_{\mathbb{F}_n}\left[G_{n}\left(\Upsilon_{n}\right)\right]}-\frac{G_{n}\left(\Upsilon_{n}\right)}{\mathbb{E}_{\mathbb{F}_n}\left[F\left(\Upsilon_{n}\right)\right]}\right| d \mathbb{F}_n\left(\Upsilon_{n}\right) \\
\leq & \lim _n \sup _{\mathbb{F}_n} \|\mathcal{H}\|_{\infty} \times\left|\frac{\mathbb{E}_{\mathbb{F}_n}\left[G_{n}\left(\Upsilon_{n}\right)-F\left(\Upsilon_{n}\right)\right]}{\mathbb{E}_{\mathbb{F}_n}\left[F\left(\Upsilon_{n}\right)\right]}\right|
\end{align*}
and that
\begin{align*}
\text{(F2)} & =\lim_n \sup_{\mathbb{F}_n}\left|\frac{\mathbb{E}_{\mathbb{F}_n}\left[\mathcal{H} \circ  P^{E\cdot j}\left(\Upsilon_{n}\right) \times G_{n}\left(\Upsilon_{n}\right)\right]-\mathbb{E}_{\mathbb{F}_n}\left[\mathcal{H} \circ  P^{E\cdot j}\left(\Upsilon_{n}\right) \times F\left(\Upsilon_{n}\right)\right]}{\mathbb{E}_{\mathbb{F}_n}\left[F\left(\Upsilon_{n}\right)\right]}\right| \\
& \leq \lim_n \sup _{\mathbb{F}_n} \|\mathcal{H}\|_{\infty} \times\left|\frac{\mathbb{E}_{\mathbb{F}_n}\left[G_{n}\left(\Upsilon_{n}\right)-F\left(\Upsilon_{n}\right)\right]}{\mathbb{E}_{\mathbb{F}_n}\left[F\left(\Upsilon_{n}\right)\right]}\right|
\end{align*}
Finally, since the condition in Assumption~\ref{assump:2} implies that
\begin{align*}
 & \lim_n \sup_{\mathbb{F}_n} \left|\frac{\mathbb{E}_{\mathbb{F}_n}\left[G_{n}\left(\Upsilon_{n}\right)-F\left(\Upsilon_{n}\right)\right]}{\mathbb{E}_{\mathbb{F}_n}\left[F\left(\Upsilon_{n}\right)\right]} \right|=0,
\end{align*}
we have that 
\[
\lim _n \sup _{\mathbb{F}_n \in \mathcal{F}_n} \left| \mathbb{E}_{\mathbb{F}_{n}}\rbracket{\mathcal{H} \circ P^{E\cdot j}(\Upsilon_{n}) | \cbracket{\widehat{S}_n= S, \widehat{V}_n^{E\cdot j} = V^{E\cdot j}}} - \frac{\mathbb{E}_{\mathbb{F}_n}\left[\mathcal{H} \circ P^{E\cdot j}(\Upsilon_{n}) \times F(\Upsilon_{n})\right]}{\mathbb{E}_{\mathbb{F}_n}\left[F(\Upsilon_{n})\right]} \right| = 0.
\]
This leads us to our claim. 
\end{proof}


\subsubsection{Asymptotic Bounds}

Throughout this section, recall $b_{i,n}$, $a_{i,n} = K^{-1/2}\nbracket{b_{i,n}-\mathbb{E}_{\mathbb{F}_n}\rbracket{b_{i,n}}}$, and $\zeta_{n}$ from Supplement~\ref{app:SIproofs} and Proposition~\ref{prop:asymptotic} of the main paper, and let 
\begin{equation}
    e_{i,n} = a_{i,n} + \frac{\Delta_{n,1}}{\sqrt{n}}.
\label{defn:ei}
\end{equation}
Note that, with $e_{i,n}$ as in \eqref{defn:ei},
\[
\zeta_{n} = \frac{1}{\sqrt{n}}\sum_{i=1}^{n}a_{i,n}, \text{ and } \Upsilon_{n}= \frac{1}{\sqrt{n}}\sum_{i=1}^{n}e_{i,n} = \frac{1}{\sqrt{n}}\sum_{i=1}^{n}a_{i,n} + \Delta_{n,1}.
\]

\begin{proposition}
\label{prop:sixthmoment}
    Under Assumption~\ref{assump:1}, it holds that
     \[\sup _n \sup _{\mathbb{F}_n \in \mathcal{F}_n} \max_{i\in[n]} \mathbb{E}_{\mathbb{F}_n}\left[\left\|a_{i, n}\right\|^6\right] < \infty.\]
\end{proposition}

\begin{proof}
Recall $b_{i,n} = X_{i}^{\top}\nabla\psi\nbracket{X_{i,E}\targetE ; Y_{i}}$ from Proposition~\ref{prop:asymptotic}, and write $\bar b_{i,n}= b_{i,n}-\mathbb{E}_{\mathbb{F}_n}\rbracket{b_{i,n}}$ for the centered score.
Assumption \ref{assump:1} states that, for some constant $B_{0}>0$,
\[\lim_n \sup_{\mathbb{F}_n\in \mathcal{F}_n} \mathbb{E}_{\mathbb{F}_{n}}\rbracket{\exp \nbracket{t\norm{\bar b_{i,n}}}} \leq \exp(B_{0}t^{2})\]
for all $t\in \mathbb{R}$; that is, $\norm{\bar b_{i,n}}$ is uniformly sub-Gaussian.
By definition $a_{i,n} = K^{-1/2}\bar b_{i,n}$, so $\norm{a_{i,n}} \leq \norm{K^{-1/2}}\norm{\bar b_{i,n}}$ and therefore, for all $t\in\mathbb{R}$,
\begin{align*}
\lim_n \sup_{\mathbb{F}_n\in \mathcal{F}_n} \mathbb{E}_{\mathbb{F}_{n}}\rbracket{\exp \nbracket{t\norm{a_{i,n}}}} &\leq \lim_n \sup_{\mathbb{F}_n\in \mathcal{F}_n} \mathbb{E}_{\mathbb{F}_{n}}\rbracket{\exp \nbracket{t\norm{K^{-1/2}}\norm{\bar b_{i,n}}}} \leq \exp(B_{1}t^{2}), \\ \quad B_{1} &= \norm{K^{-1/2}}^{2}B_{0},    
\end{align*}
where the constant picks up $\norm{K^{-1/2}}^{2}$, and not $\norm{K^{-1/2}}$, because scaling $\norm{\bar b_{i,n}}$ by $\norm{K^{-1/2}}$ rescales the argument $t$ of the moment generating function.
Hence $a_{i,n}\in\mathbb{R}^{p}$ is uniformly sub-Gaussian, and consequently, for every $i\in\{1,2,\ldots,n\}$,
\[\sup _n \sup _{\mathbb{F}_n \in \mathcal{F}_n} \mathbb{E}_{\mathbb{F}_n}\left[\left\|a_{i, n}\right\|^6\right] < \infty.\]


\end{proof}



\begin{lemma}
\label{lem:stein}
Let $\zeta_{n}[-i]=\zeta_{n}-\frac{a_{i, n}}{\sqrt{n}}$ be a leave-one out variable, after dropping the $i^{\text{th}}$ variable $a_{i,n}$. 
Let
\[
\mathcal{W}_{\alpha, \kappa}=\zeta_{n}[-1]+\frac{\alpha}{\sqrt{n}} a_{1, n}+\frac{\kappa}{\sqrt{n}} a_{1, n}^*,
\]
where $a_{i, n}^*$ is an independent copy of $a_{i, n}$.
Suppose that $G: \mathbb{R}^p \rightarrow \mathbb{R}$ is a Lebesgue-almost surely three times differentiable function such that $\mathbb{E}_{\mathcal{N}}[|G(Z)|]<\infty$.
It holds that \begin{align*}
\left|\mathbb{E}_{\mathbb{F}_n}\left[G(\zeta_{n})\right]-\mathbb{E}_{\mathcal{N}}[G(Z)]\right| \lesssim& \frac{1}{\sqrt{n}} \sum_{\substack{\lambda, \gamma \in \mathbb{N}: 
\lambda+\gamma \leq 3}} \sum_{i_1, i_2, i_3 \in[p]} \mathbb{E}_{\mathbb{F}_n}\Big[\left\|a_{1, n}\right\|^\lambda\left\|a_{1, n}^*\right\|^\gamma \times \\
 &\sup _{\alpha, \kappa \in[0,1]} \int_0^1 \frac{\sqrt{t}}{2} \mathbb{E}_{\mathcal{N}}\Big[\left|\partial_{i_1, i_2, i_3}^3 G\left(\sqrt{t} \mathcal{W}_{\alpha, \kappa}+\sqrt{1-t} Z\right)\right|\Big] d t\Big].
\end{align*}
\end{lemma} 

\begin{proof}
The Stein bound in Lemma 2 from \cite{snigdhacarving} yields:
\[
\left|\mathbb{E}_{\mathbb{F}_n}\left[G\left(\zeta_{n}\right)\right]-\mathbb{E}_{\mathcal{N}}[G(Z)]\right| \lesssim \frac{1}{\sqrt{n}} \sum_{\substack{\lambda, \gamma \in \mathbb{N}: i_1, i_2, i_3 \\ \lambda+\gamma \leq 3}} \mathbb{E}_{\mathbb{F}_n}\left[\left\|a_{1, n}\right\|^\lambda\left\|a_{1, n}^*\right\|^\gamma \sup _{\alpha, \kappa \in[0,1]}\left|\partial_{i_1, i_2, i_3}^3 \mathcal{D}_G\left(\mathcal{W}_{\alpha, \kappa}\right)\right|\right] .
\]
where
\[
\mathcal{D}_G(Z^{*})=\int_0^1 \frac{1}{2 t}\left(\mathbb{E}_{\mathcal{N}}[G(\sqrt{t} Z^{*}+\sqrt{1-t} Z)]-\mathbb{E}_{\mathcal{N}}[G(Z)]\right) d t.
\]
Then by taking derivatives we directly get:
\[
\left|\partial_{i_1, i_2, i_3}^3 \mathcal{D}_G\left(\mathcal{W}_{\alpha, \kappa}\right)\left[i_1, i_2, i_3\right]\right|=\int_0^1 \frac{\sqrt{t}}{2} \mathbb{E}_{\mathcal{N}}\left[\left|\partial_{i_1, i_2, i_3}^3 G\left(\sqrt{t} \mathcal{W}_{\alpha, \kappa}+\sqrt{1-t} Z\right)\right|\right] d t
\]
which gives us the desired Stein's bound on the difference in the two expectations.  
\end{proof}

\subsubsection{Pivot and Weight Derivatives}
For the notations in this section, we denote the partial derivative of a multivariate function $f: \mathbb{R}^p \rightarrow \mathbb{R}$, evaluated at $x=\left(x_1, x_2, \ldots, x_p\right)^{\top} \in \mathbb{R}^p$, by
\[
\partial_{i_1, \ldots, i_m}^m f(x)=\frac{\partial^m f(x)}{\partial x_{i_1} \ldots \partial x_{i_m}}.\]

\begin{proposition}
\label{pivotderivative}
For $P^{E\cdot j}(Z)$, $\widetilde{P}$ and $T_n$ defined in \eqref{StandardizedPivotUpsilon} and \eqref{StandardizedNotations}, it holds that
\[
\left|\partial_{i_1, i_2, i_3}^3 P^{E\cdot j}(Z)\right| \leq \sum_{l=0}^3 c_l\left\|\widetilde{P}Z+T_n\right\|^l,
\]
where $c_0, \ldots, c_3$ are constants that do not depend on $n$. 

\end{proposition}

\begin{proof}
Recall,
\[  P^{E\cdot j}(Z) = \dfrac{\int_{-\infty}^{A_{1}Z + \sqrt{n}\targetj}\phi(x; \sqrt{n}\targetj, \sigmaj)F(x, A_{2}Z + \sqrt{n}\gammaej)dx}{\int_{-\infty}^{\infty}\phi(x; \sqrt{n}\targetj, \sigmaj)F(x, A_{2}Z + \sqrt{n}\gammaej)dx}\]
where $F$ is the two-argument function of \eqref{StandardizedPivot},
\[
F\left(x_1, x_2\right)=\int_{I_{-}^{E\cdot j}}^{\infty} \phi\left(L t+P^{E\cdot j}_{1} x_1+P^{E\cdot j}_{2} x_2+ T ; 0_p, \Omega\right) dt.
\]
Applying the Leibniz integral rule, we have that
\begin{equation}
\begin{split}
\partial P^{E\cdot j}(Z)= & \frac{\int_{-\infty}^{A_{1}Z + \sqrt{n}\targetj} \phi\left(x ; \sqrt{n}\targetj, \sigmaj\right) \partial F\left(x, A_{2}Z + \sqrt{n}\gammaej\right) d x}{\int_{-\infty}^{\infty} \phi\left(x ; \sqrt{n}\targetj, \sigmaj\right) F\left(x, A_{2}Z + \sqrt{n}\gammaej\right) d x} \\
& +\frac{\phi\left(A_{1}Z + \sqrt{n}\targetj ; \sqrt{n}\targetj, \sigmaj\right) F\left(A_{1}Z + \sqrt{n}\targetj, A_{2}Z + \sqrt{n}\gammaej\right) A_1}{\int_{-\infty}^{\infty} \phi\left(x ; \sqrt{n}\targetj, \sigmaj\right) F\left(x,A_{2}Z + \sqrt{n}\gammaej\right) d x} \\
& -P^{E\cdot j}(Z) \times \frac{\int_{-\infty}^{\infty} \phi\left(x ; \sqrt{n}\targetj, \sigmaj\right) \partial F\left(x, A_{2}Z + \sqrt{n}\gammaej\right) d x}{\int_{-\infty}^{\infty} \phi\left(x ; \sqrt{n}\targetj, \sigmaj\right) F\left(x, A_{2}Z + \sqrt{n}\gammaej\right) d x}.
\end{split}
\label{first:deriv}
\end{equation}

Read \eqref{first:deriv} coordinate wise, with $\partial=\partial_i$.  Its first
and third terms combine as the covariance of
$\mathbbm{1}_{\{x\leq A_1Z+\sqrt n\targetj\}}$ and
$\partial_i\log F(x,A_2Z+\sqrt n\gammaej)$ under the density proportional to
$\phi(x;\sqrt n\targetj,\sigmaj)F(x,A_2Z+\sqrt n\gammaej)$.  By the definition
of $F$, this is the $x$-marginal of the joint density proportional to
\[
\phi(x;\sqrt n\targetj,\sigmaj)
\phi(Lt+P_1x+P_2A_2Z+P_2\sqrt n\gammaej+T;0_p,\Omega).
\]
This is a Gaussian density in $(x,t)$ with covariance independent of $Z$ and
$n$, truncated only in $t$.  Differentiating its normalized integral centers
its affine Gaussian score.  The resulting moments through order three are
uniformly bounded over the truncation point: after centering and scaling the
marginal $t$ density, a nonpositive truncation has probability at least $1/2$;
for a positive truncation $a>1$, shifting to zero gives a density proportional
to $e^{-ar-r^2/2}$ on $r\geq0$, whose $k$th moment is at most
$k!a^{-(k+1)}/(e^{-3/2}a^{-1})\leq e^{3/2}k!$ for $k\leq3$; the case $0<a\leq1$ is
immediate.  The conditional distribution of $x$ given $t$ has fixed variance
and affine mean, so the same bound holds for the joint affine scores.  Hence
the normalized-integral terms and their first two derivatives are bounded.

It remains to control the moving-boundary terms.  At
$x=A_1Z+\sqrt n\targetj$, the argument of the Gaussian kernel in $F$ is
\[
Lt+P_1A_1Z+P_2A_2Z+P_1\sqrt n\targetj
   +P_2\sqrt n\gammaej+T
=Lt+\widetilde PZ+T_n.
\]
After completing the square in $x$ conditional on $t$, the normalized boundary
term is a mixture of Gaussian densities with fixed positive variance.
Its first two derivatives contain fixed-variance Gaussian derivatives and
centered affine scores, controlled above.  At the boundary the remaining
kernel factors are polynomials in $Lt+\widetilde PZ+T_n$ times that Gaussian
kernel.  Since $\sup_y\|y\|^k e^{-c\|y\|^2}<\infty$ for $k\leq2$ and some
$c>0$, completion of the square in $t$ and the product rule show that these
boundary contributions have degree at most three in
$\widetilde PZ+T_n$.  Together with the bounded normalized-integral terms, this
proves
\[
\left|\partial_{i_1,i_2,i_3}^3P^{E\cdot j}(Z)\right|
\leq\sum_{l=0}^3c_l\|\widetilde PZ+T_n\|^l.
\]

\end{proof}

\begin{proposition}
\label{weightderivative}
With $\Exp(\cdot,\cdot)$ as in \eqref{defn:Exp}, we have that
\[
\left|\partial_{i_1, i_2, i_3}^3 F(Z)\right| \leq \sum_{l=0}^3 c_l^{\top}\left\|\widetilde{P} Z+T_{n}\right\|^l \times \Exp\left(\widetilde{P} Z+T_{n}, \Theta\right)
\]
with $\Theta=\Theta(\Omega)$ as in \eqref{defn:Theta}, and constants $c_0^{\top}, \ldots, c_3^{\top}$ that do not depend on $n$.    
\end{proposition}

\begin{proof}
 Observe that
\begin{align*}
F(Z) \propto \int_{I_{-}^{E\cdot j}}^{\infty} \exp & \left\{ -\frac{1}{2}\left(L t+\widetilde{P} Z+T_{n}\right)^{\top} \Omega^{-1}\left(L t+\widetilde{P} Z+T_{n}\right)\right\} d t \\
\propto \int_{I_{-}^{E\cdot j}}^{\infty} \exp & \left\{-\frac{1}{2}\left(\widetilde{P} Z+T_{n}\right)^{\top} \Theta\left(\widetilde{P} Z+T_{n}\right)\right\}\\
& \times \exp \left\{-\frac{L^{\top} \Omega^{-1} L}{2}\left[t+\frac{L^{\top} \Omega^{-1}\left(\widetilde{P} Z+T_{n}\right)}{L^{\top} \Omega^{-1} L}\right]^2\right\} d t
\end{align*}

Applying a change of variables $\widetilde{t} = t + \ell(Z)$ to the previous integral, where 
$
\ell(Z)=({L^{\top} \Omega^{-1} L})^{-1} {L^{\top} \Omega^{-1}\left(\widetilde{P} Z+T_{n}\right)},
$
we have that
\[
F(Z) \propto \Exp\left(\widetilde{P} Z+T_{n}, \Theta\right) \times \int_{I_{-}^{E\cdot j}+\ell(Z)}^{\infty} \exp \left\{-\frac{L^{\top} \Omega^{-1} L}{2} \widetilde{t}^2\right\} d \widetilde{t}.
\]

An order-$m\leq3$ derivative of the first factor is that factor times a
polynomial of degree at most $m$ in $\widetilde PZ+T_n$.  By the Leibniz rule,
an order-$k\geq1$ derivative of the integral is its Gaussian kernel at
$I_-^{E\cdot j}+\ell(Z)$ times a polynomial of degree $k-1$ in the same
argument.  Since $\sup_s |s|^r e^{-cs^2}<\infty$ for $r\leq2$, the product
rule gives
\[
\left|\partial_{i_1, i_2, i_3}^3 F(Z)\right|
\leq \sum_{l=0}^3 c_l^{\top}\left\|\widetilde{P} Z+T_{n}\right\|^l
\times \Exp\left(\widetilde{P} Z+T_{n}, \Theta\right).
\]
The same calculation gives the bounds for $m=0,1,2$.  For $m=3$, the term with
one derivative on the moving limit and two on the first factor contains its
constant and quadratic parts, which give $l=0$ and $l=2$.

\end{proof}

\subsubsection{Other Key Results}
\begin{proposition}
\label{prop:A21}
The following assertions hold. 
\begin{enumerate}
 \item  Suppose that $L^{\top} \Omega^{-1} b<0$. Let $\mathcal{D}_n=\left[-d_0 r_n\cdot \mathbf{1}_p,\, d_0 r_n\cdot \mathbf{1}_p\right]$ for a positive constant $d_0$ such that
\[
d_0<\frac{1}{2} \frac{\left|L^{\top} \Omega^{-1} b\right|}{\sqrt{p}\left\|\widetilde{P}^{\top} \Omega^{-1} L\right\|}.
\]
Then, there exists a Lebesgue-almost everywhere differentiable function $\mathcal{K}^{1}_n$ such that for $m \in\{0,1,2,3\}$:
    \[
F(Z)=\Exp\left(\widetilde{P} Z-r_n b, \Omega^{-1}\right) \times \mathcal{K}^{1}_n(Z), \text { and } \sup _{\mathbb{F}_n \in \mathcal{F}_n} \sup _{Z \in \mathcal{D}_n}\left|r_n\right|\left|\partial_{i_1, \ldots i_m}^m \mathcal{K}^{1}_n(Z)\right|<\infty.
\]

\item Suppose that $L^{\top} \Omega^{-1} b>0$. Then, there exists a Lebesgue-almost everywhere differentiable function $\mathcal{K}^{2}_n$ such that for $m \in\{0,1,2,3\}$:
\[
F(Z)=\Exp\left(\widetilde{P} Z-r_n b, \Theta\right) \times \mathcal{K}^{2}_n(Z) \text { and } \sup _{\mathbb{F}_n \in \mathcal{F}_n} \sup _{Z \in \mathbb{R}^p}\left|\partial_{i_1, \ldots i_m}^m \mathcal{K}^{2}_n(Z)\right|<\infty .
\]
 \end{enumerate}
\end{proposition}

\begin{proof}
First, consider the case when $L^{\top} \Omega^{-1} b<0$.
We begin by noting that
\begin{equation}
\begin{split}
F(Z) & \propto \Exp\left(\widetilde{P}Z+T_n, \Theta\right) \times \int_{I_{-}^{E\cdot j}}^{\infty} \exp \left\{-\frac{L^{\top} \Omega^{-1} L}{2}\left[t+\frac{L^{\top} \Omega^{-1}\left(\widetilde{P}Z+T_n\right)}{L^{\top} \Omega^{-1} L}\right]^2\right\} d t \\
& \propto \Exp\left(\widetilde{P}Z+T_n+L I_{-}^{E\cdot j}, \Theta\right) \times \int_{I_{-}^{E\cdot j}}^{\infty} \exp \left\{-\frac{L^{\top} \Omega^{-1} L}{2}\left[t+\frac{L^{\top} \Omega^{-1}\left(\widetilde{P}Z+T_n\right)}{L^{\top} \Omega^{-1} L}\right]^2\right\} d t.
\end{split}
\label{re:weight}
\end{equation}

We can rewrite this function as
\[
F(Z)=\Exp\left(\widetilde{P}Z+T_n+L I_{-}^{E\cdot j}, \Omega^{-1}\right) \times \mathcal{K}^{1}_n(Z)
\] 
where
\begin{align*}
\mathcal{K}^{1}_n(Z)=d_2 \exp & \left\{\frac{1}{2}\left(\widetilde{P}Z+T_n+L I_{-}^{E\cdot j}\right)^{\top} \frac{\Omega^{-1} L L^{\top} \Omega^{-1}}{L^{\top} \Omega^{-1} L}\left(\widetilde{P}Z+T_n+L I_{-}^{E\cdot j}\right)\right\} \\
& \times \int_{I_{-}^{E\cdot j}}^{\infty} \exp \left\{-\frac{L^{\top} \Omega^{-1} L}{2}\left[t+\frac{L^{\top} \Omega^{-1}\left(\widetilde{P}Z+T_n\right)}{L^{\top} \Omega^{-1} L}\right]^2\right\} d t
\end{align*} for a constant $d_2$.
By the variable substitution $\widetilde{t}=t-I_{-}^{E\cdot j}$ in the integral involved in the expression of $\mathcal{K}^{1}_n(Z)$, we have that
\begin{align*}
\mathcal{K}^{1}_n(Z) \propto \exp & \left\{\frac{1}{2}\left(\widetilde{P}Z-r_n b\right)^{\top} \frac{\Omega^{-1} L L^{\top} \Omega^{-1}}{L^{\top} \Omega^{-1} L}\left(\widetilde{P}Z-r_n b\right)\right\} \\
\times & \int_0^{\infty} \exp \left\{-\frac{1}{2}\left(\widetilde{P}Z-r_n b+L \widetilde{t}\right)^{\top} \frac{\Omega^{-1} L L^{\top} \Omega^{-1}}{L^{\top} \Omega^{-1} L}\left(\widetilde{P}Z-r_n b+L \widetilde{t}\right)\right\} d \widetilde{t}.
\end{align*} Using simple algebra, we simplify the expression for $\mathcal{K}^{1}_n(Z)$ which leads to the following bound:
\begin{align*}
\mathcal{K}^{1}_n(Z) & \propto \int_0^{\infty} \exp \left\{-\frac{1}{2} \widetilde{t} L^{\top} \Omega^{-1} L \widetilde{t}\right\} \times \exp \left\{-\widetilde{t} L^{\top} \Omega^{-1}\left(\widetilde{P}Z-r_n b\right)\right\} d \widetilde{t} \\
& \leq \int_0^{\infty} \exp \left\{-\widetilde{t} L^{\top} \Omega^{-1}\left(\widetilde{P}Z-r_n b\right)\right\} d \widetilde{t}.
\end{align*} Furthermore, for $Z\in \mathcal{D}_n$, we have
\[
\mathcal{K}^{1}_n(Z) \lesssim \frac{1}{L^{\top} \Omega^{-1}\left(\widetilde{P}Z-r_n b\right)} \lesssim \frac{1}{r_n\left|L^{\top} \Omega^{-1} b\right|}
\]
This proves our claim for $m=0$.

Now, for $m=1$, note that the first derivative of $\mathcal{K}^{1}_n(\cdot)$, we have
\[
\partial^1 \mathcal{K}^{1}_n(Z) \propto \int_0^{\infty}-\widetilde{t} L^{\top} \Omega^{-1} L \exp \left\{-\frac{1}{2} \widetilde{t} L^{\top} \Omega^{-1} L \widetilde{t}\right\} \times \exp \left\{-\widetilde{t} L^{\top} \Omega^{-1}\left(\widetilde{P}Z-r_n b\right)\right\} d \widetilde{t},
\]
which implies the following inequality
\begin{align*}
\left|\partial_{i_1}^1 \mathcal{K}^{1}_n(Z)\right| & \lesssim \int_0^{\infty}-\widetilde{t} \exp \left\{-\widetilde{t} L^{\top} \Omega^{-1}\left(\widetilde{P}Z-r_n b\right)\right\} d \widetilde{t} \\
& =\frac{1}{\left[L^{\top} \Omega^{-1}\left(\widetilde{P}Z-r_n b\right)\right]^2}.
\end{align*}
As a result, by definition of sets $\mathcal{D}_n$, whenever $Z\in \mathcal{D}_n$ our claim holds. A similar strategy is applied to obtain the conclusions for the second and third order partial derivatives $\left|\partial_{i_1, i_2}^2 \mathcal{K}^{1}_n(Z)\right|$ and $\left|\partial_{i_1, i_2, i_3}^3 \mathcal{K}^{1}_n(Z)\right|$.

Now we turn to the case when $L^{\top} \Omega^{-1} b>0$.
We revisit \eqref{re:weight}.
If we define
\[
\mathcal{K}^{2}_n(Z)= C_{2} \int_{I_{-}^{E\cdot j}}^{\infty} \exp \left\{-\frac{L^{\top} \Omega^{-1} L}{2}\left[t+\frac{L^{\top} \Omega^{-1}\left(\widetilde{P}Z+T_n\right)}{L^{\top} \Omega^{-1} L}\right]^2\right\} dt,
\] 
 then we can write
\[
F(Z)=\Exp\left(\widetilde{P}Z+T_n+L I_{-}^{E\cdot j}, \Theta\right) \times \mathcal{K}^{2}_n(Z)
\] 
for the constant $C_{2}$. Now the uniform bounds on the derivative of  $\mathcal{K}^{2}_n$ follow directly after we apply the
Leibniz integral rule.

\end{proof}

\begin{proposition}
\label{prop:A22}
For 
\[\mathcal{L}_n\sim\mathcal{N}\nbracket{-\rbracket{L^{\top}\nbracket{\Omega + \widetilde{P}\widetilde{P}^{\top}}^{-1}L}^{-1}L^{\top}\nbracket{\Omega + \widetilde{P}\widetilde{P}^{\top}}^{-1}T_{n}, \rbracket{L^{\top}\nbracket{\Omega + \widetilde{P}\widetilde{P}^{\top}}^{-1}L}^{-1}},\]
 we have that
\begin{align*}
\mathbb{E}_{\mathcal{N}}[F(Z)] &= \Exp \nbracket{LI_{-}^{E\cdot j} + T_{n}, \nbracket{\Omega + \widetilde{P}\widetilde{P}^{\top}}^{-1} - \frac{\nbracket{\Omega + \widetilde{P}\widetilde{P}^{\top}}^{-1}L L^{\top}\nbracket{\Omega + \widetilde{P}\widetilde{P}^{\top}}^{-1}}{L^{\top}\nbracket{\Omega + \widetilde{P}\widetilde{P}^{\top}}^{-1}L} } \\
&\qquad\qquad\times \mathbb{P}\rbracket{\mathcal{L}_n \geq I_{-}^{E\cdot j}}.
\end{align*}
\end{proposition}

\begin{proof}
Interchanging the order between expectation and integration, we obtain
\begin{align*}
\mathbb{E}_{\mathcal{N}}\left[F(Z)\right] & =\mathbb{E}_{\mathcal{N}}\left[\int_{I_{-}^{E\cdot j}}^{\infty} \phi\left(L t+\widetilde{P}Z+T_n ; 0_p, \Omega\right) d t\right] \\
& =\int_{I_{-}^{E\cdot j}}^{\infty} \mathbb{E}_{\mathcal{N}}\left[\phi\left(L t+\widetilde{P}Z+T_n ; 0_p, \Omega\right)\right] dt
\end{align*}
Observe that the expectation in the integrand equals
\begin{align*}
& \mathbb{E}_{\mathcal{N}}\left[\phi\left(L t+\widetilde{P}Z+T_n ; 0_p, \Omega\right)\right] \\
& =\int_{\mathbb{R}^p} \phi\left(L t+\widetilde{P}Z+T_n ; 0_p, \Omega\right) \phi\left(Z; 0_p, I_{p, p}\right) d Z\\
& \propto \int_{\mathbb{R}^p} \exp \left\{-\frac{1}{2}\left(L t+\widetilde{P}Z+T_n\right)^{\top} \Omega^{-1}\left(L t+\widetilde{P}Z+T_n\right)-\frac{1}{2} Z^{\top} Z\right\} d Z\\
& \propto \exp \left\{-\frac{1}{2}\left(L t+T_n\right)^{\top}\left[\Omega^{-1}-\Omega^{-1}\widetilde{P}\left(I_{p, p}+\widetilde{P}^{\top} \Omega^{-1} \widetilde{P}\right)^{-1}\widetilde{P}^{\top} \Omega^{-1}\right]\left(L t+T_n\right)\right\} \\
& \quad \times \int_{\mathbb{R}^p} \exp \left\{-\frac{1}{2}\left(Z-\Phi_n\right)^{\top}\left(I_{p, p}+\widetilde{P}^{\top} \Omega^{-1} \widetilde{P}\right)\left(Z-\Phi_n\right)\right\} d Z,
\end{align*}
where
\[
\Phi_n=-\left(I_{p, p}+\widetilde{P}^{\top} \Omega^{-1} \widetilde{P}\right)^{-1} \widetilde{P}^{\top} \Omega^{-1}\left(L t+T_n\right).
\] Then direct application of the Woodbury matrix identity gives us:
\[
\Omega^{-1}-\Omega^{-1} \widetilde{P}\left(I_{p, p}+\widetilde{P}^{\top} \Omega^{-1} \widetilde{P}\right)^{-1} \widetilde{P}^{\top} \Omega^{-1}=\left(\Omega + \widetilde{P}\widetilde{P}^{\top}\right)^{-1}.
\]
which is \eqref{defn:marginal} with $\Xi=\Omega^{-1}$ and $s=1$ in the exponent. This implies that
\[
\mathbb{E}_{\mathcal{N}}\left[\phi\left(L t+\widetilde{P}Z+T_n ; 0_p, \Omega\right)\right] \propto \exp \left\{-\frac{1}{2}\left(L t+T_n\right)^{\top}\left(\Omega + \widetilde{P}\widetilde{P}^{\top}\right)^{-1}\left(L t+T_n\right)\right\}.
\]

Then plugging this expression into our integral, we have that
\begin{align*}
\mathbb{E}_{\mathcal{N}}\left[F(Z)\right] & \propto \int_{I_{-}^{E\cdot j}}^{\infty} \exp \left\{-\frac{1}{2} t^2 L^{\top}\left(\Omega + \widetilde{P}\widetilde{P}^{\top}\right)^{-1} L-t L^{\top}\left(\Omega + \widetilde{P}\widetilde{P}^{\top}\right)^{-1} T_n\right. \\
& \left.\quad \quad\quad\quad\quad\quad\quad\quad\quad\quad\quad\quad\quad\quad\quad\quad-\frac{1}{2} T_n^{\top}\left(\Omega + \widetilde{P}\widetilde{P}^{\top}\right)^{-1} T_n\right\} d t \\
& \propto \int_{I_{-}^{E\cdot j}}^{\infty} \exp \left\{-\frac{1}{2} L^{\top}\left(\Omega + \widetilde{P}\widetilde{P}^{\top}\right)^{-1} L\left(t+\frac{L^{\top}\left(\Omega + \widetilde{P}\widetilde{P}^{\top}\right)^{-1}}{L^{\top}\left(\Omega + \widetilde{P}\widetilde{P}^{\top}\right)^{-1} L} T_n\right)^2\right\} \\
& \times \exp \left\{-\frac{1}{2} T_n^{\top}\Theta\left(\Omega + \widetilde{P}\widetilde{P}^{\top}\right) T_n\right\} d t .
\end{align*} 
At last, we conclude our proof with the observation that
\begin{align*}
& \Exp\left(T_n,\Theta\left(\Omega + \widetilde{P}\widetilde{P}^{\top}\right)\right) \\
& =\Exp\left(L I_{-}^{E\cdot j}+T_n,\Theta\left(\Omega + \widetilde{P}\widetilde{P}^{\top}\right)\right),
\end{align*}
since $\Theta(\Sigma) L=0_p$ for every $\Sigma$ by \eqref{defn:Theta}, so that shifting the argument along $L$ leaves $\Exp$ unchanged. This yields the claimed expression for $\mathbb{E}_{\mathcal{N}}\left[F(Z)\right]$.
\end{proof}

\begin{proposition}
\label{prop:A23}
For sufficiently large $n$, it holds that:
\begin{enumerate}
    \item  $\mathbb{E}_{\mathcal{N}}\left[F(Z)\right] \gtrsim \Exp\left(-r_n b,\Theta\left(\Omega + \widetilde{P}\widetilde{P}^{\top}\right)\right)$ for $L^{\top} \Omega^{-1} b<0$;
    \item $\mathbb{E}_{\mathcal{N}}\left[F(Z)\right] \gtrsim r_n^{-1} \Exp\left(-r_n b,\left[\Omega + \widetilde{P}\widetilde{P}^{\top}\right]^{-1}\right)$ for $L^{\top} \Omega^{-1} b>0$.
    
\end{enumerate}
\end{proposition} 

\begin{proof}
    
Starting with the case of $L^{\top} \Omega^{-1} b<0$, recall we have the form

\[
F(Z) \propto \Exp\left(\widetilde{P}Z-r_n b, \Theta\right) \times \int_0^{\infty} \exp \left\{-\frac{L^{\top} \Omega^{-1} L}{2}\left[t+\frac{L^{\top} \Omega^{-1}\left(\widetilde{P}Z-r_n b\right)}{L^{\top} \Omega^{-1} L}\right]^2\right\} d t.
\]

Then define the set $\mathcal{A}_n=\left\{Z: L^{\top} \Omega^{-1}\left(\widetilde{P}Z-r_n b\right)<0\right\}$. Since

\[
\int_0^{\infty} \exp \left\{-\frac{L^{\top} \Omega^{-1} L}{2}\left[t+\frac{L^{\top} \Omega^{-1}\left(\widetilde{P}Z-r_n b\right)}{L^{\top} \Omega^{-1} L}\right]^2\right\} d t \gtrsim \frac{1}{2} \quad \text { for } Z \in \mathcal{A}_n
\]

it holds that

\[
F(Z) \gtrsim \Exp\left(\widetilde{P}Z-r_n b, \Theta\right) \times \mathbbm{1}_{Z \in \mathcal{A}_n}
\]

Then, we conclude that

\begin{align*}
\mathbb{E}_{\mathcal{N}}\left[F(Z)\right] & \gtrsim \mathbb{E}_{\mathcal{N}}\left[\Exp\left(\widetilde{P}Z-r_n b, \Theta\right) \mathbbm{1}_{Z \in \mathcal{A}_n}\right] \\
& \gtrsim \mathbb{E}_{\mathcal{N}}\left[\Exp\left(\widetilde{P}Z-r_n b, \Theta\right)\right]-\mathbb{E}_{\mathcal{N}}\left[\Exp\left(\widetilde{P}Z-r_n b, \Theta\right) \mathbbm{1}_{Z \in \mathcal{A}_n^c}\right] \\
& \gtrsim \frac{1}{2} \mathbb{E}_{\mathcal{N}}\left[\Exp\left(\widetilde{P}Z-r_n b, \Theta\right)\right].
\end{align*}
Then applying the Woodbury matrix identity after integrating with respect to $Z$ and we get

\[
\mathbb{E}_{\mathcal{N}}\left[\Exp\left(\widetilde{P}Z-r_n b, \Theta\right)\right]\propto\Exp\left(-r_n b,\Theta\left(\Omega + \widetilde{P}\widetilde{P}^{\top}\right)\right),
\]
by \eqref{defn:marginal} applied with $\Xi=\Theta$ and $s=1$, the proportionality constant $\left|I_{p,p}+\widetilde{P}^{\top}\Theta\widetilde{P}\right|^{-1/2}$ being free of $n$. Our conclusion thus follows.
For the other case, note that
\begin{align*}
&\mathbb{P}\left[\mathcal{L}_n \geq I_{-}^{E\cdot j}\right] \\  \propto &\int_{I_{-}^{E\cdot j}}^{\infty} \exp \left\{-\frac{1}{2} L^{\top}\left(\Omega + \widetilde{P}\widetilde{P}^{\top}\right)^{-1} L\left(t-I_{-}^{E\cdot j}+\frac{L^{\top}\left(\Omega + \widetilde{P}\widetilde{P}^{\top}\right)^{-1}\left(L I_{-}^{E\cdot j}+T_n\right)}{L^{\top}\left(\Omega + \widetilde{P}\widetilde{P}^{\top}\right)^{-1} L}\right)^2\right\} d t \\
\propto&  \int_0^{\infty} \exp \left\{-\frac{1}{2} L^{\top}\left(\Omega + \widetilde{P}\widetilde{P}^{\top}\right)^{-1} L\left(\widetilde{t}+\frac{L^{\top}\left(\Omega + \widetilde{P}\widetilde{P}^{\top}\right)^{-1}\left(L I_{-}^{E\cdot j}+T_n\right)}{L^{\top}\left(\Omega + \widetilde{P}\widetilde{P}^{\top}\right)^{-1} L}\right)^2\right\} d \widetilde{t}.
\end{align*}through a change of variable $\widetilde{t}=t-I_{-}^{E\cdot j}$.
When $L^{\top}\left(\Omega + \widetilde{P}\widetilde{P}^{\top}\right)^{-1}\left(L I_{-}^{E\cdot j}+T_n\right)>0$, we apply the Mill's ratio bound to note that

\begin{align*}
&\mathbb{P}\left[\mathcal{L}_n \geq I_{-}^{E\cdot j}\right] \\
\geq& \left(\frac{L^{\top}\left(\Omega + \widetilde{P}\widetilde{P}^{\top}\right)^{-1}\left(L I_{-}^{E\cdot j}+T_n\right)}{L^{\top}\left(\Omega + \widetilde{P}\widetilde{P}^{\top}\right)^{-1} L}\right)^{-1} \times\left[1-\left(\frac{L^{\top}\left(\Omega + \widetilde{P}\widetilde{P}^{\top}\right)^{-1}\left(L I_{-}^{E\cdot j}+T_n\right)}{L^{\top}\left(\Omega + \widetilde{P}\widetilde{P}^{\top}\right)^{-1} L}\right)^{-2}\right] \\
& \times \Exp\left(L I_{-}^{E\cdot j}+T_n, \frac{\left(\Omega + \widetilde{P}\widetilde{P}^{\top}\right)^{-1} L L^{\top}\left(\Omega + \widetilde{P}\widetilde{P}^{\top}\right)^{-1}}{L^{\top}\left(\Omega + \widetilde{P}\widetilde{P}^{\top}\right)^{-1} L}\right).
\end{align*} Combining this with the exact expression for $\mathbb{E}_{\mathcal{N}}\left[F(Z)\right]$ in Proposition~\ref{prop:A22} yields:
\begin{align*}
\mathbb{E}_{\mathcal{N}}\left[F(Z)\right] \geq& \left(\frac{L^{\top}\left(\Omega + \widetilde{P}\widetilde{P}^{\top}\right)^{-1}\left(L I_{-}^{E\cdot j}+T_n\right)}{L^{\top}\left(\Omega + \widetilde{P}\widetilde{P}^{\top}\right)^{-1} L}\right)^{-1} \times\left[1-\left(\frac{L^{\top}\left(\Omega + \widetilde{P}\widetilde{P}^{\top}\right)^{-1}\left(L I_{-}^{E\cdot j}+T_n\right)}{L^{\top}\left(\Omega + \widetilde{P}\widetilde{P}^{\top}\right)^{-1} L}\right)^{-2}\right] \\
& \times \Exp\left(L I_{-}^{E\cdot j}+T_n, \frac{\left(\Omega + \widetilde{P}\widetilde{P}^{\top}\right)^{-1} L L^{\top}\left(\Omega + \widetilde{P}\widetilde{P}^{\top}\right)^{-1}}{L^{\top}\left(\Omega + \widetilde{P}\widetilde{P}^{\top}\right)^{-1} L}\right) \\
& \times \Exp\left(L I_{-}^{E\cdot j}+T_n,\Theta\left(\Omega + \widetilde{P}\widetilde{P}^{\top}\right)\right) \\
=& \nbracket{\frac{L^{\top}\left(\Omega + \widetilde{P}\widetilde{P}^{\top}\right)^{-1}\left(L I_{-}^{E\cdot j}+T_n\right)}{L^{\top}\left(\Omega + \widetilde{P}\widetilde{P}^{\top}\right)^{-1} L}}^{-1} \times\left[1-\left(\frac{L^{\top}\left(\Omega + \widetilde{P}\widetilde{P}^{\top}\right)^{-1}\left(L I_{-}^{E\cdot j}+T_n\right)}{L^{\top}\left(\Omega + \widetilde{P}\widetilde{P}^{\top}\right)^{-1} L}\right)^{-2}\right] \\
& \times \Exp\left(L I_{-}^{E\cdot j}+T_n,\left[\Omega + \widetilde{P}\widetilde{P}^{\top}\right]^{-1}\right) .
\end{align*}

Using the parameterization $L I_{-}^{E\cdot j}+T_n=-r_n b$, we conclude that

\[
\mathbb{E}_{\mathcal{N}}\left[F(Z)\right] \gtrsim r_n^{-1} \Exp\left(-r_n b,\left[\Omega + \widetilde{P}\widetilde{P}^{\top}\right]^{-1}\right).
\] When $L^{\top}\left(\Omega + \widetilde{P}\widetilde{P}^{\top}\right)^{-1}\left(L I_{-}^{E\cdot j}+T_n\right)<0$, we note that \[
\mathbb{P}\left[\mathcal{L}_n \geq I_{-}^{E\cdot j}\right] \geq\frac{1}{2}.\] Hence, for sufficiently large $n$, we have
\begin{align*}
\mathbb{E}_{\mathcal{N}}\left[F(Z)\right] & \gtrsim \frac{1}{2} \Exp\left(-r_n b,\Theta\left(\Omega + \widetilde{P}\widetilde{P}^{\top}\right)\right) \\
& \gtrsim \Exp\left(-r_n b,\left(\Omega + \widetilde{P}\widetilde{P}^{\top}\right)^{-1}\right) .
\end{align*}

\end{proof}

\begin{proposition}
\label{prop:A24}
Let $\mathcal{W}_{\alpha, \kappa}$ be as defined in Lemma~\ref{lem:stein}. Then under Assumptions \ref{assump:1},\ref{assump:2},\ref{assump:3}, we have
\[
\sup _n \sup _{\mathbb{F}_n \in \mathcal{F}_n} \frac{\mathbb{E}_{\mathbb{F}_n}\left[\left\|a_{1, n}\right\|^\lambda\left\|a_{1, n}^*\right\|^\gamma \sup _{\alpha, \kappa \in[0,1]}\left|\int_0^1 \sqrt{t} \Exp\left(\widetilde{P} \sqrt{t} \mathcal{W}_{\alpha, \kappa}-r_n b,\left[\Omega + (1-t)\widetilde{P}\widetilde{P}^{\top}\right]^{-1}\right) d t\right|\right]}{\Exp\left(-r_n b,\left[\Omega + \widetilde{P}\widetilde{P}^{\top}\right]^{-1}\right)}=O(1)
\]for $\lambda, \gamma \in \mathbb{N}$ such that $\lambda+\gamma \leq 3$.   
\end{proposition} 
\begin{proof}
Fixing some notations for this proof, let $e_{\max}$ be the largest eigenvalue of $\left(\Omega + \widetilde{P}\widetilde{P}^{\top}\right)$. Define \[
\Pi(t)=\left\{t \widetilde{P}^{\top}\left[\Omega + (1-t)\widetilde{P}\widetilde{P}^{\top}\right]^{-1} \widetilde{P} +I_{p, p}\right\}^{-1} \widetilde{P}^{\top}\left[\Omega + (1-t)\widetilde{P}\widetilde{P}^{\top}\right]^{-1},
\]and let $\Pi_k(t)$ denote the $k$-th row of $\Pi(t)$, and let $\left\|\Pi(t)\right\|_{\max }=\max _{k \in[p]}\left\|\Pi_k(t)\right\|_2$. For each $t\in[0,1]$ fix \[
c_{t}>\max \left(e_{\max}^{1 / 2} \times(\|b\|+1), \left\|\Pi(t)\right\|_{\max } \times(\|b\|+1)\right).\] Observe that
\begin{align*}
& \mathbb{E}_{\mathbb{F}_n}\left[\left\|a_{1, n}\right\|^\lambda\left\|a_{1, n}^*\right\|^\gamma \sup _{\alpha, \kappa \in[0,1]}\left|\int_0^1 \sqrt{t} \Exp\left(\widetilde{P} \sqrt{t} \mathcal{W}_{\alpha, \kappa}-r_n b,\left[\Omega + (1-t)\widetilde{P}\widetilde{P}^{\top}\right]^{-1}\right) d t\right|\right] \\
 \leq &\mathbb{E}_{\mathbb{F}_n}\Big[\left\|a_{1, n}\right\|^\lambda\left\|a_{1, n}^*\right\|^\gamma \sup _{\alpha, \kappa \in[0,1]}\Big|\int_0^1 \sqrt{t} \Exp\left(\widetilde{P} \sqrt{t} \mathcal{W}_{\alpha, \kappa}-r_n b,\left[\Omega + (1-t)\widetilde{P}\widetilde{P}^{\top}\right]^{-1}\right)\\
 &\;\;\;\;\times \mathbbm{1}_{\mathcal{R}_t}\left(r_n^{-1} \zeta_{n}[-1]\right) d t \Big|\Big] +\mathbb{E}_{\mathbb{F}_n}\left[\left\|a_{1, n}\right\|^\lambda\left\|a_{1, n}^*\right\|^\gamma \mathbbm{1}_{\mathcal{R}_0}\left(r_n^{-1} \zeta_{n}[-1]\right)\right],
\end{align*}
where $\mathcal{R}_t=\left[-c_{t} \cdot \mathbf{1}_p, c_{t} \cdot \mathbf{1}_p\right] \subseteq \mathbb{R}^p$, for $t \in(0,1]$, and $\mathcal{R}_0=\mathcal{R}_1^c$.
Note that for some positive constant $\chi >0 $, the bound on the right-hand side further simplifies as: 
\begin{align*}
& \mathbb{E}_{\mathbb{F}_n}\left[\left\|a_{1, n}\right\|^\lambda\left\|a_{1, n}^*\right\|^\gamma \sup _{\alpha, \kappa \in[0,1]}\left|\int_0^1 \sqrt{t} \Exp\left(\widetilde{P} \sqrt{t} \mathcal{W}_{\alpha, \kappa}-r_n b,\left[\Omega + (1-t)\widetilde{P}\widetilde{P}^{\top}\right]^{-1}\right) d t\right|\right] \\
& \leq \mathbb{E}_{\mathbb{F}_n}\left[\left\|a_{1, n}\right\|^\lambda\left\|a_{1, n}^*\right\|^\gamma \exp \left(\chi\left\|a_{1, n}\right\|\right)\right] \\
& \quad \times \int_0^1 \sqrt{t} \mathbb{E}_{\mathbb{F}_n}\left[\Exp\left(\sqrt{t} \widetilde{P}\zeta_{n}[-1]-r_n b,\left[\Omega + (1-t)\widetilde{P}\widetilde{P}^{\top}\right]^{-1}\right) \mathbbm{1}_{\mathcal{R}_t}\left(r_n^{-1} \zeta_{n}[-1]\right)\right] d t \\
& \quad+\mathbb{E}_{\mathbb{F}_n}\left[\left\|a_{1, n}\right\|^\lambda\left\|a_{1, n}^*\right\|^\gamma\right] \mathbb{E}_{\mathbb{F}_n}\left[\mathbbm{1}_{\mathcal{R}_0}\left(r_n^{-1} \zeta_{n}[-1]\right)\right].
\end{align*} 
Define the function \[\Phi_t(Z)= \begin{cases}\frac{1}{2}\left(\sqrt{t} \widetilde{P}Z-b\right)^{\top}\left[\Omega + (1-t)\widetilde{P}\widetilde{P}^{\top}\right]^{-1}\left(\sqrt{t} \widetilde{P}Z-b\right), & \text { if } t \in(0,1] \\ 0, & \text { if } t=0\end{cases}.\] 
Then, we obtain that
\begin{align*}
\frac{\int_0^1 \sqrt{t} \mathbb{E}_{\mathbb{F}_n}\left[\Exp\left(\sqrt{t} \widetilde{P} \zeta_{n}[-1]-r_n b,\left[\Omega + (1-t)\widetilde{P}\widetilde{P}^{\top}\right]^{-1}\right) \mathbbm{1}_{\mathcal{R}_t}\left(r_n^{-1} \zeta_{n}[-1]\right)\right] d t}{\Exp\left(-r_n b,\left[\Omega + \widetilde{P}\widetilde{P}^{\top}\right]^{-1}\right)} \\
\leq \sup _n \sup _{\mathbb{F}_n \in \mathcal{F}_n} \sup _{t \in(0,1]} \frac{\mathbb{E}_{\mathbb{F}_n}\left[\exp \left(-r_n^2 \Phi_t\left(r_n^{-1} \zeta_{n}[-1]\right)\right) \mathbbm{1}_{\left[-c_{t} \cdot \mathbf{1}_p, c_{t} \cdot \mathbf{1}_p\right]}\left(r_n^{-1} \zeta_{n}[-1]\right)\right] \int_0^1 \sqrt{t} d t}{\Exp\left(-r_n b,\left[\Omega + \widetilde{P}\widetilde{P}^{\top}\right]^{-1}\right)}.  
\end{align*}

Recall that \[\zeta_{n} = \frac{1}{\sqrt{n}}\sum_{i=1}^{n}a_{i,n}\] has a finite moment generating function near the origin, since we know that the $\{a_{i,n}: i\in [n]\}$ are sub-Gaussian variables under Assumption~\ref{assump:1}.
Thus, $\zeta_{n}$ satisfies a large deviation principle with rate function $I(Z)=\|Z\|^2 / 2$.
Varadhan's large deviation principle ensures that for any $\mathbb{F}_n \in \mathcal{F}_n$ satisfying Assumptions~\ref{assump:2} and~\ref{assump:3}, and for sufficiently large $n$, the following holds: 
\begin{align*}
r_n^{-2} \log \mathbb{E}_{\mathbb{F}_n}\left[\exp \left(-r_n^2 \Phi_t\left(\frac{\zeta_{n}}{r_n}\right)\right) \mathbbm{1}_{\mathcal{R}_t}\left(r_n^{-1} \zeta_{n}\right)\right] & \leq \sup _{Z \in \mathcal{R}_t}\left(-\frac{\|Z\|^2}{2}-\Phi_t(Z)\right) \\
& =-\inf _{Z \in \mathcal{R}_t}\left(\frac{1}{2} Z^{\top} Z+\Phi_t(Z)\right)
\end{align*} 
for $t\in [0,1]$, where $\mathcal{R}_t=\left[-c_{t} \cdot \mathbf{1}_p, c_{t} \cdot \mathbf{1}_p\right]$ for $c_{t}>0$ and $t \in(0,1]$, and $\mathcal{R}_0$ is the complement of $\left[-c_{t} \cdot \mathbf{1}_p, c_{t} \cdot \mathbf{1}_p\right]$.

Therefore, we conclude that
\[
\sup _n \sup _{\mathbb{F}_n \in \mathcal{F}_n} \sup _{t \in(0,1]} \frac{\mathbb{E}_{\mathbb{F}_n}\left[\exp \left\{-r_n^2 \Phi_t\left(r_n^{-1} \zeta_{n}[-1]\right)\right\} \mathbbm{1}_{\left[-c_{t} \cdot \mathbf{1}_p, c_{t} \cdot \mathbf{1}_p\right]}\left(r_n^{-1} \zeta_{n}[-1]\right)\right]}{\Exp\left(-r_n b,\left[\Omega + \widetilde{P}\widetilde{P}^{\top}\right]^{-1}\right)}<\infty.
\]
\end{proof}

\section{Extensions}
\label{app:extension}

\subsection{Extensions to other penalties}

In this section, we extend our approach to other penalized approaches for regression.
More precisely, we show that the construct of our pivot easily generalizes to a broader framework of penalized regression, requiring only minor adjustments.

Let
\begin{equation}
\label{rand:lreg}
\elasso = \underset{b\in \mathbb{R}^p}{\text{argmin}} \left\{\frac{1}{\sqrt{n}}\sum_{i=1}^{n}\psi(X_{i}b; Y_{i}) + \sum_{j=1}^{p}\lP(|b_{j}|) -\sqrt{n}\omega_{n}^{\top}b \right\}
\end{equation}
represent the randomized estimator regularized with a general penalty function, $\lP: [0,\infty) \to [0, \infty)$, where $\lambda\in \mathbb{R}^d$ represents the tuning parameters involved in the approach.
As done in \eqref{rand:lasso}, $\sqrt{n}\omega_{n}$ denotes a $p$-dimensional randomization variable drawn from $\mathcal{N}\left(0_p, \Omega\right)$, independent of the observed data. 

Suppose the selected set of moderators, corresponding to the indices of the non-zero entries of $\elasso$, is denoted by
\[\widehat{E}=\left\{j\in \{1,2,\ldots, p\}: |\sign(\elassoj)|= 1\right\}.\] 
Now, say that we observe $\{\widehat{E}=E\}$, where $E$ represents the observed value of the selected set $\widehat{E}$ and $E'$ represents the complement of $E$. Using the selected set $E$, we define our master statistics and the covariance matrices
$H$ and $K$, exactly as in the case where $E$ was selected using the lasso penalty. 
Since $\widehat{E}$ is selected using $\lP$-penalty, the penalized estimators based on $E$ implicitly depend on the choice of the penalty function $\lP$. 
However, with slight notational abuse, we continue to use the same symbols as before.

Following the approach used for the randomized lasso, we begin with the K.K.T. conditions of stationarity and reformulate them, analogous to Proposition \ref{prop:KKTmaster}, in the form:
\begin{align}
\label{genKKT} 
P_{1}^{E\cdot j} \sqrt{n}\estj  + P_{2}^{E\cdot j}\sqrt{n}\gammaj + H_{E}\sqrt{n}\elassoE + 
  \widehat{D}^{\lambda}     = \sqrt{n}\widetilde{\omega}_{n}.
\end{align} 
Here $\sqrt{n}\widetilde{\omega}_{n}=\sqrt{n}\omega_{n}+R_{n,2}$, where $R_{n,2}=o_p(1)$ denotes a remainder term that converges in probability to zero, and both $\elassoE$ and $\widehat{D}^{\lambda}$ are derived from the penalized estimator after solving \eqref{rand:lreg}.
This formulation allows us to identify the conditioning event, which corresponds to a proper subset of $\{\widehat{E}=E\}$, as sign constraints on the penalized estimator $\elassoE$. In what follows, we provide examples of penalized regression using the smoothly clipped absolute deviations penalty (SCAD) and minimax concave penalty (MCP), identifying the representation in \eqref{genKKT} for both cases.

\subsubsection{SCAD}

The SCAD penalty function, proposed by \cite{SCAD}, is defined as
\[
\lP(|x|)= \begin{cases}\lambda_{1} |x| & \text { if } |x| \leq \lambda_{1}, \\ \dfrac{2 \lambda_{2} \lambda_{1} |x|-x^2-\lambda_{1}^2}{2(\lambda_{2}-1)} & \text { if } \lambda_{1}<|x| \leq \lambda_{2} \lambda_{1}, \\ \dfrac{\lambda_{1}^2(\lambda_{2}+1)}{2} & \text { if } |x|>\lambda_{2} \lambda_{1},\end{cases}
\] 
where $\lambda = \nbracket{\lambda_{1}, \lambda_{2}}$ are the tuning parameters such that $\lambda_{1}>0, \lambda_{2}>2$.  

Consider solving \eqref{rand:lreg} with the SCAD penalty function.
Define the sets $E_0, E_1, E_2$ as the observed values of :
\begin{align*}
\widehat{E}_{0} = \cbracket{j: |\elassoj|>\lambda_{2} \lambda_{1}}, \widehat{E}_{1} = \cbracket{j: \lambda_{1} < |\elassoj|\leq \lambda_{2} \lambda_{1}}, \widehat{E}_{2} = \cbracket{j: 0<|\elassoj|\leq \lambda_{1}},
\end{align*}
which form a partition of the selected set $E$.

\begin{proposition}
\label{ext:SCAD}
The K.K.T stationarity conditions at the randomized SCAD equal
\[P_{1}^{E\cdot j} \sqrt{n}\estj  + P_{2}^{E\cdot j}\sqrt{n}\gammaj +  \begin{bmatrix}
    H_{E_{0},E} \\
    H_{E_{1},E} - \dfrac{1}{\lambda_{2}-1}{R_{E_{1}}}\\
    H_{E_{2},E} \\
    H_{E',E}
\end{bmatrix}\sqrt{n}\elassoE + \widehat{D}^{\lambda} = \sqrt{n}\widetilde{\omega}_{n}\] where $P_{1}^{E\cdot j}, P_{2}^{E\cdot j}$ are as defined in Proposition~\ref{prop:KKTmaster}, 
\[
\widehat{D}^{\lambda} =\begin{bmatrix}
   0_{|E_0|}\\
   \dfrac{\lambda_{2}\lambda_{1}}{\lambda_{2}-1}S_{E_1}\\
    \lambda_{1}S_{E_{2}}\\
 \lambda_{1}S_{E'}
    \end{bmatrix},
\] 
and $\sqrt{n}\widetilde{\omega}_{n}=\sqrt{n}\omega_{n}+R_{n,2}$ where $R_{n,2} =o_p(1)$.
\end{proposition}

\begin{proof}[Proof of Proposition~\ref{ext:SCAD}]
For $I \subseteq [p]$, define \[\nabla\lP\nbracket{\lvert\widehat{\beta}^{(\mathcal{P}, \lambda)}_{n,I}\rvert} = \rbracket{\nabla\lP\nbracket{\lvert\widehat{\beta}^{(\mathcal{P}, \lambda)}_{n,j}\rvert}: j \in I}^{\top}\in \mathbb{R}^{|I|}.\]
Since the gradient of the SCAD penalty w.r.t $|x|$ is \[\nabla \lP(|x|)= \begin{cases}\lambda_{1} & \text { if }|x| \leq \lambda_{1} \\ \frac{\lambda_{2} \lambda_{1}-|x|}{\lambda_{2}-1} & \text { if } \lambda_{1}<|x|\leq\lambda_{2} \lambda_{1} \\ 0 & \text { if }|x| > \lambda_{2} \lambda_{1}\end{cases},\]
we note that the sub-gradient vector at the randomized solution equals 
\begin{align*}
    \begin{bmatrix}
\Dg{\nbracket{S_{E_0}}}\nabla\lP\nbracket{\lvert\widehat{\beta}^{(\mathcal{P}, \lambda)}_{n,E_{0}}\rvert}\\
\Dg{\nbracket{S_{E_1}}}\nabla\lP\nbracket{\lvert\widehat{\beta}^{(\mathcal{P}, \lambda)}_{n,E_{1}}\rvert}\\
\Dg{\nbracket{S_{E_2}}}\nabla\lP\nbracket{\lvert\widehat{\beta}^{(\mathcal{P}, \lambda)}_{n,E_{2}}\rvert}\\
\Dg{\nbracket{S_{E'}}}\nabla\lP\nbracket{\lvert\widehat{\beta}^{(\mathcal{P}, \lambda)}_{n,E'}\rvert}
    \end{bmatrix} =     \begin{bmatrix}
   0_{|E_0|}\\
   \dfrac{\lambda_{2}\lambda_{1}S_{E_1}-\Dg{\nbracket{S_{E_1}}}\lvert\widehat{\beta}^{(\mathcal{P}, \lambda)}_{n,E_{1}}\rvert}{\lambda_{2}-1}\\
    \lambda_{1}S_{E_{2}}\\
 \lambda_{1}S_{E'}
    \end{bmatrix} = \begin{bmatrix}
   0_{|E_0|}\\
   \dfrac{\lambda_{2}\lambda_{1}S_{E_1}-\widehat{\beta}^{(\mathcal{P}, \lambda)}_{n,E_{1}}}{\lambda_{2}-1}\\
    \lambda_{1}S_{E_{2}}\\
 \lambda_{1}S_{E'}
    \end{bmatrix}.
\end{align*}  
By grouping the $\widehat{\beta}^{(\mathcal{P}, \lambda)}_{n,E_{1}}$ terms from the subgradient vector with those derived from the Taylor expansion of the loss function, we obtain the desired representation.
\end{proof}
The construct of our pivot is then straightforward by conditioning on the event \[\cbracket{\widehat{S}_{n,E} =S_{E},\widehat{S}_{n,E'} =S_{E'}, \widehat{E_{1}}=E_{1}, \widehat{E_{2}}=E_{2}} \subset \cbracket{\widehat{S}_{n} =S}\] and it takes the same form as seen for the randomized lasso in Section \ref{sec:SI}. 

\subsubsection{MCP}
The MCP penalty function, introduced in \citep{MCP} is defined as
\[
\lP(|x|;\lambda_{1}, \lambda_{2})= \begin{cases}\lambda_{1}|x|-\frac{x^2}{2 \lambda_{2}}, & \text { if }|x| \leq \lambda_{2} \lambda_{1} \\ \frac{1}{2} \lambda_{2} \lambda_{1}^2, & \text { if }|x|>\lambda_{2} \lambda_{1}\end{cases},
\]
where $\lambda = \nbracket{\lambda_{1},\lambda_{2}}$ are tuning parameters satisfying $\lambda_{1}>0,\lambda_{2}>1$. 

Consider solving \eqref{rand:lreg} with the MCP penalty function.
Define the sets $E_0, E_1$ as the observed values of 
\begin{align*}
\widehat{E}_{0} = \cbracket{j: |\elassoj|>\lambda_{2} \lambda_{1}},
\widehat{E}_{1} = \cbracket{j: 0<|\elassoj|\leq \lambda_{2} \lambda_{1}},\end{align*}
which form a partition of the selection set.

\begin{proposition}
\label{ext:MCP}
The K.K.T stationarity conditions for the randomized MCP equal 
\[P_{1}^{E\cdot j} \sqrt{n}\estj  + P_{2}^{E\cdot j}\sqrt{n}\gammaj + \begin{bmatrix}
    H_{E_{0},E} \\
    H_{E_{1},E} - \dfrac{1}{\lambda_{2}}{R_{E_{1}}}\\
    H_{E',E}
\end{bmatrix}\sqrt{n}\elassoE + \widehat{D}^\lambda = \sqrt{n}\widetilde{\omega}_{n}\] 
    where $P_{1}^{E\cdot j}, P_{2}^{E\cdot j}$ are as defined in Proposition~\ref{prop:KKTmaster},
    \[
    \widehat{D}^\lambda = \begin{bmatrix}
   0_{|E_0|} \\
   \lambda_{1}S_{E_1}\\
 \lambda_{1}S_{E'}
    \end{bmatrix},
    \]
    and $\sqrt{n}\widetilde{\omega}_{n}=\sqrt{n}\omega_{n}+R_{n,2}$. 
\end{proposition}

\begin{proof}[Proof of Proposition~\ref{ext:MCP}]
Since the gradient of the MCP penalty with respect to $|x|$ is equal to
\[
\nabla \lP(|x|;\lambda_{1}, \lambda_{2})= \begin{cases}\lambda_{1}-\frac{|x|}{\lambda_{2}}, & \text { if }|x| \leq \lambda_{2} \lambda_{1} \\ 0, & \text { if }|x|>\lambda_{2} \lambda_{1}\end{cases},
\]
the subgradient vector at the randomized solution is given by
\begin{align*}
    \begin{bmatrix}
\Dg{\nbracket{S_{E_0}}}\nabla\lP\nbracket{\lvert\widehat{\beta}^{(\mathcal{P}, \lambda)}_{n,E_{0}}\rvert}\\
\Dg{\nbracket{S_{E_1}}}\nabla\lP\nbracket{\lvert\widehat{\beta}^{(\mathcal{P}, \lambda)}_{n,E_{1}}\rvert}\\
\Dg{\nbracket{S_{E'}}}\nabla\lP\nbracket{\lvert\widehat{\beta}^{(\mathcal{P}, \lambda)}_{n,E'}\rvert}
    \end{bmatrix} =     \begin{bmatrix}
   0_{|E_0|}\\
   \dfrac{\lambda_{2}\lambda_{1}S_{E_1}-\Dg{\nbracket{S_{E_1}}}\lvert\widehat{\beta}^{(\mathcal{P}, \lambda)}_{n,E_{1}}\rvert}{\lambda_{2}}\\
 \lambda_{1}S_{E'}
    \end{bmatrix} = \begin{bmatrix}
   0_{|E_0|}\\
   \dfrac{\lambda_{2}\lambda_{1}S_{E_1}-\widehat{\beta}^{(\mathcal{P}, \lambda)}_{n,E_{1}}}{\lambda_{2}}\\
 \lambda_{1}S_{E'}
    \end{bmatrix}.
\end{align*} 
By grouping the $\widehat{\beta}^{(\mathcal{P}, \lambda)}_{n,E_{1}}$ terms from the subgradient vector with those derived from the Taylor expansion of the loss function, we obtain the desired representation.
\end{proof}

Similar to SCAD, conditioning on the event $\{\widehat{S}_{n,E} =S_{E},\widehat{S}_{n,E'} =S_{E'}, \widehat{E_{1}}=E_{1}\}$ we can obtain the pivot in the same form. Using this further conditioning, in the context of graphical models \cite{asymptoticinference2025graphicalmodels} derives the polyhedral representations of non-randomized SCAD and MCP conditioning events.

\subsection{Linear Risk}
For the linear contrast, a population risk function can be defined as
\[
R^{lin}(\beta) := \mathbb{E} \left[ \sum_{t=1}^T \widetilde{\sigma}^2_t (M_{i,t}) \left \{ (f_t(M_{i,t})^\top \beta)^2 - 2 f_t(M_{i,t})^\top \beta \left( \mu_t(H_{i,t}, 1) - \mu_t (H_{i,t},0) \right) \right \}  \right],
\]
where $\mu_t(h,a) := \mathbb{E}_{\mathbf{p}}[ Y_{i,t+1} | H_{i,t} = h, A_{i,t} = a]$ denotes conditional mean proximal outcome given history~$h$ and action~$a$, and $\widetilde{\sigma}^2_t (M_{i,t}) = \widetilde{p}_{t}(1|M_{i,t}) \nbracket{1 - \widetilde{p}_{t}(1|M_{i,t})}$.
Recalling that $\beta_{\mathbf{p}}(t ; s) := \mathbb{E}[ \mu_t(H_{i,t}, 1) - \mu_t(H_{i,t}, 0) | M_{i,t} =s]$, minimizing the risk  $R^{lin}(\beta)$ is equivalent to minimizing
\begin{align*}
&\sum_{t=1}^T \mathbb{E} \left[ \widetilde{\sigma}^2_t (M_{i,t}) \left \{ (f_t(M_{i,t})^\top \beta)^2 - 2 f_t(M_{i,t})^\top \beta \beta(t;M_{i,t}) + \beta (t;M_{i,t})^2  \right \} \right] \\
=&\sum_{t=1}^T \mathbb{E} \left[  \widetilde{\sigma}^2_t (M_{i,t}) \left \{f_t(M_{i,t})^\top \beta - \beta (t; M_{i,t}) \right \}^2  \right] \\
=&\sum_{t=1}^T \mathbb{E} \left[  \widetilde{\sigma}^2_t (M_{i,t}) \left \{f_t(M_{i,t})^\top \beta - f_t(M_{i,t})^\top \beta^\star \right \}^2  \right].
\end{align*}
implying that under the linear causal model being correctly specified, i.e., $\beta_{\mathbf{p}}(t ; s) = f_t(M_{i,t})^\top \beta^\star$,  
the risk function is minimized by $\beta = \beta^\star$. Under model misspecification, the risk minimizer can be thought of as a weighted $L_2$-projection of the true causal model as shown in previous work~\cite{dempsey2020}.

Computing the population risk requires taking an expectation over the unknown distribution~$\mathcal{P}$ and knowing the conditional mean $\mu_t (h,a)$. 
First, we replace the expectation by an empirical-version and construct an initial estimate $\widehat{\mu}_t(h,a)$ of the conditional mean, $\mu_t(h,a)$. In particular, we consider a doubly-robust initial estimator
\[
\widehat{\mu}_t^{(DR)} (H_{i,t}, a) = \widehat{\mu}_t (H_{i,t}, a) + \frac{1[A_{i,t} = a]}{p(A_{i,t}|H_{i,t})} \left( Y_{i, t+1} - \widehat{\mu}_{t} (H_{i,t}, a) \right),
\]
Motivated by recent work on  empirical risk 
minimization~\cite{vanderlaan2024combining}, we consider a one-step debiased estimation procedure to ensure minimizing the empirical risk yield oracle efficiency similar to orthogonal learning strategies~\cite{shi2023metalearning}.  Given the initial estimate, we propose to construct a refined estimator~$ \mu_t^\star(h,a) = \widehat{\mu}_t(h,a) + (a - \widehat{p}_t (1 | M_{i,t}) ) f_t(M_{i,t})^\top \widehat{\theta}$ where $\widehat{\theta}$ is the minimizer of 
\[
\sum_{i=1}^{N}\left [ \sum_{t=1}^T W_{i,t} (Y_{i,t+1} - \widehat{\mu}_t^{(DR)}(H_{i,t}, A_{i,t}) - (A_{i,t} - \widetilde{p}_t(1| M_{i,t})) f_t(M_{i,t})^\top \theta )^2 \right]
\]
where~$W_{i,t} = \widetilde{p}_t(A_{i,t} \mid M_{i,t}) / \widehat{p}_t (A_{i,t} \mid H_{i,t})$.  
This guarantees the refined estimator satisfies similar Neyman-orthogonality constraints for consistent causal estimation. This is desirable because we want $\widehat{\mu}^\star$ to satisfy 
\begin{align*}
\sum_{i=1}^{N}\left [ \sum_{t=1}^T W_{i,t} (A_{i,t} - \widetilde{p}_t(1|M_{i,t})) (Y_{i,t+1}- \mu_t^\star(H_{i,t},A_{i,t})) f_t(M_{i,t}) \right] = 0.
\end{align*}
The final estimate~$\widehat{\beta}$ minimizes the empirical version of the population risk with the refined estimator serving as a plug-in estimator for $\mu_t(h,a)$, i.e.,
\[
\sum_{i=1}^{N}\left[ \sum_{t=1}^T \widetilde{\sigma}^2_t (M_{i,t}) \left \{ (f_t(M_{i,t})^\top \beta)^2 - 2 f_t(M_{i,t})^\top \beta \left( \mu^\star_t(H_{i,t}, 1) - \mu^\star_t (H_{i,t},0) \right) \right \} \right] \]
which is equivalent to minimizing the loss
\begin{equation} \label{LRest}
\sum_{i=1}^{N}\rbracket{ \sum_{t=1}^T  \widetilde{\sigma}^2_t (M_{i,t}) \left \{f_t(M_{i,t})^\top \beta - \beta^\star (t; M_{i,t}) \right \}^2},
\end{equation}
where $\widetilde{\mu}^\star(t; M_{i,t}) := \mu^\star_t(H_{i,t}, 1) - \mu^\star_t (H_{i,t},0)$.

\subsection{Relative Risk}

For the relative risk contrast, a population-level risk function can be defined as
\[
\mathbb{E} \left[ \sum_{t=1}^T \widetilde{\sigma}_{i,t}^2 (M_{i,t}) \left \{ \left( \mu_t(H_{i,t}, 1) + \mu_t(H_{i,t},0) \right) \log ( 1 + e^{f_t (M_{i,t})^\top \beta}) - \mu_t (H_{i,t}, 1) f_t(M_{i,t})^\top \beta \right \} \right],
\]where $\widetilde{\sigma}^2_t (M_{i,t}) = \widetilde{p}_{t}(1|M_{i,t}) \nbracket{1 - \widetilde{p}_{t}(1|M_{i,t})}$.
Under the linear relative risk causal excursion model, there exists~$\beta \in \mathbb{R}^d$ such that $\mathbb{E}[ \mu_t (H_{i,t},1) | M_{i,t}] = e^{f_t(M_{i,t})^\top \beta}\mathbb{E}[ \mu_t(H_{i,t}, 0) | M_{i,t}]$.  Thus the population risk can be re-written as
\[
\sum_{t=1}^T \mathbb{E} \left[ \widetilde{\sigma}_{i,t}^2 (M_{i,t}) \times \mathbb{E} [ \mu_t (H_{i,t}, 0) | M_{i,t} ] \cbracket{ \left( 1 + e^{f_t(M_{i,t})^\top \beta} \right) \log \nbracket{ 1 + e^{f_t (M_{i,t})^\top \beta)}} - e^{f_t (M_{i,t})^\top \beta} f_t(M_{i,t})^\top \beta }\right].
\]
Differentiating with respect to $\beta$, it can be shown that the population risk is minimized at the true parameter value of $\beta$, as desired.
For the relative risk, the debiasing term is given by
\begin{align*}
&\sum_{i=1}^{n}\left[ \sum_{t=1}^T \frac{1}{p(A_{i,t} | H_{i,t})} \widetilde{\sigma}^2_t (M_{i,t}) 
\left \{ 
\log \left( 1 + e^{f_t (M_{i,t})^\top \beta} \right) - A_{i,t} f_t (M_{i,t})^\top \beta 
\right \}  (Y_{i,t+1} - \mu^\star (H_{i,t+1}, A_{i,t}))  \right] 
\end{align*}
Unlike the linear risk setting, an exact solution cannot be derived.  Instead we construct a basis~$g(M_{i,t}) \in \mathbb{R}^d \supset f_t(M_{i,t})$ such that $\log \left( 1 + e^{f_t (M_{i,t})^\top \beta} \right)$ can be well approximated by $g(M_{i,t})^\top \alpha$ for some $\alpha \in \mathbb{R}^d$.  Then the debiasing term yields the following constraints:
\begin{align*}
&\sum_{i=1}^{n}\left[ \sum_{t=1}^T \frac{1}{p(A_{i,t} | H_{i,t})} \widetilde{\sigma}^2_t (M_{i,t}) 
(Y_{i,t+1} - \mu^\star(H_{i,t+1}, A_{i,t}))  
\left \{ 
\begin{array}{c}
g(M_{i,t}) \\
- A_{i,t} g(M_{i,t}) 
\end{array} 
\right \}
\right] 
= 0.
\end{align*}
The first set of constraints imply the action-centering adjustments from the linear setting continue to hold.
We then again apply the additive adjustment but now using $\mu_t^\star (H_{i,t}, A_{i,t}) = \widehat{\mu}_t (H_{i,t}, A_{i,t}) + (g(M_{i,t}), A_{i,t} g(M_{i,t}))^\top \widehat{\theta}$ where $\widehat{\theta}$ solves the above estimating equation.  This does not reflect constraints on the outcome.


\section{Simulation details and additional results}\label{app:addsim}

\paragraph{Data-generating mechanism \& parameter values.}
We start by describing the data generating mechanism and parameter settings used in our empirical analysis. The moderator process for individual $i$ follows
\[
M_{i,t} = \rho_{\mathrm{state}}\, M_{i,t-1} + e_{i,t}, \qquad e_{i,t} \sim \mathcal{N}(0,\,\sigma_{\mathrm{state}}^2\,\Sigma_{\mathrm{corr}}),
\]
where $(\Sigma_{\mathrm{corr}})_{jk} = r^{|j-k|}$ is a Toeplitz cross-moderator correlation matrix. Each coordinate evolves as an independent AR(1) process when $r=0$. The centered moderators used for WCLS estimation are $\widetilde{M}_{i,t} = M_{i,t} - \mathbb{E}[M_{i,t}\mid H_{i,t-1}, A_{i,t-1}] = e_{i,t}$, i.e., the innovations of the AR process. Treatment is assigned via a logistic propensity model,
\[
\mathrm{logit}\;p_t(1\mid H_{i,t}) = \eta_1\, A_{i,t-1} + \eta_2\,\bar{M}_{i,t},
\]
where $\bar{M}_{i,t} = p^{-1}\mathbf{1}_p^\top M_{i,t}$ is the cross-sectional average of the raw moderators. Outcomes follow
\[
Y_{i,t} = \theta_1^\top \widetilde{M}_{i,t} + \bigl(A_{i,t} - p_t(1\mid H_{i,t})\bigr)\,\beta^{*\top}\widetilde{M}_{i,t} + \epsilon_{i,t},
\]
where $\epsilon_{i,t}$ is an AR(1) process with autocorrelation $\rho_{\mathrm{error}}$ and marginal standard deviation $\sigma_{\mathrm{error}}$. To study robustness to nuisance misspecification, we also consider an augmented outcome model that adds the non-linear term $\gamma\bigl(p^{-1}\sum_{j=1}^p e_{i,t,j}^2 - \sigma_{\mathrm{state}}^2\bigr)$ to the outcome, where $\gamma \geq 0$ controls the severity; the linear nuisance estimator cannot capture this centered quadratic component.

We set $p=50$, $|E^*|=5$ (the first five moderators), $\theta_1 = 0.8\cdot\mathbf{1}_p$, $\sigma_{\mathrm{state}}=1.2$, $\sigma_{\mathrm{error}}=1.2$, $(\eta_1,\eta_2)=(-0.8,\,0.8)$, and $\rho_{\mathrm{state}}=\rho_{\mathrm{error}}=0.5$ unless otherwise varied. Moderators are independent across coordinates ($r=0$) unless otherwise stated. The nuisance coefficient $\theta_1$ is estimated by OLS of $Y$ on $\widetilde{M}$ using a held-out third of the individuals; the residualized outcome is then used for lasso selection and inference. The nuisance estimates are treated as fixed, plug-in inputs into a single lasso selection and selective-inference step on the analysis fold. For the randomization covariance, we set $\Omega = \frac{1}{2}\widehat{\sigma}^2 I_p$, i.e., the low randomization level ($\tau=0.5$) adopted throughout the paper (Figure~1 in the main text).

\paragraph{Evaluation metrics.}
For each Monte Carlo replication with selected set $\widehat{E}$, we report:
\begin{itemize}
    \item \emph{Recall:} $|\widehat{E}\cap E^*|/|E^*|$, measuring selection quality.
    \item \emph{Coverage:} the empirical coverage rate, $|\{j \in \widehat{E}: \beta^{\widehat{E}\cdot j} \in \widehat{C}^{\widehat{E}\cdot j}\}| / \max(|\widehat{E}|,1)$.
    \item \emph{CI length:} the average length of the confidence intervals $\{\widehat{C}^{\widehat{E}\cdot j}: j \in \widehat{E}\}$.
\end{itemize}
The selected set $\widehat{E}$ varies across replications; all metrics are evaluated relative to $\widehat{E}$ in each iteration. For polyhedral, we only consider the replications in which all produced intervals are finite and additionally report the proportion of replications in which all intervals are finite.
 
\begin{figure}[t]
    \centering
    \includegraphics[width=\textwidth]{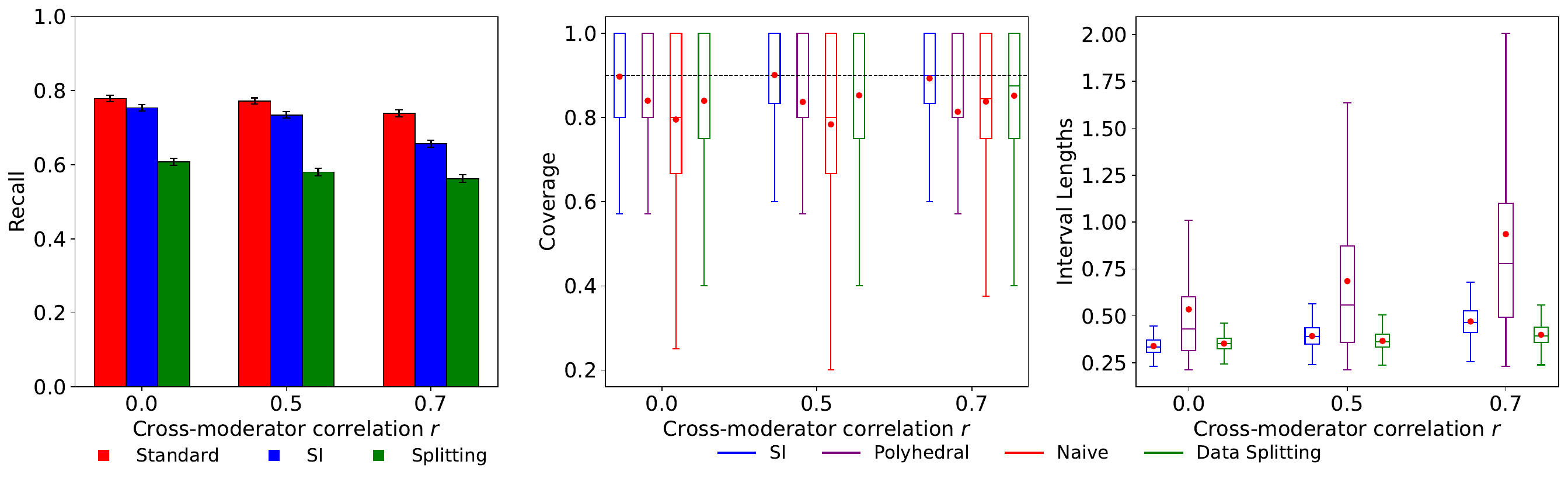}
    \caption{Average Recall (left), Average Coverage (center), Average Length of bounded CIs (right) for cross-moderator correlation with $r \in \{0, 0.5, 0.7\}$. \textbf{\textcolor{red}{Polyhedral}} produces infinitely long intervals in \textbf{\textcolor{red}{14.2\%}}, \textbf{\textcolor{red}{13.6\%}}, \textbf{\textcolor{red}{23\%}} of replications for $r=0,0.5,0.7$ respectively}
    \label{corr_plot}
\end{figure}
 
\noindent\textbf{Cross-moderator correlation.}
In Figure~\ref{corr_plot}, we vary the Toeplitz cross-moderator correlation $r \in \{0, 0.5, 0.7\}$ at $n=60$ with $\beta_{11}=0.2$. SI maintains valid coverage across all correlation levels. The polyhedral method exhibits slight undercoverage as $r$ increases, likely because the effective degrees of freedom in the polyhedral constraint decrease with stronger dependence. Interestingly, the naive method's coverage improves at high correlation---when moderators are highly correlated, many models are nearly interchangeable, so the selection event carries less information and the post-selection bias diminishes---though it still undercovers relative to the nominal level.

\begin{figure}[t]
    \centering
    \includegraphics[width=\textwidth]{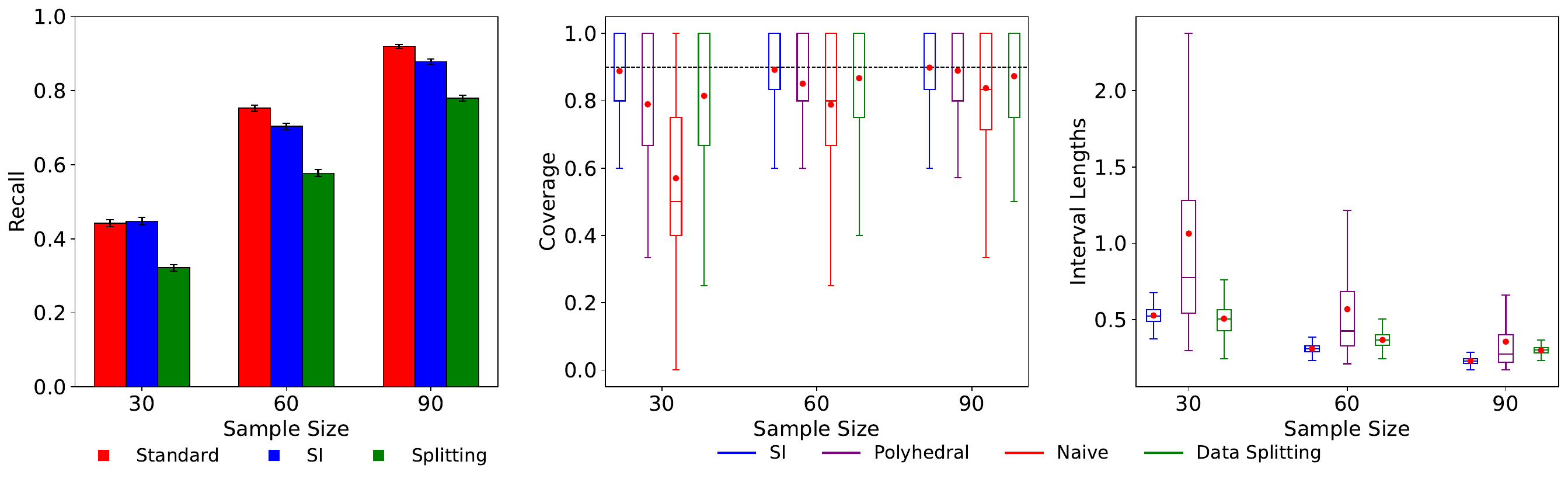}
    \caption{Average Recall (left), Average Coverage (center), Average Length of bounded CIs (right) under nuisance misspecification ($\gamma=1.0$) for varying sample size with $\beta_{11}=0.2$ fixed. \textbf{\textcolor{red}{Polyhedral}} produces infinitely long intervals in \textbf{\textcolor{red}{16\%}}, \textbf{\textcolor{red}{11.8\%}}, \textbf{\textcolor{red}{7.2\%}} of replications for $n=30,60,90$ respectively}
    \label{misspec_plot}
\end{figure}

\noindent\textbf{Nuisance misspecification.}
In Figure~\ref{misspec_plot}, we vary $n \in \{30,60,90\}$ under nuisance misspecification ($\gamma = 1.0$, $\beta_{11}=0.20$). The unmodeled quadratic component inflates residual variance, and the naive method undercovers more severely than in the well-specified case. Data splitting also falls below the nominal level at low-to-moderate sample sizes: the 30\% test split leaves too few observations for reliable inference when residual variance is inflated by misspecification, so that the sandwich variance estimator underestimates uncertainty. SI maintains valid coverage throughout.

\begin{figure}[ht] 
\centering
\includegraphics[width=1\textwidth]{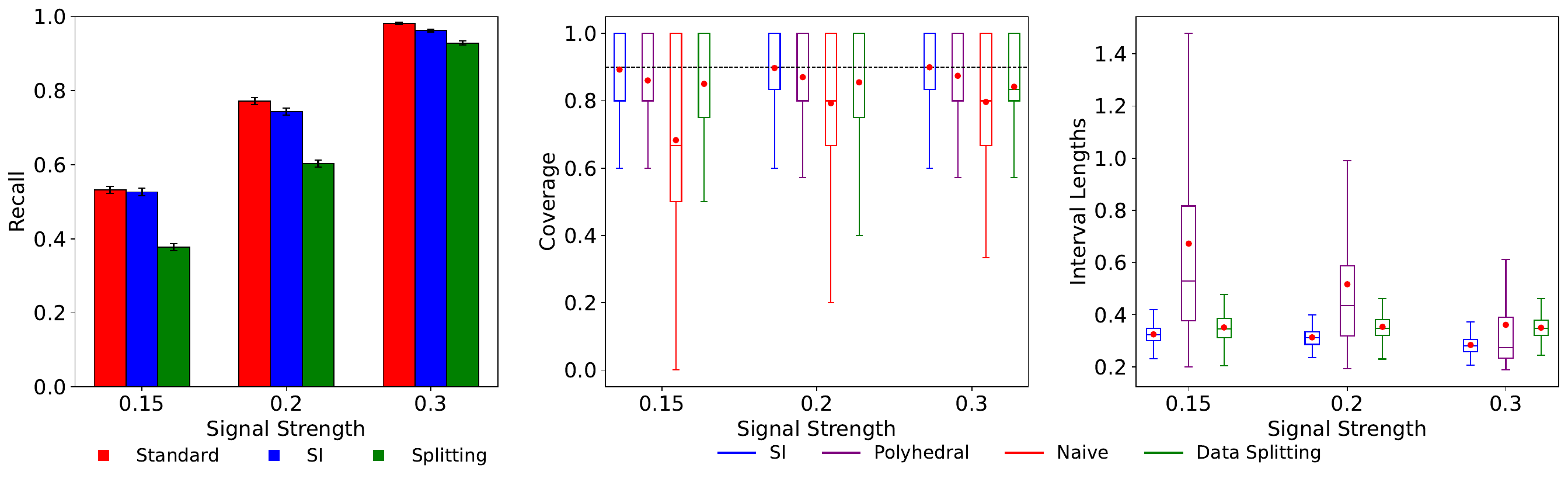}
\caption{Average Recall (left), Average Coverage (center), Average Length of bounded CIs (right) for varying signal strength $\beta_{11} \in \{0.15,0.20,0.30\}$ under Laplace errors. \textbf{\textcolor{red}{Polyhedral}} produces infinitely long intervals in \textbf{\textcolor{red}{13.6\%}}, \textbf{\textcolor{red}{11\%}}, \textbf{\textcolor{red}{7.8\%}} of replications for $\beta_{11}=0.15,0.20,0.30$ respectively
}
\label{laplace_plot}
\end{figure}

\begin{figure}[ht] 
\centering
\includegraphics[width=1\textwidth]{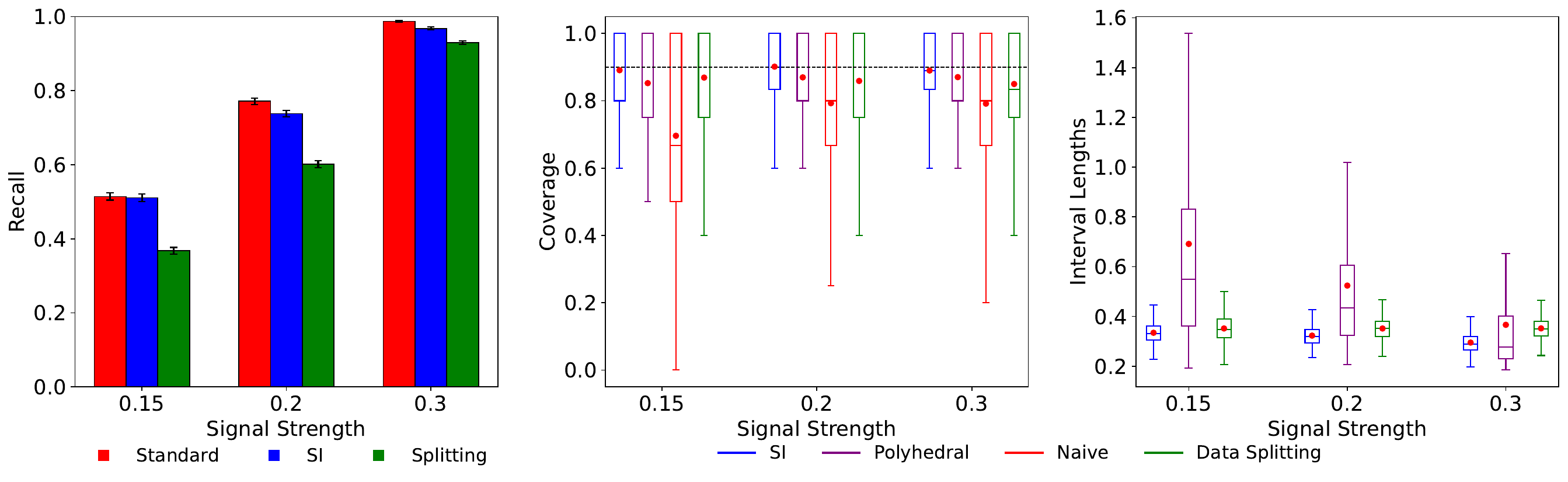}
\caption{
Average Recall (left), Average Coverage (center), Average Length of bounded CIs (right) for varying signal strength $\beta_{11} \in \{0.15,0.20,0.30\}$ under exponential errors. \textbf{\textcolor{red}{Polyhedral}} produces infinitely long intervals in \textbf{\textcolor{red}{14.2\%}}, \textbf{\textcolor{red}{13\%}}, \textbf{\textcolor{red}{8.2\%}} of replications for $\beta_{11}=0.15,0.20,0.30$ respectively}
\label{exp_plot}
\end{figure}

\noindent\textbf{Non-Gaussian errors.}
 In Figures~\ref{laplace_plot}--\ref{exp_plot}, we replace the Gaussian error distribution with Laplace and centered exponential distributions, respectively. The qualitative findings are unchanged, confirming robustness to non-Gaussian errors.

\begin{figure}[t]
    \centering
    \includegraphics[width=0.45\textwidth]{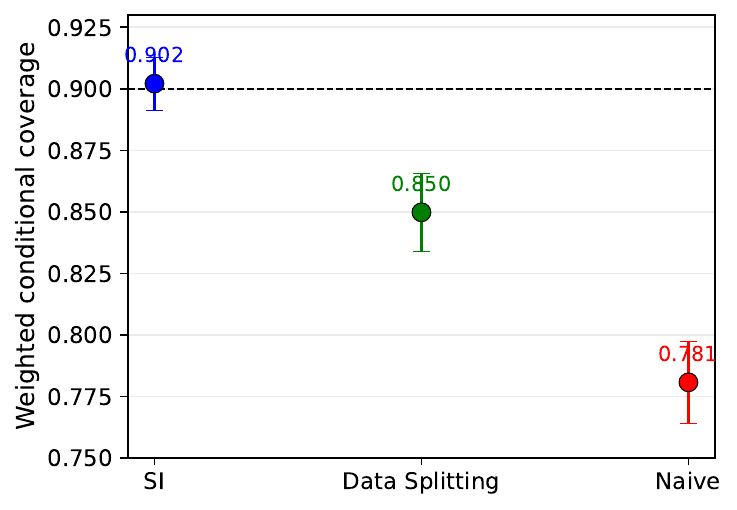}
    \par\medskip
    \includegraphics[width=\textwidth]{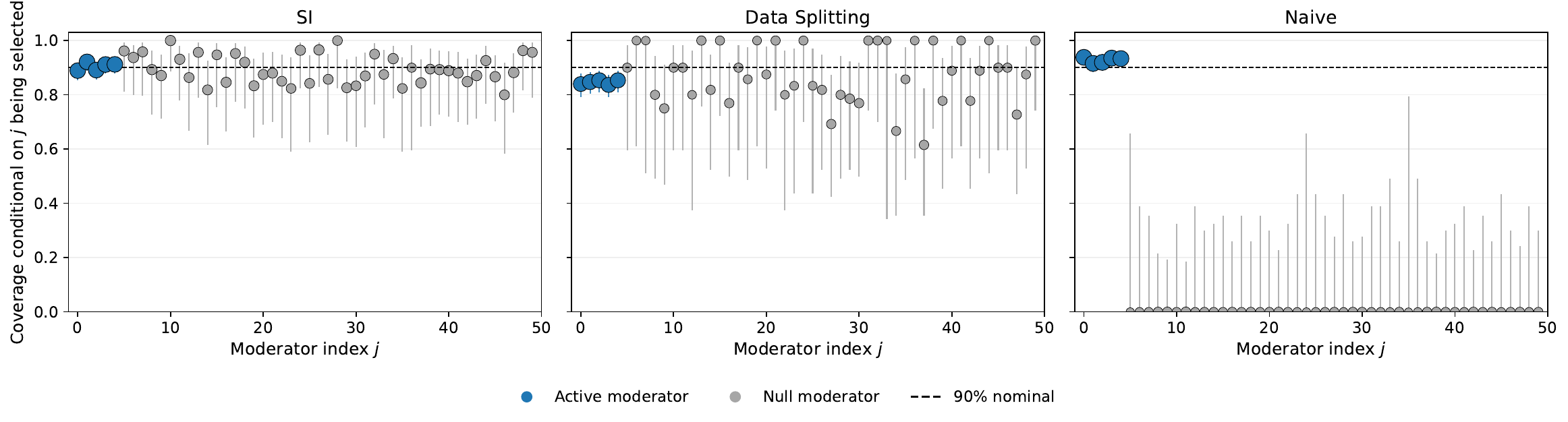}
    \caption{Coverage conditional on selection at $n=60$ and $\beta_{11}=0.2$. Upper panel: selection-frequency weighted conditional coverage, with 95\% Monte Carlo error bars and a truncated vertical axis. Lower panels: moderator-specific conditional coverage, with Wilson 95\% intervals. Blue and gray points denote active and null moderators; marker size is proportional to selection frequency. Dashed lines mark the nominal 90\% level.}
    \label{fig:cond_coverage}
\end{figure}

\noindent\textbf{Conditional coverage analysis.} \ The coverage metric reported above is $1-$empirical FCR, a marginal summary that averages the within-replication coverage proportion over selected sets $\widehat E$ that vary across replications. To assess whether this marginal performance masks poor coverage for particular moderators when selected, Figure~\ref{fig:cond_coverage} reports only conditional-coverage summaries at $n=60$ and $\beta_{11}=0.2$.

For each moderator $j$, we estimate
\begin{equation}
\widehat{\operatorname{cov}}(j)
=
\frac{\sum_{b=1}^{B}\mathbbm{1}\{j\in\widehat E_b\}\mathbbm{1}\{\beta^{\widehat E_b\cdot j}\in\widehat C_b^{\widehat E_b\cdot j}\}}
{\sum_{b=1}^{B}\mathbbm{1}\{j\in\widehat E_b\}},
\end{equation}
the empirical coverage among replications in which $j$ is selected. The upper panel is the selection-frequency weighted mean of these estimates, so each selected moderator--replication pair receives equal weight. It is therefore a conditional-coverage summary, distinct from empirical $1-\mathrm{FCR}$, which weights replications equally. The lower panels display each $\widehat{\operatorname{cov}}(j)$ with a Wilson 95\% interval: with $x_j$ covered intervals in $N_j$ selections, this is the 95\% Wilson score interval for the binomial proportion $x_j/N_j$ ($z=1.96$); active and null moderators are distinguished by color, and marker size is proportional to selection frequency.

SI has weighted conditional coverage $0.902$; its five active-moderator estimates range from $0.889$ to $0.921$. The naive method has weighted conditional coverage $0.781$ but never covers a null moderator when selected. Data splitting has weighted conditional coverage $0.850$ and active-moderator coverage between $0.837$ and $0.854$. Because null moderators are selected in only about $1$--$7\%$ of replications, their moderator-specific intervals are wide.



\begin{figure}[t]
    \centering
    \includegraphics[width=\textwidth]{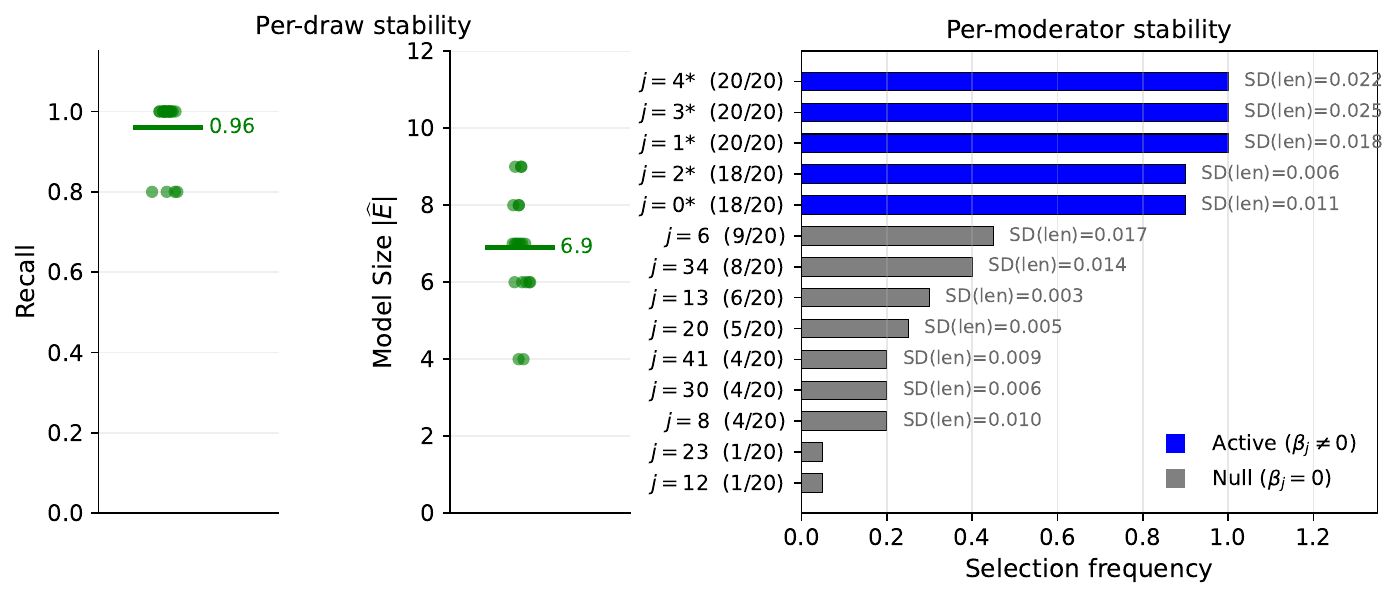}
    \caption{Randomization stability ($K=20$ draws, $n=60$, $\beta_{11}=0.2$). Left: recall (blue, left axis) and selected set size (green, right axis) per draw; dotted lines show perfect recall and $|E^*|=5$. Right: selection frequency per moderator; asterisks denote active moderators, annotations show SD of CI length.}
    \label{fig:seed_stability}
\end{figure}

 \noindent\textbf{Stability of our randomized selection.} \
Finally, we assess the stability of the randomized procedure by fixing a single dataset at $n=60$, $\beta_{11}=0.20$ and repeating the procedure $K=20$ times with independent draws of $\omega$.
Figure~\ref{fig:seed_stability} summarizes the results. The left panel shows the recall and selected model size for each draw: the active moderators are consistently recovered, and the total number of selected variables is stable across draws. The right panel displays the per-moderator selection frequency; active moderators appear in nearly all draws, while variation is confined to borderline null moderators whose lasso gradients are near the penalty threshold. CI lengths for stably selected moderators vary minimally across draws (SD of order $10^{-2}$), confirming that randomization with $\tau=0.5$, i.e., $\Omega=\frac{1}{2}\widehat{\sigma}^2 I_p$, produces practically stable inference.

\begin{figure}[t]
    \centering
    \includegraphics[width=\textwidth]{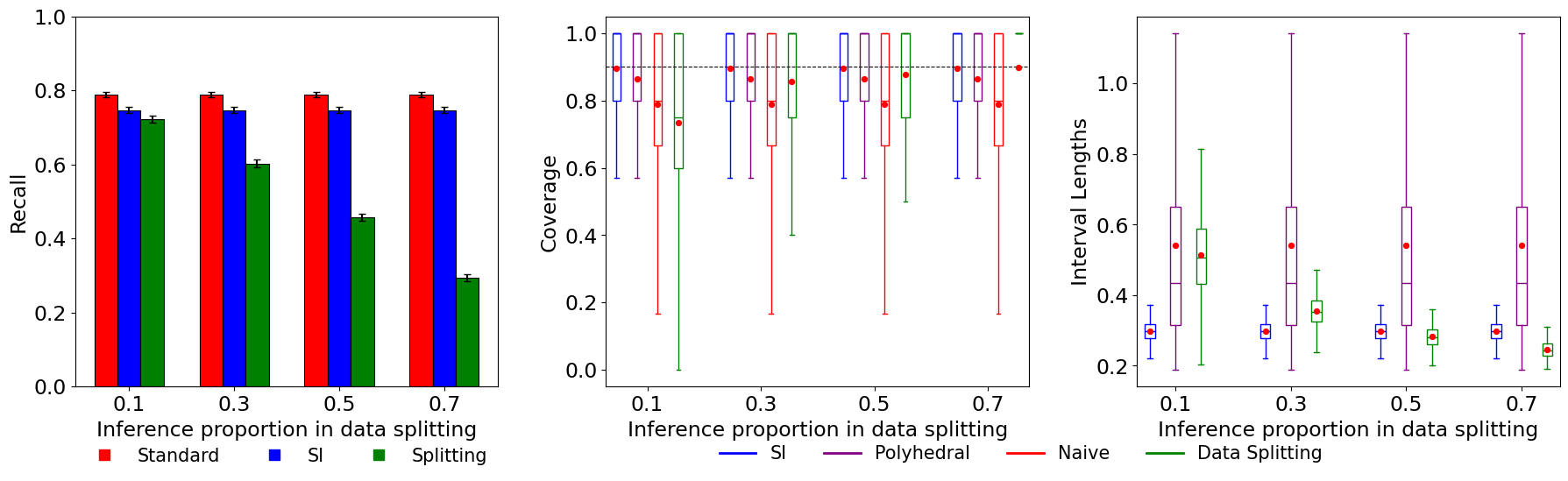}
    \caption{Recall (left), coverage (center), and bounded CI length (right) as the data-splitting inference proportion varies over $\{0.1,0.3,0.5,0.7\}$. The proposed SI procedure uses the full analysis sample for selection and inference and is shown as a reference at each split ratio.}
    \label{fig:split_ratio}
\end{figure}

\noindent\textbf{Data-splitting ratio.}
Figure~\ref{fig:split_ratio} varies the proportion of individuals reserved for the independent data-splitting inference sample, holding $n=60$, $\beta_{11}=0.2$, and the SI randomization level fixed. Reserving more observations for splitting inference can shorten its intervals, but reduces recall because fewer observations remain for selection. Conversely, allocating more observations to selection improves recall but yields longer splitting intervals. The SI intervals are shorter than, or comparable to, those from the default $70\%/30\%$ split while retaining higher recall.

\section{VALENTINE study: Data application details}
\label{app:DA}

VALENTINE is a remotely managed randomized trial assessing whether mHealth
interventions increase exercise adherence in low-to-medium-risk cardiac-rehab
patients. Participants receive a compatible smartwatch alongside standard care,
and Just-in-Time Adaptive Interventions micro-randomize contextually tailored
notifications to study their short-term effect on activity. Notifications are of
two functional content types: \emph{activity-suggestion} messages (e.g.\ prompting a walk), comprising about $78\%$ of delivered notifications, and \emph{anti-sedentary} messages (e.g.\ prompting a brief stretch or stand-up break), comprising the remaining $22\%$; notifications are delivered at one of four times of day, each with probability $0.25$, and are tailored based on weather, day of week, time of day, and rehabilitation phase. The scientific goal
is to identify the contexts under which notifications are most effective---effect
modification rather than marginal effect estimation.

\subsection{Data processing and moderator sets}
We analyze Apple Watch users. After excluding rows with missing values, the data
comprise $23{,}462$ observations from $62$ participants ($36$--$592$ each). From
$23$ history variables (Table~\ref{tab:datavariables}) we build three
progressively larger candidate moderator sets, splitting the enrollment-time
trend by phase: (i) \emph{baseline}---raw context variables, $25$ moderators,
no interactions; (ii) \emph{phase--time interactions}---adding
phase\,$\times$\,moderator terms, $58$ moderators; (iii) \emph{extended
interactions}---additionally adding demographic\,$\times$\,behavioral terms, $94$
moderators.

\paragraph{Nuisance model.}
On the $30\%$ nuisance fold we fit an OLS regression of the log step count on the
history features (baseline steps, log step count in the $30$ minutes before
notification, time/phase in study, notification type, weekend, and
indoor/weather indicators). Its residual standard deviation is $\widehat{\sigma}$,
which sets the Negahban penalty level and the randomization covariance 
$\Omega=\frac{1}{2}\widehat{\sigma}^2 I_p$; on the analysis fold the
treatment-centered design columns are standardized to unit scale so a single
$\lambda$ treats the heterogeneous-scale moderators comparably.

\subsection{Experiments: baseline and extended settings}
\begin{figure}[h]
    \centering
\includegraphics[width=0.7\textwidth]{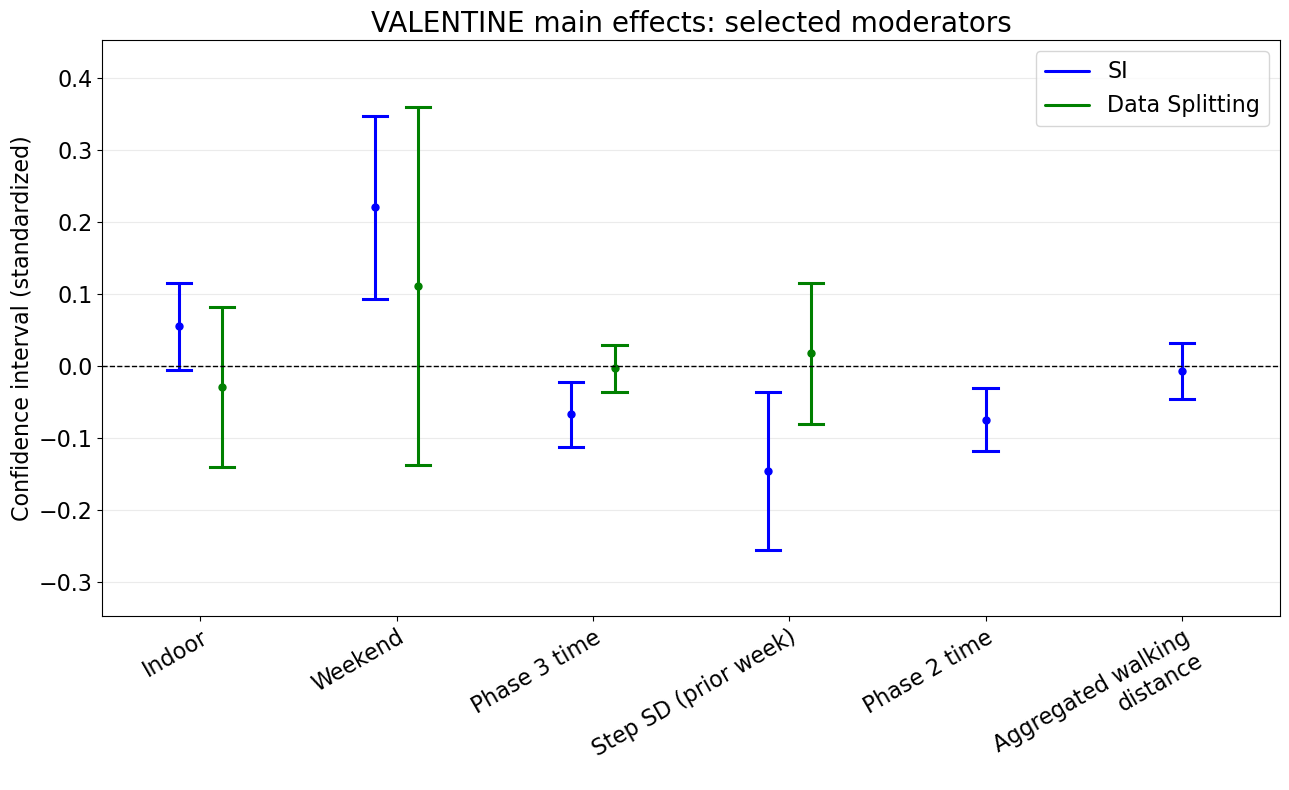}
    \caption{Baseline main-effects setting ($25$ candidate moderators): $90\%$
selective confidence intervals for the selected moderators. Our method selects
$6$ moderators with average length $0.142$; data splitting selects $9$ moderators
with average length $0.323$; polyhedral produces infinitely long intervals and is
hence omitted.}
    \label{fig:main25}
\end{figure}

\begin{figure}[h]
    \centering
    \includegraphics[width=0.7\textwidth]{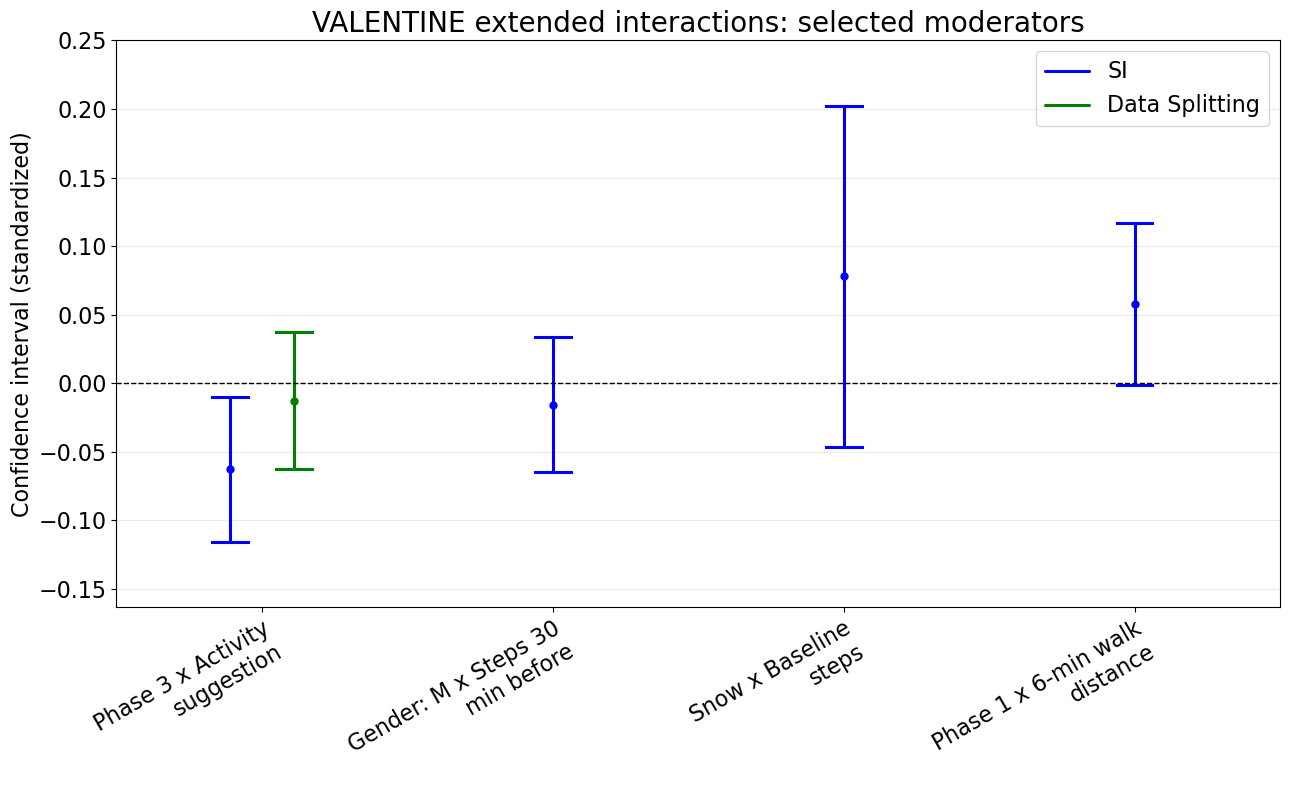}
\caption{Extended interaction setting ($94$ candidate moderators): $90\%$
selective confidence intervals for the selected moderators. Our method selects
$4$ moderators with average length $0.143$; data splitting selects $7$ moderators
with average length $0.259$; polyhedral produces infinitely long intervals and is
hence omitted.}
    \label{fig:ext94}
\end{figure}

The main-paper comparison (polyhedral intervals infinite; our method finite and
shorter on average than splitting) holds in both remaining settings; the robust
significant moderators for all three are collected in
Table~\ref{tab:valentine_significant}. In the \emph{baseline} $25$-moderator set
(\Figref{fig:main25}), included as an interpretable case where selection is not
strictly necessary, our method flags four significant moderators---weekend $(+)$,
study phases $2$ and $3$ $(-)$, and prior-week step variability $(-)$---whereas
data splitting distinguishes none of the shared moderators from zero, a direct
consequence of its smaller inference fold. In the \emph{extended} $94$-moderator
set (\Figref{fig:ext94}) our method isolates the negative
Phase\,3\,$\times$\,activity-suggestion effect---significant under our method in
both the phase and extended settings---while data splitting flags a different set
of context interactions; this within-method recurrence shows that the procedure
remains stable and parsimonious as the candidate set grows.

\begin{table}[t]
\centering
\caption{Robust significant moderators ($90\%$ selective confidence intervals
excluding zero; sign of the moderated effect in parentheses), by candidate
moderator set. Polyhedral intervals are omitted as they are infinite here.}
\label{tab:valentine_significant}
\begin{tabular}{l p{5.2cm} p{4.6cm}}
\toprule
 & \multicolumn{2}{c}{Robust significant moderators} \\
\cmidrule(lr){2-3}
Setting & By SI & By data splitting \\
\midrule
Main effects (25) &
  Weekend $(+)$, Step-SD prior week $(-)$, Phase 2 time $(-)$, Phase 3 time $(-)$ &
  --- \\
Phase--time (58) &
  Phase 1 $\times$ Weekend $(+)$, Phase 3 $\times$ Activity $(-)$ &
  Phase 1 $\times$ Weekend $(+)$ \\
Extended (94) &
  Phase 3 $\times$ Activity $(-)$ &
  Indoor $\times$ Agg.\ walking distance $(+)$,
  Indoor $\times$ Agg.\ prior-week steps $(-)$,
  Snow $\times$ Agg.\ prior-week steps $(-)$ \\
\bottomrule
\end{tabular}
\end{table}

\paragraph{Reliability and sensitivity.}
Three caveats apply. First, we analyze complete cases; excluding observations or
participants with missing values can bias estimates if missingness is related to
context, availability, or the outcome, so the selected set is conditional on
availability. Second, selection depends on the single $30\%/70\%$ split and on
$\lambda$; because several effects sit near the threshold, a different split or
penalty can change borderline moderators, so effects stable across settings and
methods (Table~\ref{tab:valentine_significant}) should be weighted most heavily.
Third, data splitting uses a smaller inference fold, so its intervals are wider
and, for rare-context moderators, its robust variance can be unstable, whereas
our full-data selective adjustment avoids both.

\begin{longtable}[htbp]{lp{0.7\textwidth}}
\caption{Descriptions of variables in the dataset.} \label{tab:datavariables} \\
\toprule
\textbf{Variable} & \textbf{Description} \\
\midrule
\endfirsthead
\toprule
\textbf{Variable} & \textbf{Description} \\
\midrule
\endhead
\midrule
\multicolumn{2}{r}{\textit{Continued on next page...}} \\
\midrule
\endfoot
\bottomrule
\endlastfoot
\texttt{ParticipantIdentifier} & Unique participant identifier. \\
\texttt{Value\_tran} & Log-transformed post-notification step count (outcome). \\
\texttt{Notification\_c} & Centered treatment (notification) indicator. \\
\texttt{TimeEnrolled\_days} & Days enrolled in the study. \\
\texttt{Phases} & Rehabilitation / study phase. \\
\texttt{Value\_30min\_before} & Step count in the $30$ minutes before notification. \\
\texttt{NotificationType} & Notification category (e.g.\ morning, afternoon). \\
\texttt{Baseline\_steps} & Steps during the baseline period. \\
\texttt{IsWeekend} & Indicator (1 = weekend). \\
\texttt{IsIndoor} & Indicator for indoor activity (1 = indoor). \\
\texttt{IsLossFramed} & Indicator for loss-framed message (1 = loss-framed). \\
\texttt{IsSnow} & Indicator for snow (1 = snow). \\
\texttt{IsActivity} & Indicator for message content type: $1$ = activity-suggestion message (e.g.\ prompting a walk), $0$ = anti-sedentary message (e.g.\ prompting a brief stretch or stand-up break). \\
\texttt{Value\_30min\_before\_tran} & Log step count $30$ min before notification. \\
\texttt{AgeEnrollment\_years} & Age at enrollment (years). \\
\texttt{Gender} & Participant gender. \\
\texttt{Race} & Participant race/ethnicity. \\
\texttt{ExerciseTimeAgg\_min} & Aggregated exercise time (minutes). \\
\texttt{WalkDistanceAgg\_m} & Aggregated walking distance (meters). \\
\texttt{StepsAgg\_priorweek} & Aggregated step count over the prior week. \\
\texttt{Value\_tran\_sd\_week} & SD of transformed step counts, current week. \\
\texttt{Value\_tran\_sd\_priorweek} & SD of transformed step counts, prior week. \\
\texttt{Distance\_m\_0} & $6$-minute walk distance at baseline (meters). \\
\end{longtable}

\bibliographystyle{plainnat}
\bibliography{references}

\end{document}